%% file: Paper-Preprint.tex
\documentclass[aip,jap,reprint,floatfix]{revtex4-1}

\usepackage{amsmath,amsfonts, mathrsfs, graphicx, amsthm, geometry}
\usepackage{multirow}

\newcommand{\grad}{\nabla}	
\newcommand{\D}[1]{\,\mathrm{d}#1}
\newcommand{\DD}[2]{\frac{\mathrm{d}#1}{\mathrm{d}#2}}
\newcommand{\DDP}[2]{\frac{\partial #1}{\partial #2}}

\begin{document}


\title{Tuning gain and bandwidth of traveling wave tubes using metamaterial beam-wave interaction structures}

\author{Robert Lipton}
\email[]{lipton@math.lsu.edu}
\affiliation{Department of Mathematics, Louisiana State University}

\author{Anthony Polizzi}
\email[]{polizzi@math.lsu.edu}
\affiliation{Department of Mathematics, Louisiana State University}

\date{\today}

\begin{abstract}
We employ metamaterial beam-wave interaction structures for tuning the gain and bandwidth of short traveling wave tubes. The interaction structures are made from metal rings of uniform cross section, which are periodically deployed along the length of the traveling wave tube. The aspect ratio of the ring cross sections are adjusted to control both gain and bandwidth. The frequency of operation is controlled by the filling fraction of the ring cross section with respect to the period. 
\end{abstract}


\maketitle

\section{Introduction}

From an operational viewpoint, a traveling wave tube  amplifier can be thought of as a 
cylindrical dielectrically loaded waveguide with an electron beam running through its center. The electron motion is parallel to the waveguide and confined by a strong uniform magnetic field applied along the beam. The beam is surrounded by a dielectric jacket separated from the beam by vacuum. When the dielectric constant is larger than unity, it is possible to get amplification from the traveling wave tube (TWT)  \cite{Schachter,Nation}.
Unfortunately, most dielectric materials are insufficient for high power applications and break down after a few operational cycles. On the other hand  Shiffler,  Luginsland, and Watrous\cite{Shiffler} propose  sub-wavelength all-metal interaction structures that effectively act as a dielectric medium with dielectric constant greater than unity. This provides the opportunity for design of TWTs with metal beam-wave interaction structures.
 
In this paper, we investigate the influence of metal interaction structures on the anisotropy of the effective dielectric tensor and the tune-ability of gain, bandwidth, and frequency of operation for short TWT amplifiers. Along the way, it is shown that effective dielectric properties arise naturally and in a systematic way by applying a two-scale asymptotic expansion for the solution of Maxwell's equations describing the beam-wave interaction inside the TWT. 

\begin{figure}[htbp]
	\begin{center}
	\scalebox{.8}{\includegraphics{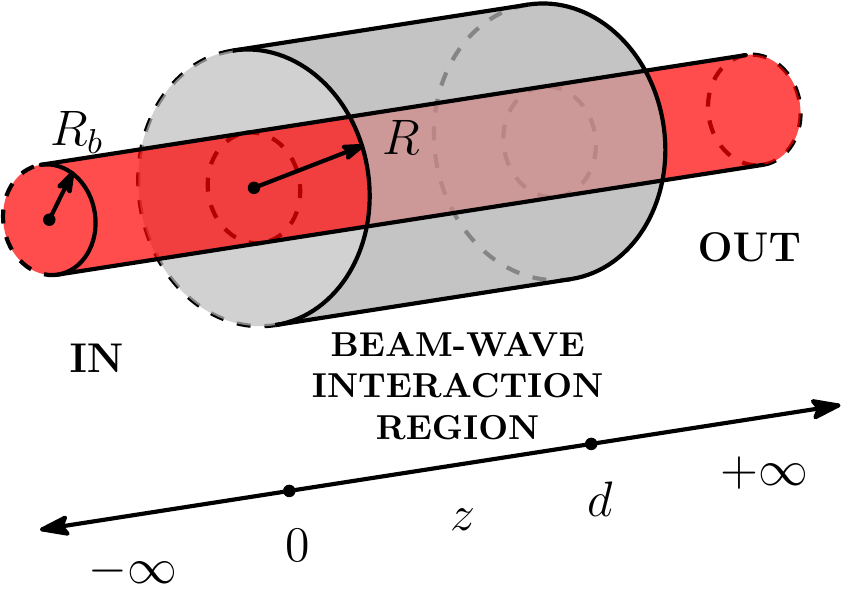}}
	\caption{Traveling wave tube amplifier with length $d$ fed with an electron beam of radius $R_b$ traveling longitudinally from $z=-\infty$}
	\label{device}
	\end{center}
	\end{figure}

Here we study the TWT  described in the work of Sch\"achter,  Nation, and Kerslick\cite{Nation}. This TWT is a short Cerenkov system comprised  of three components: a feeding waveguide, a finite length TWT amplifier region, and an output waveguide (see Figure \ref{device}). The entire system is excited by a generator. Each of these components have different characteristic impedances, and reflections can occur at both input and output ends of the amplifier. The objective is to characterize the influence of the geometry of the interaction structure on the transmission pattern as well as its effect on the gain and bandwidth of transmission peaks. We follow Sch\"acter {\em{et. al.}}\cite{Nation} and transmission patterns are calculated using the incident and reflected waves in the feeding waveguide, the dominant interacting modes inside the TWT amplifier region, and the space-charge waves in the output waveguide emitted from the amplifier region. In Section \ref{plots}, the transmission coefficient is depicted as a function of frequency for a collection of different
all-metal beam-wave interaction structures. To fix ideas, we study two classes of geometries associated with the interaction structure: rings with lonzenge-shaped cross sections, and rings with ellipsoidal cross sections. Inclusion geometries are indexed by their filling fraction relative to the period cell and their aspect ratio related to the eccentricity of their shape (see Figure \ref{crossSections} below). For reference we calculate the transmission coefficient using the isotropic dielectric constant chosen in  Sch\"acter {\em{et. al.}}\cite{Nation} (see Figure \ref{isotropic}) and use it as a benchmark to demonstrate the effect of varying the metal interaction structure on the performance of the TWT, see Figures \ref{isotropic} through \ref{ellipsoids80}.

The numerical calculations suggest the following trends.  
The amplifier's  frequency of operation can be strongly influenced  by altering the filling fraction of metallic rings. Here, higher filling fractions are seen to lower the operational frequencies as well as reducing the frequency range over which the TWT functions as an amplifier.  
It is found that the  aspect ratio associated with the cross-sectional shape of the rings can be  used to tune the gain and bandwidth of  the device. Higher gain is found to be associated with cross-sectional shapes with eccentricity along the longitudinal direction parallel to the beam. 
It is seen that the band width of the gain region drops with the eccentricity.

Our approach is organized as follows.  In the third section we ignore finite length effects and carry out the dispersion analysis for an infinitely long TWT amplifier region loaded with a sub wave length all-metal interaction structure. Here we apply the methods of two-scale asymptotic analysis \cite{papanicolaou,sanchezpalencia} to the Maxwell system used to model beam wave interaction inside the infinitely long amplifier. We focus on TM modes, and the asymptotic analysis delivers a leading order theory, from which we recover the leading order dispersion relation for the amplifier. This dispersion relation is expressed in terms of the anisotropic effective dielectric properties associated with the all-metal interaction structure. This dispersion relation is used to calculate the wave impedances associated with the spatially growing space-charge wave, the spatially decaying space-charge wave, and the oscillating space-charge wave inside the TWT beam-wave interaction region. In the fourth section we formulate the electrodynamic problem for the short TWT within the transmission line approximation. Here the transmission line approximation is posed in terms of  wave impedances derived from the  dispersion relations for the  space-charge waves obtained in Section \ref{dispersionRelations}. The resulting transmission patterns for the all-metal beam-wave interaction structures are displayed and analyzed in Section \ref{plots}.

\section{Device description}
\begin{figure}[htbp]
	\begin{center}
	\scalebox{.7}{\includegraphics{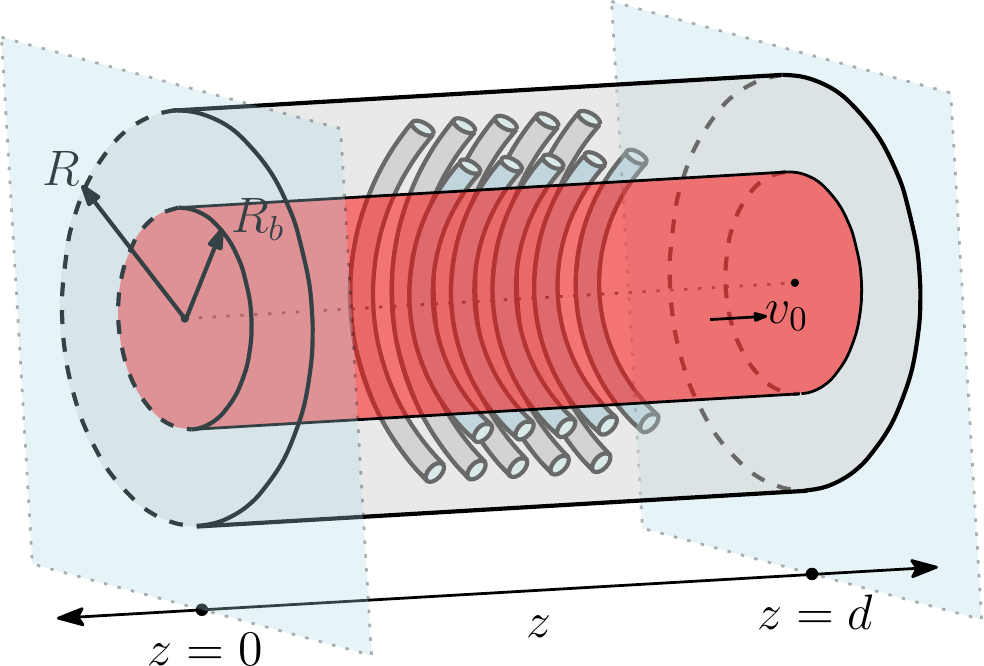}}
	\caption{The outer region, $R_b<r<R$ contains the metal beam-wave interaction structure given by the periodic distribution of rings of constant cross section.}
	\label{InputOutput}
	\end{center}
	\end{figure}

The short TWT amplifier consists of three regions, the input waveguide  $-\infty<z<0$, the beam-wave interaction region $0\leq z\leq d$ and the output waveguide $d<z<\infty$. Each region is a circular cylindrical waveguide enclosed by perfectly conducting walls (see Figure \ref{device}).  
Both the input waveguide and the output waveguide have electromagnetic properties  associated with vacuum, $\epsilon_0 = 8.85 \times 10^{-12}$ F/m and $\mu_0 = 4\pi \times 10^{-7}$ H/m and are of radius $R_b$.  
The interaction region, $0 \leq z \leq d$ has outer radius $R>R_b$ and contains the metal beam-wave interaction structure described by a periodic arrangement of metal rings of constant cross section (see Figure \ref{InputOutput}). The interaction structure is confined to the annular region  $R_b<r<R$. Away from the rings the  electromagnetic properties are given by the vacuum values.

Far to the left $(z\ll 0)$ a generator excites the lowest symmetric transverse magnetic (TM) mode in the input waveguide at a given frequency $\omega$ above cutoff, \emph{i.e.},\, $\omega > cp_1/R$, where $c$ is the vacuum speed of light, and $p_1$ is the first zero of the zero-order Bessel function of first kind, $J_0(p_1)=0$. 
In this region, \emph{i.e.},\, $-\infty < z < 0$, the azimuthal component of the steady-state magnetic field  is given (in cylindrical polar coordinates) by
\begin{equation}
H_{\theta}(r,z,\omega) =  A_0 J_1(\frac{p_1}{R_b}r)(e^{-ik_1 z}), 
\end{equation}
from which the component electric fields $E_r$, $E_z$ are recovered by means of the relations 
\begin{eqnarray}
E_r &=& \frac{i}{\omega \epsilon_0} \DDP{}{z}H_{\theta},  \\* 
E_z &=& -\frac{i}{\omega \epsilon_0} \frac{1}{r} \DDP{}{r}(r H_{\theta}), 
\end{eqnarray}
where $k_1 = \sqrt{(\omega/c)^2-(p_1/R)^2}$ denotes the wavenumber of the incident wave and $A_0$ is the amplitude.

At the entrance to the input waveguide $(z=-\infty)$ a beam of electrons is injected into the system with average velocity $v_0$ and average density $n_0$. The spatial distribution of electrons is uniform along the transverse direction and a strong uniform magnetic field is applied along the beam restricting the electron velocity to be parallel to the longitudinal direction. 
Here it is assumed that the longitudinal momentum distribution is narrow so the electron dynamics is described by the hydrodynamic approximation \cite{Chu,Pierce,Nation,Schachter}. 
Within the hydrodynamic approximation, the current density is related to the electric field  by
\begin{align}
\vec{J} &=J_z \vec{e}_z  = -i\omega \epsilon_0 \frac{{\omega_p}^2}{\gamma^3(\omega - v_ok)^2}E_z\vec{e}_z,
\label{hydromax}
\end{align}
where $\beta = v_0/c$, $\gamma = (1-\beta^2)^{-1/2}$, and $\omega_p$ is the plasma frequency, ${\omega_p}^2=e^2n_0/(m\epsilon_0)$ and $m$ is the electron mass \cite{Pierce}.

The short TWT described here is considered in earlier work \cite{Nation} under the assumption that the dielectric properties inside the interaction region were specified by the dielectric constant $\epsilon=3.5$ for $R_b<r<R$. The analysis presented here is limited to the case where the electron beam almost fills $R_b$ so that we can compare the transmission patterns displayed here with those of Sch\"acter \emph{et. al.} \cite{Nation}.  The electrodynamic problem for the short TWT is formulated within the transmission line approximation and presented in Section \ref{transmissionLine}. The next section develops the dispersion analysis for an infinitely long beam-wave interaction structure. This analysis provides the wave impedances used to model beam-wave interaction within the  transmission line approximation.

\section{Dispersion analysis ignoring end effects: An infinitely long beam-wave interaction structure} \label{dispersionRelations}

\begin{figure}[htbp]
	\begin{center}
	\scalebox{.8}{\includegraphics{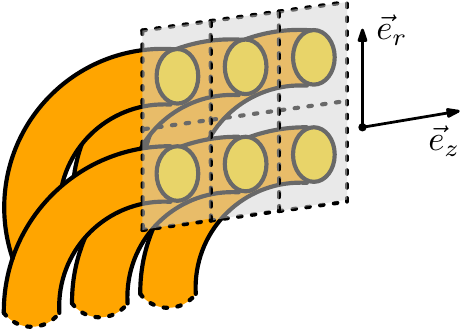}}
	\caption{Periodic, concentric, metallic rings embedded in a host material whose dielectric properties are those of a vaccum}
	\label{inclusions}
	\end{center}
	\end{figure}

We now perform a dispersion analysis on an infinitely long TWT amplifier with a sub wavelength metal beam-wave
interaction structure. Here Maxwell's equations inside the beam $0<r<R_b$, $-\infty< z<\infty$ are given by

\begin{align}
                 \label{maxinbeam}
		\grad \times \vec{E} &= -i\omega \vec{B} \\
		\grad \times \vec{B} &= i\omega \mu_0 \epsilon_0 \vec{E} + \mu_0 \vec{J} \\
		\grad \cdot \vec{B} & = 0
	\end{align}
where the beam current density $\vec{J}$ is given by \eqref{hydromax}.
In the beam-wave interaction region $R_b<r<R$, $-\infty<z<\infty$,  Maxwell's equations outside the metal rings
are given by
\begin{align}
                 \label{maxinwave}
		\grad \times \vec{E} &= -i\omega \vec{B} \\
		\grad \times \vec{B} &= i\omega \mu_0 \epsilon_0 \vec{E}  \\
		\grad \cdot \vec{B} & = 0
	\end{align}
on the boundary of the metal rings $\vec{n}\times\vec{E}=0$, where $\vec{n}$ is the unit inward pointing normal on the ring surface. On $r=R$ we have $\vec{n}\times\vec{E}=0$, where $\vec{n}$ is the unit normal pointing out of the waveguide. The ring geometry is symmetric with respect to the $\theta$ variable and periodic in the variables $(r,z)$ see, Figure \ref{inclusions}. 

We find the dispersion relation for TM modes of the form $\vec{B}=\vec{e}_\theta \psi_\theta(r,z)e^{-ikz}$. Substitution of this mode into \eqref{maxinbeam} and \eqref{maxinwave} delivers the equation for $\vec{B}$ in the beam, $0<r<R_b$,
\begin{align}
                 \label{maxbinbeam}
		\grad \times\mathbf{\epsilon}^{-1}(\omega,k) \grad \times\vec{B} &= \frac{\omega^2}{c^2}\vec{B},
\end{align}
with the dielectric tensor
\begin{eqnarray}
\label{timespacedispersion}
&&\mathbf{\epsilon}^{-1}(\omega,k)=\vec{e}_r\otimes\vec{e}_r+\vec{e}_\theta\otimes\vec{e}_\theta+\notag\\
&&+\vec{e}_z\otimes\vec{e}_z\left(1-\frac{\omega_p^2}{\gamma^3(\omega-v_0k)^2}\right)^{-1},
\end{eqnarray}
and in the interaction region $R_b<r<R$,
\begin{align}
                 \label{maxbinwave}
		\grad \times \grad \times \vec{B} &= \frac{\omega^2}{c^2}\vec{B},
\end{align}
with $\vec{n}\times\nabla\times\vec{B}=0$ on the surface of the metal rings and on $r=R$.

Here the goal is to identify the pairs $\omega,k$ associated with gain inside the TWT, \emph{i.e.},\, complex wave numbers $k$  with positive imaginary part for real frequencies $\omega$. It is clear from \eqref{timespacedispersion} that the beam is represented by a dielectric tensor associated with both spatial and temporal dispersion.

In the next subsection we develop two-scale asymptotic expansions for TM modes and recover the leading order theory
for dispersion inside TWTs containing sub wavelength periodic metallic interaction structures. These are expressed in terms of the components of an anisotropic effective dielectric tensor. In the following subsection we simplify these to obtain a Pierce like dispersion relation for characterizing growing TM modes associated with gain inside the TWT.

 \subsection{Anisotropic effective  properties and the dispersion relation}
 \label{anisoeffect}
 In this section we recover the leading order behavior for TM waves $\vec{B}=\vec{e}_\theta B_\theta=\vec{e}_\theta\psi(r,z)e^{-ikz}$ of \eqref{maxbinbeam} and \eqref{maxbinwave} using two-scale asymptotic expansions. 
 
Within the interaction region, $R_b<r<R$ and for any choice of, $0\leq \theta< 2\pi$, the ring cross sections are the same and are distributed periodically inside the rectangle $R_b<r<R$, $-\infty<z<\infty$ see, Figure \ref{inclusions}. The interaction structure is constructed so that the ratio between the side length of the period and $R$ is given by  $\varepsilon=1/n$ where $n$ is an integer, $n>1$. Here we suppose $\varepsilon$ is small with respect to $R$ and the metamaterial interaction structure is constructed so that the interval $R_b<r<R$ contains an integral number of periods with side length $p=\varepsilon R$. To carry out the two-scale expansion we rescale both $r$ and $z$ coordinates by $1/R$ and introduce the dimensionless coordinates $\tilde{r}=r/R$, $\tilde{z}=z/R$. The coordinates of the center of each period cell of side length $\varepsilon$ contained within $R_b/R<\tilde{r}<1$, $-\infty<\tilde{z}<\infty$ is given by $(\varepsilon\left[\frac{\tilde{r}}{\varepsilon}\right],\varepsilon\left[\frac{\tilde{z}}{\varepsilon}\right])$, where $\left[s\right]$ denotes the integer part of the number $s$ (the nearest integer greater than $s$). Any point $(\tilde{r},\tilde{z})$ within the rectangle $R_b/R<\tilde{r}<1$, $-\infty <\tilde{z}< \infty$, can be written as $(\varepsilon\left[\frac{\tilde{r}}{\varepsilon}\right]+\varepsilon \rho,\varepsilon\left[\frac{\tilde{z}}{\varepsilon}\right]+\varepsilon y)$ where $(\rho,y)=(\frac{\tilde{r}}{\varepsilon}-\left[\frac{\tilde{r}}{\varepsilon}\right],\frac{\tilde{z}}{\varepsilon}-\left[\frac{\tilde{z}}{\varepsilon}\right])$  lie inside the unit period cell $Y$ given by $-1/2<\rho<1/2$, $-1/2<y<1/2$. 
 
Writing $\tilde{\psi}(\tilde{r},\tilde{z})=\psi(R\tilde{r},R\tilde{z})$ we develop the 
two-scale expansion for $\tilde{\psi}(\tilde{r},\tilde{z})$  given  by
 \begin{eqnarray}
 \label{twoscalefield}
\tilde{\psi}(\tilde{r},\tilde{z}) &=& \tilde{\psi}(\tilde{r},\rho,y)  = \tilde{\psi}_0(\tilde{r})+\varepsilon\tilde{\psi}_1(\tilde{r},\rho,y)  \nonumber \\*
&+& \varepsilon^2\tilde{\psi}_2(\tilde{r},\rho,y)+\cdots,
 \end{eqnarray}
 where  $\tilde{\psi}(\tilde{r},\rho,y)$ is periodic in $(\rho,y)$ with period cell $Y$.
 For $\omega$ fixed the wave number is expanded as
 \begin{equation}
 \label{vavenumexansion}
 k=k_0+\varepsilon k_1+\varepsilon^2 k_2+\cdots.
 \end{equation}
 Changing coordinates $r\rightarrow\tilde{r}$, $z\rightarrow\tilde{z}$, in \eqref{maxbinbeam}, \eqref{maxbinwave}, substitution of the expansions \eqref{twoscalefield}, \eqref{vavenumexansion}, and equating like powers of $\varepsilon$ delivers the homogenized differential equation for $\tilde{\psi}_0(\tilde{r})$.  Changing back to $(r,z)$ coordinates, writing $\psi_0(r)=\tilde{\psi_0(r/R)}$ delivers the leading order behavior for $\vec{B}$ given by
 \begin{eqnarray}
  \vec{B}=\vec{e}_\theta \psi_0(r)e^{-ik_0z}+O(\varepsilon). 
  \label{leadingorder}
  \end{eqnarray}
 Where the equations and boundary conditions satisfied by  $\psi_0(r)$ are given by
 \begin{eqnarray}
                 \label{maxbinbeamh}
		\grad \times\mathbf{\epsilon}^{-1}(\omega,k_0) \grad \times\vec{e}_\theta\psi_0(r)e^{-ik_0z}  \nonumber \\*
		 =\frac{\omega^2}{c^2}\vec{e}_\theta\psi_0(r)e^{-ik_0z}
\end{eqnarray}
within the beam, $0<r<R_b$,
where the dielectric tensor is 
\begin{eqnarray}
\label{timespacedispersionh}
&& \mathbf{\epsilon}^{-1}(\omega,k_0) \nonumber  = \vec{e}_r\otimes\vec{e}_r  \\ &&+  \vec{e}_z\otimes\vec{e}_z\left(1-\frac{\omega_p^2}{\gamma^3(\omega-v_0k_0)^2}\right)^{-1},
\end{eqnarray}
and in the interaction region, $R_b<r<R$,
\begin{eqnarray}
                 \label{maxbinwaveh}
		\grad \times (\mathbf{\epsilon}^{\rm eff})^{-1}\grad \times \vec{e}_\theta\psi_0(r)e^{-ik_0z}  \nonumber \\*
 = \frac{\omega^2}{c^2}\vec{e}_\theta\psi_0(r)e^{-ik_0z}.
\end{eqnarray}
$\vec{B}$ also satisfies the boundary conditions that $\vec{n}\times(\mathbf{\epsilon}^{\rm eff})^{-1}\nabla\times(\vec{e}_\theta\psi_0(r)e^{-ik_0z})$ vanish at $r=R$ and at $r=R_b$:
\begin{eqnarray}
  \vec{n}\times(\mathbf{\epsilon}^{\rm eff})^{-1}\nabla & \times & \left. (\vec{e}_\theta\psi_0(r)e^{-ik_0z}) \right\vert_{r=R_b^+} =  \nonumber \\*
   \vec{n} \times\mathbf{\epsilon}^{-1}(\omega,k_0)\nabla  & \times & \left. (\vec{e}_\theta\psi_0(r)e^{-ik_0z}) \right\vert_{r=R_b^-}.
\label{beaminteractinterface}
\end{eqnarray}
The effective dielectric tensor $\mathbf{\epsilon}^{\rm eff}$ is defined by local field problems defined on the unit period cell $Y$ containing the ring cross section.  The boundary of the ring cross section is denoted by $\partial P$, and the part of the unit cell containing vacuum surrounding the ring is denoted by $M$. The unit outward pointing normal to $\partial P$ is denoted by
$\vec{n}=n_r\vec{e}_r+n_y\vec{e}_z$. The effective dielectric tensor is written as
\begin{eqnarray}
\label{efftensor}
\mathbf{\epsilon}^{\rm eff}={\epsilon}^{\rm eff}_{rr}\vec{e}_r\otimes\vec{e}_r+{\epsilon}^{\rm eff}_{zz}\vec{e}_z\otimes\vec{e}_z.
\end{eqnarray}
Here, the components of ${\epsilon}^{\rm eff}$ are given by
\begin{eqnarray}
&&{\epsilon}^{\rm eff}_{rr}=\int_M(\nabla\varphi^\rho(\rho,y)+\vec{e}_r) \D{\rho} \D{y} \label{err},\\
&&{\epsilon}^{\rm eff}_{zz}=\int_M(\nabla\varphi^y(\rho,y)+\vec{e}_z) \D{\rho} \D{y} \label{eyy},
\end{eqnarray}
where $\varphi^\rho$ and $\varphi^y$ are periodic with unit period $Y$ and are, respectively, solutions of
\begin{eqnarray}
\nabla^2\varphi^\rho=0 & \hbox { and } & \nabla^2\varphi^y=0,
\label{equlib}
\end{eqnarray}
in the vacuum region $M$ and
\begin{eqnarray}
&& \left. \vec{n}\cdot(\nabla\varphi^\rho+\vec{e}_r)\right\vert_{\partial P}=0, \label{boundary1}\\
&&\left. \vec{n}\cdot(\nabla\varphi^y+\vec{e}_z)\right\vert_{\partial P} =0,
\label{boundary2}
\end{eqnarray}
on the surface of the ring cross sections $\partial P$. Here $\nabla=\vec{e}_r\partial_\rho+\vec{e}_z\partial_y$.
The effective coefficients $\epsilon^{\rm eff}_{rr}$ and $\epsilon^{\rm eff}_{zz}$ are computed numerically for the different cross-sectional shapes.
An outline of the two-scale approach used to recover the homogenized problem is provided in the Appendix.

The $B_\theta$  field inside  the beam
$0<r<R_b$ is given by

	\begin{equation}
		B_{\theta}=C_0J_1(\nu_b r)e^{-ik_0z}.
		\label{inthebeam}
	\end{equation}
	
	And, in the metamaterial $R_b<r<R$, we apply the boundary condition on the outer wall of the TWT to get
	\begin{equation}
		B_{\theta}=C_2T_1(\nu_{d} r)e^{-ik_0z},
		\label{inthedielectric}
	\end{equation}
       with 
	\begin{eqnarray}
		T_1(\nu_{d} r)=J_1(\nu_d r)Y_0(\nu_d R) \nonumber \\*
		-Y_1(\nu_d r)J_0(\nu_d R),
		\label{predispersion}
	\end{eqnarray}
	where
	\begin{eqnarray}
		\nu_b^2=\left(1-\frac{\omega_p^2}{v_0\gamma^3(\frac{\omega}{v_0}-k_0)^2}\right)^{-1} \nonumber \\*
		 \times \left(\frac{\omega^2}{c^2} -k_0^2\right),
	\end{eqnarray}
	and
	\begin{equation}
		\nu_d^2 =\epsilon^{\mathrm{ef{}f}}_{zz}\frac{\omega^2}{c^2}-\frac{\epsilon^{\mathrm{ef{}f}}_{zz}}{\epsilon^{\mathrm{ef{}f}}_{rr} }k_0^2.
	\end{equation}
	
	The dispersion relation follows from the transmission condition \eqref{beaminteractinterface}. From now on, we focus on the leading order behavior and write $k=k_0$. The dispersion relation between frequency $\omega$ and propagation constant $k$ is of the form
                \begin{equation}
		D_{\mathrm{act}}(\omega,k)=0,
		\label{disperse}
		\end{equation}
where $D_{\mathrm{act}}$ is defined as
	\begin{eqnarray}
		D_{\mathrm{act}}(\omega,k)=D_{\mathrm{beam}}(\omega,k)F_{\mathrm{beam}}(\omega,k) \nonumber \\*
		+ \epsilon^{\mathrm{ef{}f}}_{zz}D_{\mathrm{pass}}(\omega,k)F_{\mathrm{pass}}(\omega,k).
		\label{components}
		\end{eqnarray}
The four components of $D_{\mathrm{act}}$ are
		\begin{eqnarray}
		D_{\mathrm{beam}}=\epsilon^{\mathrm{ef{}f}}_{zz}\nu_dT_0(\nu_d R_b)Y_1(\nu_{\mathrm{vac}} R_b)  \nonumber \\*
		-\nu_{\mathrm{vac}}Y_0(\nu_{\mathrm{vac}} R_b)T_0(\nu_d R_b),
		\label{dbeam}
		\end{eqnarray}
	        \begin{eqnarray}
		F_{\mathrm{beam}}=\epsilon_b^{-1/2}J_1(\nu_b R_b)J_0(\nu_{\mathrm{vac}} R_b)  \nonumber \\*
		-J_0(\nu_b R_b)J_1(\nu_{\mathrm{vac}} R_b),
		\label{fbeam}
		\end{eqnarray}
	        \begin{eqnarray}
		D_{\mathrm{pass}}=-\nu_{\mathrm{vac}}\epsilon_{zz}^{\mathrm{ef{}f}}J_0(\nu_{\mathrm{vac}} R_b)T_1(\nu_d R_b)  \nonumber \\*
		-\nu_d J_1(\nu_{\mathrm{vac}} R_b)T_0(\nu_d R_b),
		\label{dpass}
	        \end{eqnarray}
	         \begin{eqnarray}
		F_{\mathrm{pass}}=-J_1(\nu_b R_b)Y_0(\nu_{\mathrm{vac}}R_b)  \nonumber \\*
		+\epsilon_b^{-1/2}J_0(\nu_bR_b)Y_1(\nu_{\mathrm{vac}} R_b),
		\label{fpass}
	        \end{eqnarray}
with $\epsilon_b$ defined as
		\begin{equation}
		\epsilon_b=\left(1-\frac{\omega_p^2}{v_0\gamma^3(\frac{\omega}{v_0}-k)^2}\right)^{-1} 
		\end{equation}
and 
		 \begin{eqnarray}
		 T_0(\nu_d R_b)=J_0(\nu_d R_b)Y_0(\nu_d R)  \nonumber \\*
		  -Y_0(\nu_d R_b)J_0(\nu_d R).
		 \end{eqnarray}

 \subsection{A Pierce-like approach to dispersion}

For  $\omega$ fixed and in the absence of the beam the propagation constant $k=k^{(0)}$ is the root of the passive structure
	dispersion relation
	\begin{eqnarray}
		D_{\mathrm{pass}}(\omega,k^{(0)})=0
		\label{passive}		
	\end{eqnarray}
		Set 
		\begin{eqnarray}
	         \alpha&=&\frac{\omega_p^2}{v_0^2\gamma^3(\frac{\omega}{v_0}-k)^2}.	
	         \label{alpha}
	\end{eqnarray}
Following Sch\"achter, Nation and Kerslick assume the beam does not significantly effect the fields in the waveguide and suppose $\alpha \ll 1$. Fixing $\omega$, expand $D_{\mathrm{act}}$ as a function of $\alpha$ and $k$ in a Taylor series about $\alpha=0$ and $k=k^{(0)}$. Writing $k=k^{(0)}+q$ gives
	\begin{eqnarray}
		D_{\mathrm{act}}(\alpha,k) &=& D_{\mathrm{act}}(0,k^{(0)})  +\partial_k D_{\mathrm{act}}(0,k^{(0)})q   \nonumber \\*
		&+&\partial_\alpha D_{\mathrm{act}}(0,k^{(0)})\alpha +  o(\alpha,q).
		\label{active}
	\end{eqnarray}
	Neglecting higher order terms in the expansion for $D_{\mathrm{act}}$ we get the third order equation for $q$ given by
		\begin{eqnarray}
	         (\Delta k -q)^2q=-K^3,
	         \label{delta}	
	\end{eqnarray}
	        Where $\Delta k=\frac{\omega}{v_0}-k^{(0)}$ is the slip between the phase velocity of the beam and the propagation constant in
	        the passive structure and $K^3$ is the non-normalized Pierce factor given here by
	         \begin{eqnarray}
	         K^3&=&\left(\frac{e\eta_0 I}{mc^2(\beta\gamma)^3\pi R_b^2}\right)  \nonumber \\*
	          &\times& \frac{\partial_\alpha D_{\mathrm{act}}(0,k^{(0)})}{\partial_kD_{\mathrm{act}}(0,k^{(0)})}.
	         \label{piercefactor}
		\end{eqnarray}

		This equation has three roots: one real, and one pair of complex conjugates. Denoting the roots by $q_j$, $j=1,2,3$, the wave numbers for the beam wave interaction structure are given by $\kappa_j=k^{(0)}+q_j$, $j=1,2,3$. The complex root $q$ of the third order equation with positive imaginary part $\mathrm{Im}\{q\}$   corresponds to the growing wave and  is a measure of the gain per unit length associated with 
		the infinitely long TWT  \cite{Pierce2}. 
		 The gain seen over a distance $d$ within the infinitely long TWT  is $20\log\{\exp{(\mathrm{Im}\{q\}d)}\}$ and is displayed together with the transmission coefficient for the finite length device, $\mathrm{d}=15\mathrm{cm}$, for different metamaterial interaction structures in Figures \ref{isotropic} - \ref{ellipsoids80}. 
		 
		 We conclude noting that the effect of the all-metal interaction structure is encoded into the effective dielectric properties appearing in the dispersion relation \eqref{delta}.

\section{Electrodynamics inside a finite length TWT: transmission line model}
\label{transmissionLine}

The transmission and reflection for the short TWT system is calculated accounting for the interaction of all electromagnetic and space-charge waves present in the input waveguide, the interaction region, and in the output waveguide. For generic situations, this requires the use of an infinite number of modes in order to satisfy transmission conditions between waveguides. However, the system considered here is operated over a frequency range for which an exponentially growing  mode is excited within the interaction region. It is also assumed that the energy stored in the other modes are smaller than that in the growing mode.
With these caveats in mind, the electrodynamics for the short TWT is modeled within the transmission line approximation \cite{Nation}. 
Here voltage waves $V=V(z)$ along the transmission line are equivalent to the radial component of the electric field $E_r(z)$, while current waves $I=I(z)$ are equivalent to the azimuthal component of the magnetic field $H_\theta(z)$.

It is assumed that there are no incident space-charge waves, and that the voltage wave along the input transmission line
is given by an incident wave (of unit magnitude, excited by a generator) together with a reflected wave of amplitude $\rho$, with
\begin{equation}
V(z) = e^{-ik_{\mathrm{in}}z} + \rho e^{ik_{\mathrm{in}}z}
\end{equation}
in the region $-\infty < z < 0$. The wave impedance of the input line is 
\begin{eqnarray}
Z_{\mathrm{in}}=E_r/H_\theta=\eta_0ck_{\mathrm{in}}/\omega,
\label{intimpeadence}
\end{eqnarray}
with $\eta_0=\sqrt{\mu_0/\epsilon_0}$.
The current is equivalent to  the azimuthal component of the magnetic field and is given by
\begin{equation}
I(z) = \frac{1}{Z_{\mathrm{in}}}e^{-ik_{\mathrm{in}}z}  - \frac{\rho}{Z_{\mathrm{in}}} \rho e^{ik_{\mathrm{in}}z}
\end{equation}
for $-\infty < z < 0$,

Following  Sch\"achter,  Nation, and Kerslick\cite{Nation}, the electrodynamics within the interaction region, $0 \leq z \leq d$, is modeled by four waves 
\begin{equation}
V(z) = \sum_{j=1}^3A_je^{-i\kappa_jz} + A_4e^{ik^{(0)}z}
\end{equation}
with wave numbers
\begin{eqnarray}
\kappa_j = k^{(0)} + q_j, \quad j=1,2,3.
\label{spacecharge}
\end{eqnarray}
Here $k^{(0)}$ is the wave number for the infinite beam-wave interaction structure in the absence of the beam and is the solution of the cold structure dispersion relation $D_{\mathrm{pass}}(\omega,k^{(0)})=0$. The wave numbers $\kappa_j$ are associated with the  space-charge waves in the infinite beam-wave interaction structure, where $q_j$ are the roots of \eqref{delta}.
The associated wave impedances are given by
\begin{align}
\label{interactimpead}
Z_{\mathrm{bw}}^j &=\eta_0c\kappa_j/\omega, \quad j=1,2,3, \\
Z_{\mathrm{bw}}^4 &=-\eta_0ck^{(0)}/\omega.\label{imeractcool}
\end{align}
The current inside the interaction region is given by
\begin{eqnarray}
I(z) =\sum_{j=1}^3\frac{A_j}{Z_{\mathrm{bw}}^j} e^{-i\kappa_j z}  - \frac{ A_4}{Z_{\mathrm{bw}}^4} e^{ik^{(0)}z}.
\label{interactcurrent}
\end{eqnarray}

Along the output transmission line, the voltage wave consists of two space-charge waves emitted from the interaction region and one electromagnetic mode:
\begin{equation}
V(z) =  \tau e^{-ik_{\mathrm{out}}z} +  \sum_{j=2}^3 B_je^{-i\chi_j z},
\label{outputvoltage}
\end{equation}
for $d<z<\infty$. The wave number $k_{\mathrm{out}}=k_{\mathrm{in}}$ for the electromagnetic mode, and
the wave numbers of the emitted space-charge waves $\chi_2$, $\chi_3$ are given by \cite{Nation}
\begin{gather}
\chi_2 = \frac{\omega}{v_0} +  \frac{\omega_p}{\gamma^{3/2}v_0} \sqrt{1 + \left(\frac{cp_1}{\gamma \beta \omega R_b }\right)^2}, \label{forward}\\
\chi_3 = \frac{\omega}{v_0} -  \frac{\omega_p}{\gamma^{3/2}v_0} \sqrt{1 + \left(\frac{cp_1}{\gamma \beta \omega R_b}\right)^2}.\label{otherway}
\end{gather}
Their associated wave impedances are given by
\begin{align}
\label{outimpead}
Z_{\mathrm{out}} &=Z_{\mathrm{in}}, \\
Z_{\mathrm{out}}^j &=\eta_0c\chi_j/\omega, \quad  j=2,3, \label{outem}
\end{align}
and the current wave along the output line is given by
\begin{equation}
I(z) =  \frac{B_1}{Z_{\mathrm{out}}} e^{-ik_{\mathrm{out}}z} + \sum_{j=2}^3 \frac{B_j}{Z_{\mathrm{out}}^j}e^{-i\chi_j z}
\end{equation}
for $d<z<\infty$.

We conclude by writing down the beam dynamics associated with the hydrodynamic approximation.
The beam modulation is given by the electronic oscillation $\delta v(z)$ about the average beam velocity $v_0$  and is proportional to the electric field in the longitudinal direction \cite{Pierce,Chu,Nation}, \emph{i.e.},\,
\begin{align}
\delta v(z) &= -\frac{ie}{m\gamma^3(\omega - v_0k)}E_z(z) \notag\\ 
&= - \frac{ie}{m\gamma^3(\omega - v_0k)}\frac{1}{\eta_0 c k}E_r(z)\notag\\
&=  -\frac{ie}{m\gamma^3(\omega - v_0k)}\frac{1}{\eta_0 c k}V(z),\label{velperturb}
\end{align}
The perturbation of electron density about its average value $n_0$ is given by
\begin{eqnarray}
\delta n(z) &=&  \frac{n_0 k}{\omega-v_0k}\delta v(z) \label{denspertrub}  \nonumber \\*
&=&  -\frac{ien_0}{mc\eta_0\gamma^3(\omega - v_0k)^2}V(z).
\end{eqnarray}

\subsection{Solution of the transmission line approximation} 
The voltage and current waves are determined by imposing the continuity of $V(z),I(z),\delta v(z)$, and $\delta n(z)$ at $z=0$  and $z=d$.
Imposing continuity on the voltage, $V(z=0^-) = V(z=0^+)$, yields
\begin{equation}
1 + \rho = \sum_{j=1}^3 A_j + A_4.
\label{voltinout}
\end{equation}
Continuity of the current, $I(z=0^-) = I(z=0^+)$, gives
\begin{equation}
 \frac{1}{Z_{\mathrm{in}}} - \frac{\rho}{Z_{\mathrm{in}}} = \sum_{j=1}^3 \frac{A_j}{Z_{\mathrm{bw}}^j}  -   \frac{A_4}{Z_{\mathrm{bw}}^4} . 
\end{equation}
Continuity of the electron oscillation, $\delta v(z=0^-) = \delta v(z=0^+)$ delivers the condition
\begin{eqnarray}
 \frac{1}{k_{\mathrm{in}}(\omega - v_0k_{\mathrm{in}})} &+& \frac{\rho}{k_{\mathrm{in}}(\omega + v_0 k_{\mathrm{in}})}  \nonumber \\*
 =  \sum_{j=1}^3 \frac{A_j}{\kappa_j(\omega-v_0\kappa_j)}  &+& \frac{A_4}{k^{(0)}(\omega +v_0k^{(0)})}, 
\end{eqnarray}
and continuity of electron density, $\delta n(z=0^-) = \delta n(z=0^+)$ gives
\begin{eqnarray}
\frac{1}{(\omega - v_0k_{\mathrm{in}})^2} &-& \frac{\rho}{(\omega + v_0 k_{\mathrm{in}})^2}  \nonumber \\*
   =  \sum_{j=1}^3 \frac{A_j}{(\omega - v_0\kappa_j)^2} &-& \frac{A_4}{(\omega + v_0 k^{(0)})^2}.
\end{eqnarray}

Likewise, imposing  continuity conditions at output ($z=d$) yields, for voltage,
\begin{eqnarray}
\sum_{j=1}^3  A_je^{-i\kappa_jd} &+& A_4e^{ik^{(0)}d}  \nonumber \\*
 =  \tau e^{-ik_{\mathrm{in}}d} &+& \sum_{l=2,3} B_le^{-i\chi_l d}, 
\end{eqnarray}
for current,
\begin{eqnarray}
\sum_{j=1}^3 \frac{A_je^{-i\kappa_jd}}{Z_{\mathrm{bw}}^j}   &-&   \frac{A_4e^{ik^{(0)}d}}{Z_{\mathrm{bw}}^4}  \nonumber \\*
= \frac{\tau e^{-ik_{\mathrm{out}}d}}{Z_{\mathrm{out}}^1} &+&  \sum_{l=2,3} \frac{B_le^{-i\chi_l d}}{Z_{\mathrm{out}}^l}, 
\end{eqnarray}
for  electron oscillation,
\begin{eqnarray}
\sum_{j=1}^3 \frac{A_je^{-i\kappa_jd}}{\kappa_j(\omega - v_0\kappa_j)}  &+& \frac{A_4e^{ik^{(0)}d}}{k^{(0)}(\omega +v_0k^{(0))}}  \nonumber \\*
=\frac{\tau e^{ik_{\mathrm{out}}d}}{k_{\mathrm{out}}(\omega -v_0k_{\mathrm{out}})}  &+&  \sum_{l=2,3}  \frac{B_le^{-i\chi_l d}}{\chi_l(\omega - v_0\chi_l)}
\end{eqnarray}
and, for  electron density,
\begin{eqnarray}
 \sum_{j=1}^3 \frac{A_je^{-i\kappa_jd}}{(\omega - v_0\kappa_j)^2} &-& \frac{A_4 e^{ik^{(0)}d}}{(\omega + v_0 k^{(0)})^2}    \nonumber \\* 
 = \frac{\tau e^{-ik_{\mathrm{out}}d}}{(\omega - v_0 k_{\mathrm{out}})^2} 
&+&  \sum_{l=2,3}  \frac{B_le^{-i\chi_l d}}{(\omega - v_0\chi_l)^2}. \label{lastline}
\end{eqnarray}

We apply an iterative numerical method to solve the system \eqref{voltinout}-\eqref{lastline} and depict the transmission coefficient $\tau$ as a function of frequency for different beam-wave interaction structures.

\section{Discussion of results}\label{plots}
\begin{figure}[htbp]
	\begin{center}
	\scalebox{.5}{\includegraphics{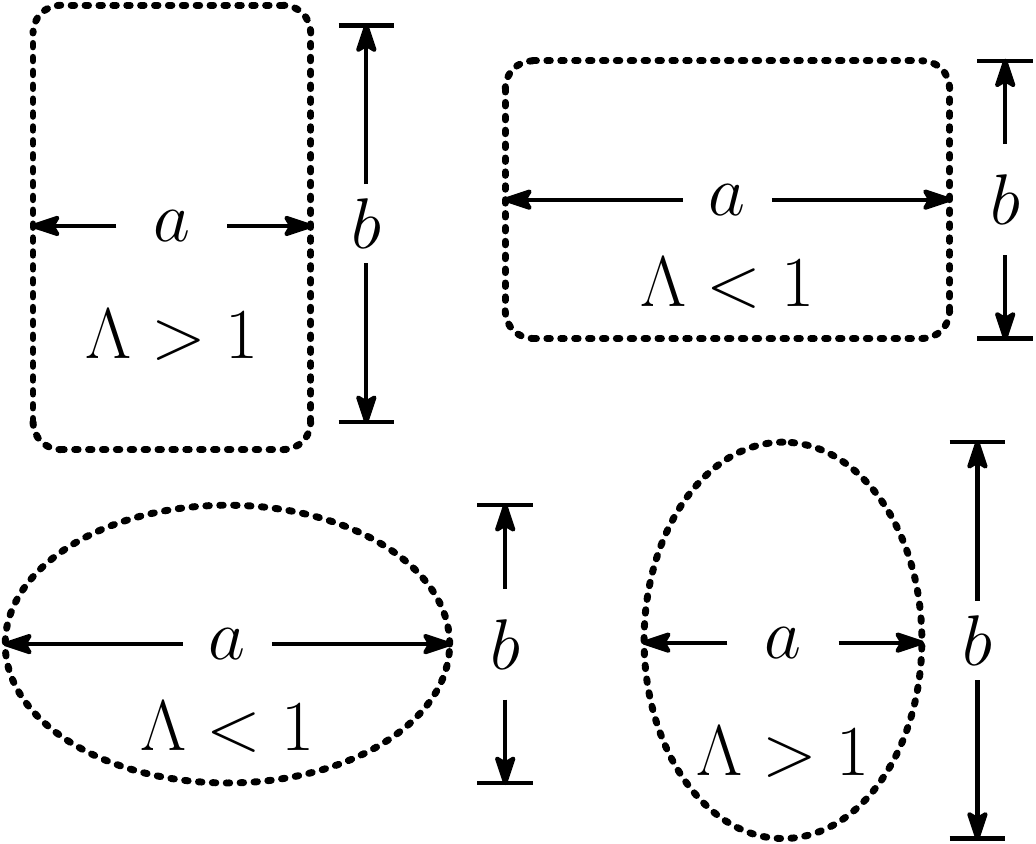}}
	\caption{Lozenge-shaped (above) and ellipsoidal (below) cross sections with different eccentricities corresponding to different aspect ratios $\Lambda = b/a$.  Lengths with value $a$ correspond to the $z$-direction.}
	\label{crossSections}
	\end{center}
\end{figure}

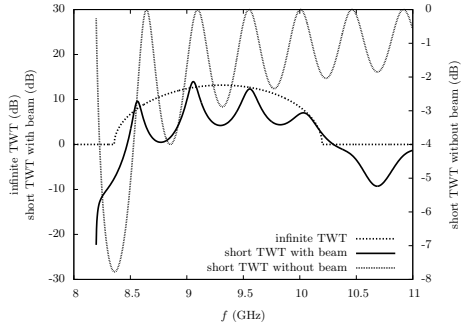
\begin{figure}[htbp]
	\begin{center}
	\scalebox{.5}{\input{Nation3.tex}}
	\caption{Transmission coefficient as a function of frequency for an isotopic dielectric  with dielectric constant $3.5$ for an amplifier driven by a 1 kA beam.
	}
	\label{isotropic}
	\end{center}
\end{figure}

\begin{figure}[htbp]
		\begin{center}
		\begin{tabular}{c}
			\scalebox{.5}{\input{EllipseLambda3.tex}}
			\\
			\scalebox{.5}{\input{EllipseLambda7.tex}} 
			\\
			\scalebox{.5}{\input{EllipseLambda9.tex}}
			\\
			\scalebox{.5}{\input{EllipseLambda11.tex}} 
		\end{tabular}
		\caption{The effect of aspect ratio $\Lambda$ for ellipsoidal inclusions with fixed filling fraction $\theta=0.5410$}
		\label{ellipsoidsLambda}
		\end{center}
	\end{figure}
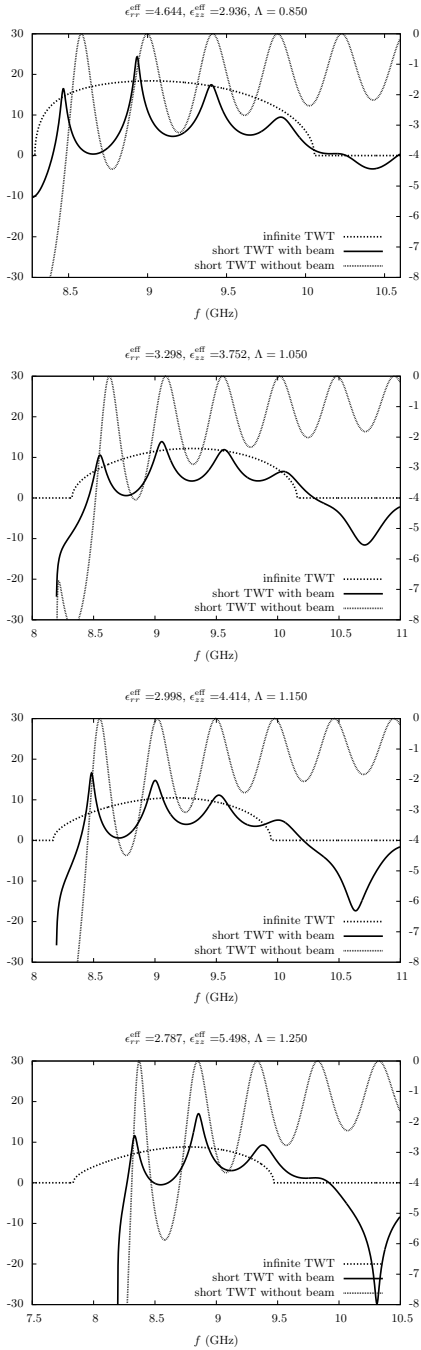

\begin{figure}[htbp]
		\begin{center}
		\begin{tabular}{c}
			\scalebox{.5}{\input{Lozenge125Theta0.tex}}
			\\
			\scalebox{.5}{\input{Lozenge125Theta1.tex}} 
			\\
			\scalebox{.5}{\input{Lozenge125Theta2.tex}}
		\end{tabular}
		\caption{The effect of filling fraction, lozenge-shaped inclusions with aspect ratio $\Lambda=1.25$}
		\label{lozenges125}
		\end{center}
	\end{figure}
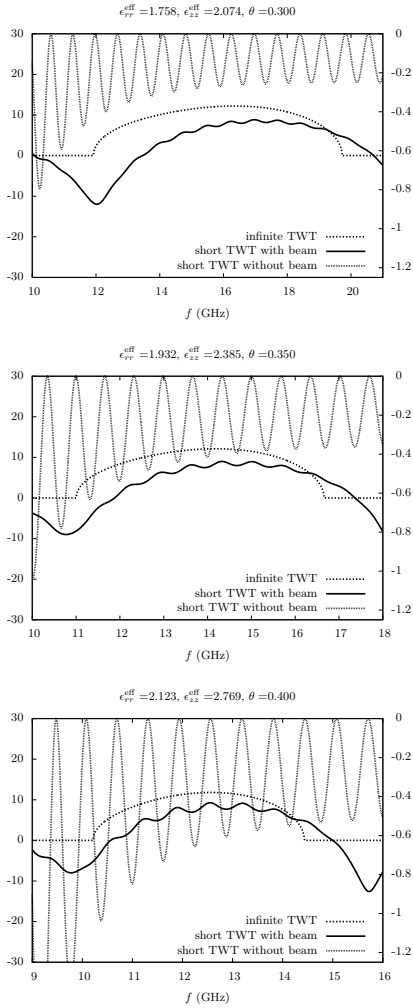

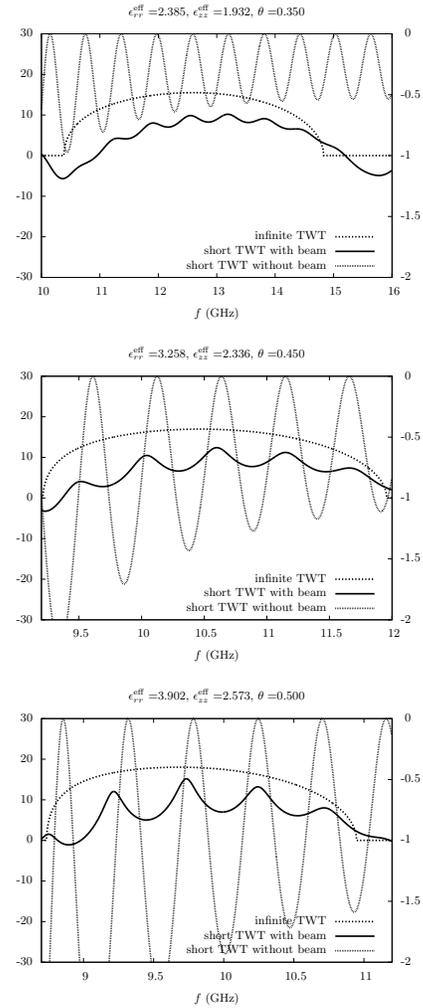
\begin{figure}[htbp]
		\begin{center}
		\begin{tabular}{c}
			\scalebox{.5}{\input{Lozenge80Theta1.tex}}
			\\
			\scalebox{.5}{\input{Lozenge80Theta3.tex}} 
			\\
			\scalebox{.5}{\input{Lozenge80Theta4.tex}}
		\end{tabular}
		\caption{The effect of filling fraction, lozenge-shaped inclusions with aspect ratio $\Lambda=0.80$}
		\label{lozenges80}
		\end{center}
	\end{figure}

\begin{figure}[htbp]
		\begin{center}
		\begin{tabular}{c}
			\scalebox{.5}{\input{Ellipse125Theta1.tex}}
			\\
			\scalebox{.5}{\input{Ellipse125Theta3.tex}} 
			\\
			\scalebox{.5}{\input{Ellipse125Theta4.tex}}
		\end{tabular}
		\caption{The effect of filling fraction, ellipsoidal inclusions with aspect ratio $\Lambda=1.25$}
		\label{ellipsoids125}
		\end{center}
	\end{figure}
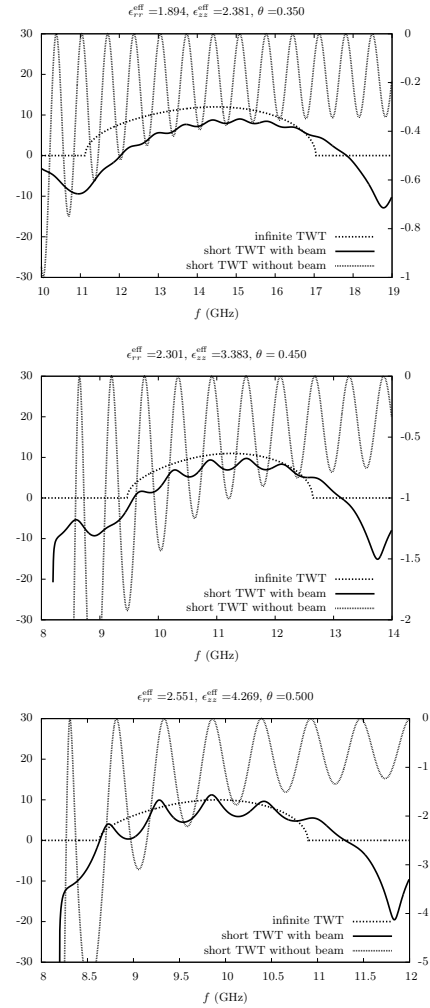

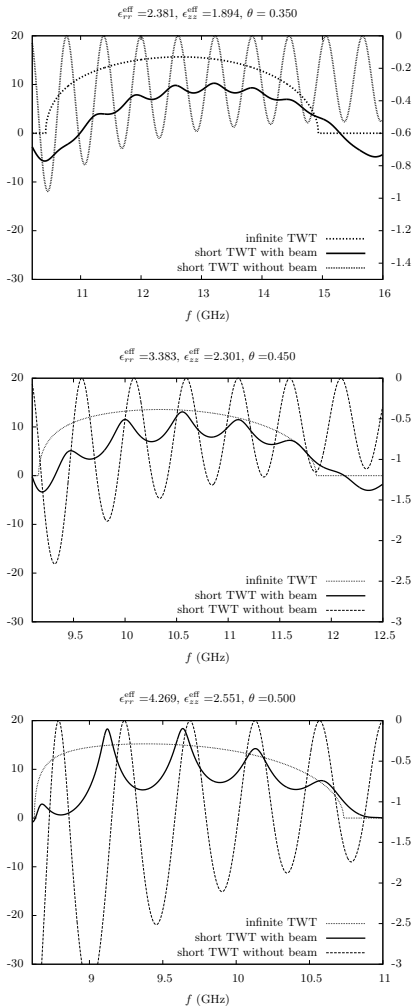
\begin{figure}[htbp]
		\begin{center}
		\begin{tabular}{c}
			\scalebox{.5}{\input{Ellipse80Theta1.tex}}
			\\
			\scalebox{.5}{\input{Ellipse80Theta3.tex}} 
			\\
			\scalebox{.5}{\input{Ellipse80Theta4.tex}}
		\end{tabular}
		\caption{The effect of filling fraction, ellipsoidal inclusions with aspect ratio $\Lambda=0.80$}
		\label{ellipsoids80}
		\end{center}
	\end{figure}

We now turn to the qualitative trends observed in the transmission patterns displayed in Figures \ref{isotropic} - \ref{ellipsoids80}.  
We characterize metamaterial geometries by the filling fraction $\theta$ (relative to a unit cell size) and aspect ratio $\Lambda$ of cross sections of a single metallic inclusion. 
The filling fraction of an inclusion is a measure of its size, and the aspect ratio measures the eccentricity of its shape.  
The aspect ratio is defined by $\Lambda = b/a$, where $b$ corresponds to cross-sectional length in the radial direction, and $a$ corresponds to the longitudinal.  
Cross sections with aspect ratio $\Lambda$ less than unity correspond to inclusions that are more eccentric in the longitudinal direction; aspect ratios exceeding unity correspond to radial eccentricity (see Figure \ref{crossSections}).  

Figures \ref{isotropic} through \ref{ellipsoids80} are compositions of three separate graphs.  
The grey dotted curve is a plot of $20\log\{\exp(\mathrm{Im}\{q\}d\}$ versus frequency.  
Here $\mathrm{Im}\{q\}$ is the gain factor associated with the infinitely long TWT. 
The left axis is a measure of $20\log\{\exp(\mathrm{Im}\{q\}d)\}$ in dB. 
The dark solid curve is a plot of the transmission coefficient $\tau$ for the short TWT device versus frequency. 
The left axis is a measure of $\tau$ in dB. 
The grey dashed curve is the transmission coefficient $\tau$ for the short TWT with no beam present and is plotted against the right axis in dB. 
Here there is no gain, and perfect transmission corresponds to 0 dB loss. 

All transmission profiles are computed and plotted for an amplifier of length $d=15$ cm, $R_b=1.4$ cm, $R=1.82$ cm driven by a $1$-KA beam current, and $\beta=v_0/c=0.9$.

The transmission peaks and valleys correspond to, respectively, constructive and destructive interference due to reflection in the short TWT configuration.

Figure \ref{isotropic} corresponds to the isotropic dielectric constant $\epsilon^{\mathrm{eff}} \equiv 3.5$ considered in the work of Sch\"achter \emph{et. al}\cite{Nation}. 
This is realized using metallic inclusions with symmetric cross sections (for instance, lozenge-shaped cross sections with filling fraction $\theta = 0.5508$ or with circular cross sections with filling fraction $\theta = 0.5410$). 
This transmission pattern is used as a benchmark to identify the effects of changing aspect ratios and filling fractions on transmission profiles.

It should be noted that all filling fractions and aspect ratios studied here are measured with respect to a reference period cell.  
Hence, for fixed filling fraction, extremely eccentric geometries are not permissible, as the their cross sections must be contained within the unit cell. 
Although the eccentricities we considered here are relatively mild ($0.8 < \Lambda < 1.3$), the effect on transmission patterns can be significant.


	\begin{figure}[htbp]
	\begin{center}
		\begin{tabular}{c}
			\scalebox{.5}{\input{LozengeLambda3.tex}}
			\\
			\scalebox{.5}{\input{LozengeLambda7.tex}} 
			\\
			\scalebox{.5}{\input{LozengeLambda9.tex}}
			\\
			\scalebox{.5}{\input{LozengeLambda11.tex}} 
		\end{tabular}
	\caption{The effect of aspect ratio $\Lambda$ for lozenge-shaped inclusions with fixed filling fraction $\theta=0.5508$}
	\label{lozengesLambda}
	\end{center}
	\end{figure}

Figure \ref{lozengesLambda} shows transmission profiles for lozenge-shaped inclusions with different aspect ratios.  
The filling fraction for each cross section is fixed at 0.5508.  
As mentioned earlier the symmetric lozenge shape ($a=b$) with filling fraction $0.5508$ delivers $\epsilon^{\mathrm{eff}}=3.5$ (the same dielectric constant used in the work of Sch\"achter \emph{et. al.}).  

Figure \ref{ellipsoidsLambda} shows the transmission patterns for the same range of aspect ratios but for ellipsoidal cross sections at fixed volume fraction $\theta=0.5410$. As pointed out earlier the circular cross section with this filling fraction delivers $\epsilon^{\mathrm{eff} }=3.5$ Both figures show a relatively constant range of operation and bandwidths that shrink slightly as aspect ratios increase.  
Both also show an increase in gain for more eccentric geometries, with a more uniform increase corresponding to longitudinal eccentricity.

	\begin{figure}[htbp]
		\begin{center}
		\begin{tabular}{c}
			\scalebox{.5}{\input{LozengeTheta0.tex}}
			\\
			\scalebox{.5}{\input{LozengeTheta1.tex}} 
			\\
			\scalebox{.5}{\input{LozengeTheta2.tex}}
		\end{tabular}
		\caption{The effect of filling fraction, symmetric lozenge-shaped inclusions}
		\label{lozengesSymmetric}
		\end{center}
	\end{figure}
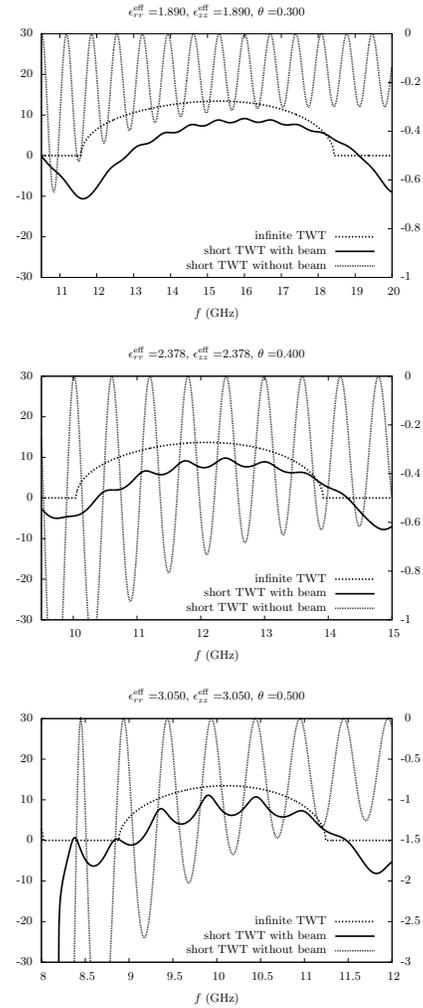
	
	\begin{figure}[htbp]
		\begin{center}
		\begin{tabular}{c}
			\scalebox{.5}{\input{EllipseTheta0.tex}}
			\\
			\scalebox{.5}{\input{EllipseTheta1.tex}} 
			\\
			\scalebox{.5}{\input{EllipseTheta2.tex}}
		\end{tabular}
		\caption{The effect of filling fraction, circular inclusions}
		\label{ellipsoidsSymmetric}
		\end{center}
	\end{figure}
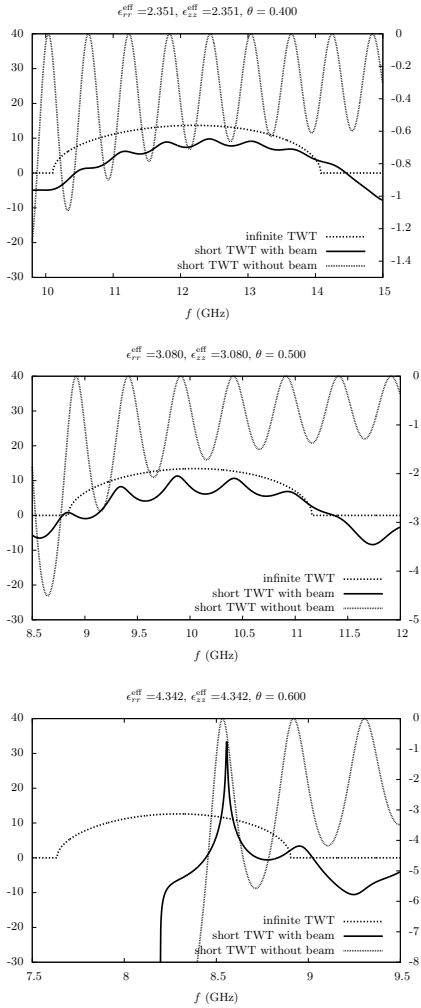
	
Figures \ref{lozengesSymmetric} - \ref{lozenges80} demonstrate the effect of varying the filling fraction of lozenge-shaped inclusions, each of them for a different aspect ratio.  In each case, the range of operation increases as the filling fraction decreases, while both gain and bandwidth remain relatively constant.  
Comparing each of the three figures, it can again be seen that more longitudinally eccentric geometries yield a higher gain, as suggested by Figures \ref{lozengesLambda} and \ref{ellipsoidsLambda}.  
The same trends are seen for ellipsoidal inclusions in Figures \ref{ellipsoidsSymmetric} - \ref{ellipsoids80}.  
For fixed aspect ratios, ellipsoidal geometries show uniformly higher gains than their lozenge-shaped counterparts. The large gain seen in Figure \ref{ellipsoidsSymmetric} for filling fraction $\theta=0.6$ stems from the fact that neighboring cross sections are separated by a distance that is a small fraction of their common diameter. From an operational perspective this geometry is not desirable due to the presence of high electric field concentrations between adjacent cross sections.
	
\section{Summary}
We conclude by pointing out that this work investigates the sensitivity of TWT performance with respect to variations of a simple interaction structure. It is seen that the geometry of the structure strongly influences the performance of the TWT and can be controlled for tuning frequency of operation, gain and bandwidth. The interaction structure considered here does not account for the support structures connecting the metal rings to the outer wall of the TWT. Indeed the support structure itself can effect the TWT performance introducing multiple hybrid modes and shifting the frequency of operation. Building on the methodology developed here, new types of interaction structures, together with their support structures, are currently under development for control of TWT performance.

\section{Acknowledgements}
This research is supported by AFOSR MURI Grant FA9550-12-1-0489 administered through the University of New Mexico, NSF grant DMS-1211066, and NSF EPSCOR Cooperative Agreement No. EPS-1003897 with additional support from the Louisiana Board of Regents.

\section{Appendix}
We provide an outline of the two-scale asymptotic method as it is applied here to identify the homogenized transmission problem for the infinitely long TWT. The method starts with the definition of $\rho$ and $y$ given in Section \ref{anisoeffect} and writing the total derivatives for terms in the expansion $\tilde{\psi}_i$, $i=1,2,\ldots$,
\begin{eqnarray}
&&\DD{}{\tilde{r}}\tilde{\psi}_i(\tilde{r},\rho,y)=\left(\frac{\partial}{\partial \tilde{r}}+\frac{1}{\varepsilon}\frac{\partial}{\partial\rho}\right)\tilde{\psi}_i \label{rho}\\
&&\DD{}{\tilde{z}} \tilde{\psi}_i(\tilde{r},\rho,y)=\frac{1}{\varepsilon}\frac{\partial}{\partial y}\tilde{\psi}_i.
\label{y}
\end{eqnarray}
Now apply the change of variables $r\rightarrow\tilde{r}$, $z\rightarrow\tilde{z}$ to the equations  \eqref{maxbinbeam}, \eqref{timespacedispersion}, \eqref{maxbinwave}, and substitute the expansions \eqref{twoscalefield}, \eqref{vavenumexansion} into these and apply \eqref{rho} , \eqref{y}. Equating like powers of $\varepsilon$  delivers a sequence of transmission  problems. Here the solution of the transmission problems $\tilde{\psi}_j$, $k_j$, with $j<i$ provide data for the transmission problem and consistency conditions used to determine $\tilde{\psi}_i$ and $k_i$. The leading order problem \eqref{maxbinbeamh}, \eqref{timespacedispersionh}, \eqref{maxbinwaveh}, \eqref{beaminteractinterface} is obtained from the consistency condition for the solution of the transmission problem for $\tilde{\psi}_2$ and changing from $(\tilde{r},\tilde{z})$ coordinates back to $(r,z)$ coordinates.  The effective properties appearing in this relation are identified as the moments of $\partial_\rho\tilde{\psi}_1$ and $\partial_y\tilde{\psi}_1$.  The linearity of the problem allows one to express $\tilde{\psi}_1$ as linear combinations of the solution of the cell problems \eqref{equlib}, \eqref{boundary1}, \eqref{boundary2} and this delivers the expressions for the effective properties given by \eqref{err} and \eqref{eyy}.

\bibliography{Paper-Preprint}

\end{document}

%% file: Nation3.tex
\begingroup
  \makeatletter
  \providecommand\color[2][]{%
    \GenericError{(gnuplot) \space\space\space\@spaces}{%
      Package color not loaded in conjunction with
      terminal option `colourtext'%
    }{See the gnuplot documentation for explanation.%
    }{Either use 'blacktext' in gnuplot or load the package
      color.sty in LaTeX.}%
    \renewcommand\color[2][]{}%
  }%
  \providecommand\includegraphics[2][]{%
    \GenericError{(gnuplot) \space\space\space\@spaces}{%
      Package graphicx or graphics not loaded%
    }{See the gnuplot documentation for explanation.%
    }{The gnuplot epslatex terminal needs graphicx.sty or graphics.sty.}%
    \renewcommand\includegraphics[2][]{}%
  }%
  \providecommand\rotatebox[2]{#2}%
  \@ifundefined{ifGPcolor}{%
    \newif\ifGPcolor
    \GPcolorfalse
  }{}%
  \@ifundefined{ifGPblacktext}{%
    \newif\ifGPblacktext
    \GPblacktexttrue
  }{}%
  \let\gplgaddtomacro\g@addto@macro
  \gdef\gplbacktext{}%
  \gdef\gplfronttext{}%
  \makeatother
  \ifGPblacktext
    \def\colorrgb#1{}%
    \def\colorgray#1{}%
  \else
    \ifGPcolor
      \def\colorrgb#1{\color[rgb]{#1}}%
      \def\colorgray#1{\color[gray]{#1}}%
      \expandafter\def\csname LTw\endcsname{\color{white}}%
      \expandafter\def\csname LTb\endcsname{\color{black}}%
      \expandafter\def\csname LTa\endcsname{\color{black}}%
      \expandafter\def\csname LT0\endcsname{\color[rgb]{1,0,0}}%
      \expandafter\def\csname LT1\endcsname{\color[rgb]{0,1,0}}%
      \expandafter\def\csname LT2\endcsname{\color[rgb]{0,0,1}}%
      \expandafter\def\csname LT3\endcsname{\color[rgb]{1,0,1}}%
      \expandafter\def\csname LT4\endcsname{\color[rgb]{0,1,1}}%
      \expandafter\def\csname LT5\endcsname{\color[rgb]{1,1,0}}%
      \expandafter\def\csname LT6\endcsname{\color[rgb]{0,0,0}}%
      \expandafter\def\csname LT7\endcsname{\color[rgb]{1,0.3,0}}%
      \expandafter\def\csname LT8\endcsname{\color[rgb]{0.5,0.5,0.5}}%
    \else
      \def\colorrgb#1{\color{black}}%
      \def\colorgray#1{\color[gray]{#1}}%
      \expandafter\def\csname LTw\endcsname{\color{white}}%
      \expandafter\def\csname LTb\endcsname{\color{black}}%
      \expandafter\def\csname LTa\endcsname{\color{black}}%
      \expandafter\def\csname LT0\endcsname{\color{black}}%
      \expandafter\def\csname LT1\endcsname{\color{black}}%
      \expandafter\def\csname LT2\endcsname{\color{black}}%
      \expandafter\def\csname LT3\endcsname{\color{black}}%
      \expandafter\def\csname LT4\endcsname{\color{black}}%
      \expandafter\def\csname LT5\endcsname{\color{black}}%
      \expandafter\def\csname LT6\endcsname{\color{black}}%
      \expandafter\def\csname LT7\endcsname{\color{black}}%
      \expandafter\def\csname LT8\endcsname{\color{black}}%
    \fi
  \fi
  \setlength{\unitlength}{0.0500bp}%
  \begin{picture}(7200.00,5040.00)%
    \gplgaddtomacro\gplbacktext{%
      \csname LTb\endcsname%
      \put(814,704){\makebox(0,0)[r]{\strut{}-30}}%
      \put(814,1383){\makebox(0,0)[r]{\strut{}-20}}%
      \put(814,2061){\makebox(0,0)[r]{\strut{}-10}}%
      \put(814,2740){\makebox(0,0)[r]{\strut{} 0}}%
      \put(814,3418){\makebox(0,0)[r]{\strut{} 10}}%
      \put(814,4097){\makebox(0,0)[r]{\strut{} 20}}%
      \put(814,4775){\makebox(0,0)[r]{\strut{} 30}}%
      \put(946,484){\makebox(0,0){\strut{} 8}}%
      \put(1798,484){\makebox(0,0){\strut{} 8.5}}%
      \put(2649,484){\makebox(0,0){\strut{} 9}}%
      \put(3501,484){\makebox(0,0){\strut{} 9.5}}%
      \put(4352,484){\makebox(0,0){\strut{} 10}}%
      \put(5204,484){\makebox(0,0){\strut{} 10.5}}%
      \put(6055,484){\makebox(0,0){\strut{} 11}}%
      \put(6187,704){\makebox(0,0)[l]{\strut{}-8}}%
      \put(6187,1213){\makebox(0,0)[l]{\strut{}-7}}%
      \put(6187,1722){\makebox(0,0)[l]{\strut{}-6}}%
      \put(6187,2231){\makebox(0,0)[l]{\strut{}-5}}%
      \put(6187,2740){\makebox(0,0)[l]{\strut{}-4}}%
      \put(6187,3248){\makebox(0,0)[l]{\strut{}-3}}%
      \put(6187,3757){\makebox(0,0)[l]{\strut{}-2}}%
      \put(6187,4266){\makebox(0,0)[l]{\strut{}-1}}%
      \put(6187,4775){\makebox(0,0)[l]{\strut{} 0}}%
      \put(176,2739){\rotatebox{-270}{\makebox(0,0){\shortstack{infinite TWT (dB) \\ short TWT with beam (dB)}}}}%
      \put(6692,2739){\rotatebox{-270}{\makebox(0,0){\strut{}short TWT without beam (dB)}}}%
      \put(3500,154){\makebox(0,0){\strut{}$f$ (GHz)}}%
    }%
    \gplgaddtomacro\gplfronttext{%
      \csname LTb\endcsname%
      \put(5068,1317){\makebox(0,0)[r]{\strut{}infinite TWT}}%
      \csname LTb\endcsname%
      \put(5068,1097){\makebox(0,0)[r]{\strut{}short TWT with beam}}%
      \csname LTb\endcsname%
      \put(5068,877){\makebox(0,0)[r]{\strut{}short TWT without beam}}%
    }%
    \gplbacktext
    \put(0,0){\includegraphics{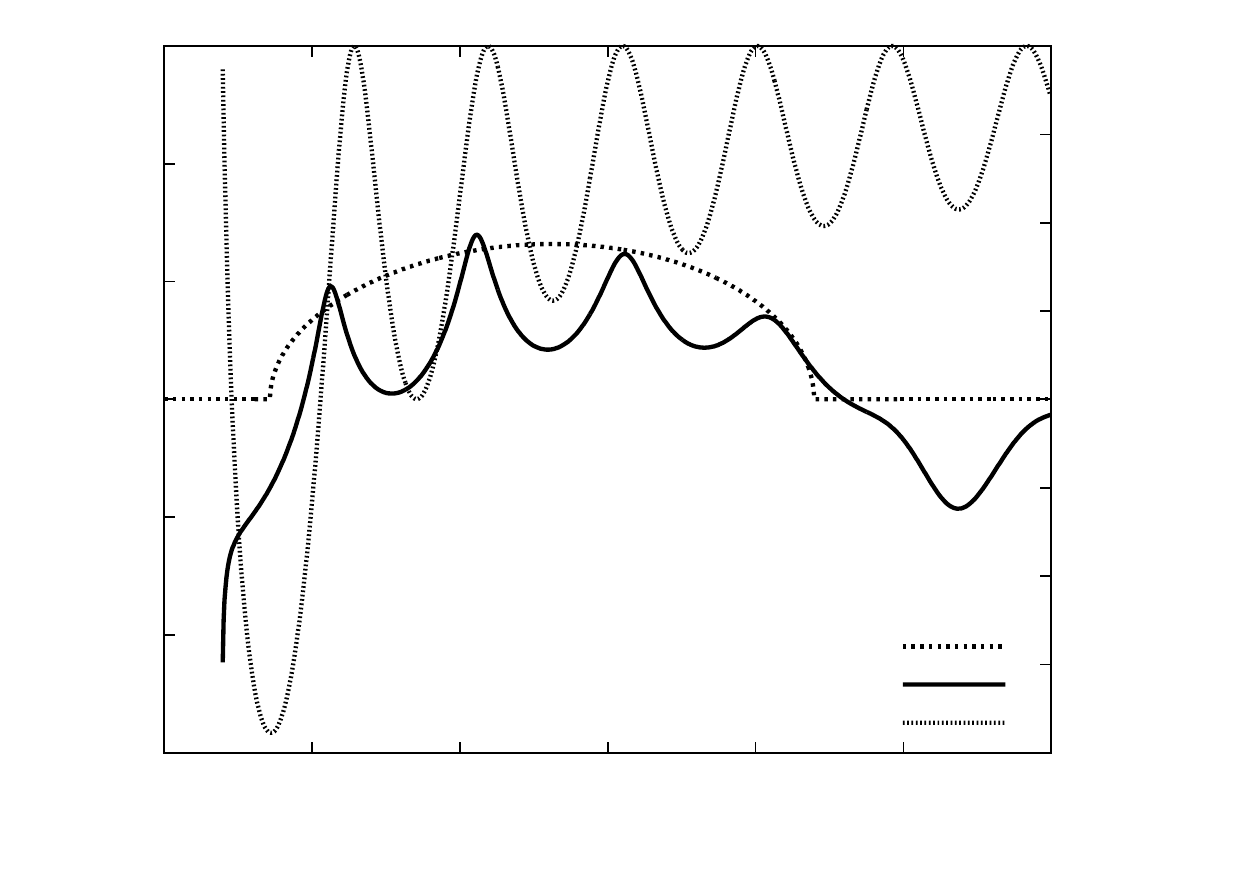}}%
    \gplfronttext
  \end{picture}%
\endgroup

%% file: LozengeLambda3.tex
\begingroup
  \makeatletter
  \providecommand\color[2][]{%
    \GenericError{(gnuplot) \space\space\space\@spaces}{%
      Package color not loaded in conjunction with
      terminal option `colourtext'%
    }{See the gnuplot documentation for explanation.%
    }{Either use 'blacktext' in gnuplot or load the package
      color.sty in LaTeX.}%
    \renewcommand\color[2][]{}%
  }%
  \providecommand\includegraphics[2][]{%
    \GenericError{(gnuplot) \space\space\space\@spaces}{%
      Package graphicx or graphics not loaded%
    }{See the gnuplot documentation for explanation.%
    }{The gnuplot epslatex terminal needs graphicx.sty or graphics.sty.}%
    \renewcommand\includegraphics[2][]{}%
  }%
  \providecommand\rotatebox[2]{#2}%
  \@ifundefined{ifGPcolor}{%
    \newif\ifGPcolor
    \GPcolorfalse
  }{}%
  \@ifundefined{ifGPblacktext}{%
    \newif\ifGPblacktext
    \GPblacktexttrue
  }{}%
  \let\gplgaddtomacro\g@addto@macro
  \gdef\gplbacktext{}%
  \gdef\gplfronttext{}%
  \makeatother
  \ifGPblacktext
    \def\colorrgb#1{}%
    \def\colorgray#1{}%
  \else
    \ifGPcolor
      \def\colorrgb#1{\color[rgb]{#1}}%
      \def\colorgray#1{\color[gray]{#1}}%
      \expandafter\def\csname LTw\endcsname{\color{white}}%
      \expandafter\def\csname LTb\endcsname{\color{black}}%
      \expandafter\def\csname LTa\endcsname{\color{black}}%
      \expandafter\def\csname LT0\endcsname{\color[rgb]{1,0,0}}%
      \expandafter\def\csname LT1\endcsname{\color[rgb]{0,1,0}}%
      \expandafter\def\csname LT2\endcsname{\color[rgb]{0,0,1}}%
      \expandafter\def\csname LT3\endcsname{\color[rgb]{1,0,1}}%
      \expandafter\def\csname LT4\endcsname{\color[rgb]{0,1,1}}%
      \expandafter\def\csname LT5\endcsname{\color[rgb]{1,1,0}}%
      \expandafter\def\csname LT6\endcsname{\color[rgb]{0,0,0}}%
      \expandafter\def\csname LT7\endcsname{\color[rgb]{1,0.3,0}}%
      \expandafter\def\csname LT8\endcsname{\color[rgb]{0.5,0.5,0.5}}%
    \else
      \def\colorrgb#1{\color{black}}%
      \def\colorgray#1{\color[gray]{#1}}%
      \expandafter\def\csname LTw\endcsname{\color{white}}%
      \expandafter\def\csname LTb\endcsname{\color{black}}%
      \expandafter\def\csname LTa\endcsname{\color{black}}%
      \expandafter\def\csname LT0\endcsname{\color{black}}%
      \expandafter\def\csname LT1\endcsname{\color{black}}%
      \expandafter\def\csname LT2\endcsname{\color{black}}%
      \expandafter\def\csname LT3\endcsname{\color{black}}%
      \expandafter\def\csname LT4\endcsname{\color{black}}%
      \expandafter\def\csname LT5\endcsname{\color{black}}%
      \expandafter\def\csname LT6\endcsname{\color{black}}%
      \expandafter\def\csname LT7\endcsname{\color{black}}%
      \expandafter\def\csname LT8\endcsname{\color{black}}%
    \fi
  \fi
  \setlength{\unitlength}{0.0500bp}%
  \begin{picture}(7200.00,5040.00)%
    \gplgaddtomacro\gplbacktext{%
      \csname LTb\endcsname%
      \put(594,704){\makebox(0,0)[r]{\strut{}-30}}%
      \put(594,1317){\makebox(0,0)[r]{\strut{}-20}}%
      \put(594,1929){\makebox(0,0)[r]{\strut{}-10}}%
      \put(594,2542){\makebox(0,0)[r]{\strut{} 0}}%
      \put(594,3154){\makebox(0,0)[r]{\strut{} 10}}%
      \put(594,3767){\makebox(0,0)[r]{\strut{} 20}}%
      \put(594,4379){\makebox(0,0)[r]{\strut{} 30}}%
      \put(1274,484){\makebox(0,0){\strut{} 8.5}}%
      \put(2465,484){\makebox(0,0){\strut{} 9}}%
      \put(3655,484){\makebox(0,0){\strut{} 9.5}}%
      \put(4846,484){\makebox(0,0){\strut{} 10}}%
      \put(6037,484){\makebox(0,0){\strut{} 10.5}}%
      \put(6407,704){\makebox(0,0)[l]{\strut{}-8}}%
      \put(6407,1163){\makebox(0,0)[l]{\strut{}-7}}%
      \put(6407,1623){\makebox(0,0)[l]{\strut{}-6}}%
      \put(6407,2082){\makebox(0,0)[l]{\strut{}-5}}%
      \put(6407,2542){\makebox(0,0)[l]{\strut{}-4}}%
      \put(6407,3001){\makebox(0,0)[l]{\strut{}-3}}%
      \put(6407,3460){\makebox(0,0)[l]{\strut{}-2}}%
      \put(6407,3920){\makebox(0,0)[l]{\strut{}-1}}%
      \put(6407,4379){\makebox(0,0)[l]{\strut{} 0}}%
      \put(3500,154){\makebox(0,0){\strut{}$f$ (GHz)}}%
      \put(3500,4709){\makebox(0,0){\strut{}$\epsilon^{\mathrm{eff}}_{rr}=$4.335, $\epsilon^{\mathrm{eff}}_{zz}=$2.990, $\Lambda=0.850$}}%
    }%
    \gplgaddtomacro\gplfronttext{%
      \csname LTb\endcsname%
      \put(5288,1317){\makebox(0,0)[r]{\strut{}infinite TWT}}%
      \csname LTb\endcsname%
      \put(5288,1097){\makebox(0,0)[r]{\strut{}short TWT with beam}}%
      \csname LTb\endcsname%
      \put(5288,877){\makebox(0,0)[r]{\strut{}short TWT without beam}}%
    }%
    \gplbacktext
    \put(0,0){\includegraphics{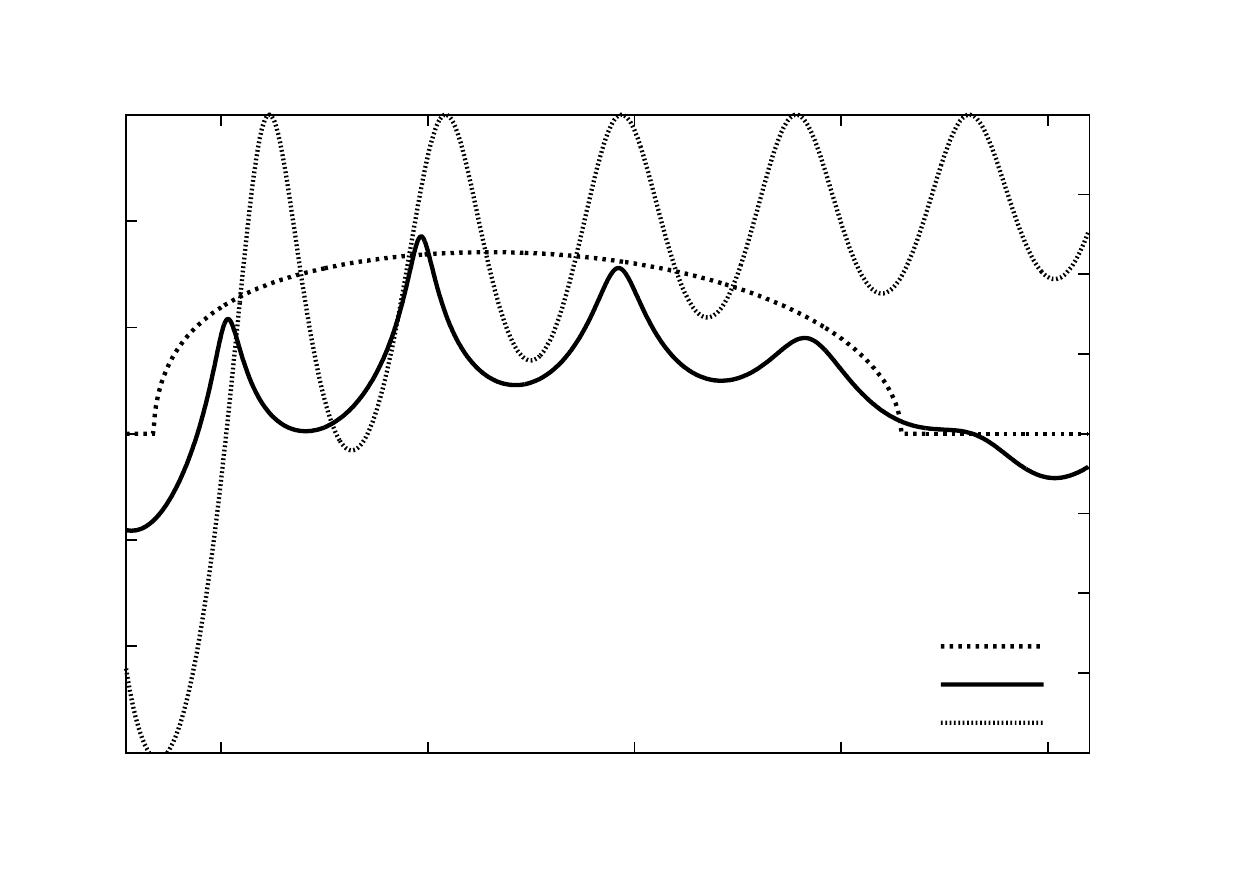}}%
    \gplfronttext
  \end{picture}%
\endgroup

%% file: LozengeLambda7.tex
\begingroup
  \makeatletter
  \providecommand\color[2][]{%
    \GenericError{(gnuplot) \space\space\space\@spaces}{%
      Package color not loaded in conjunction with
      terminal option `colourtext'%
    }{See the gnuplot documentation for explanation.%
    }{Either use 'blacktext' in gnuplot or load the package
      color.sty in LaTeX.}%
    \renewcommand\color[2][]{}%
  }%
  \providecommand\includegraphics[2][]{%
    \GenericError{(gnuplot) \space\space\space\@spaces}{%
      Package graphicx or graphics not loaded%
    }{See the gnuplot documentation for explanation.%
    }{The gnuplot epslatex terminal needs graphicx.sty or graphics.sty.}%
    \renewcommand\includegraphics[2][]{}%
  }%
  \providecommand\rotatebox[2]{#2}%
  \@ifundefined{ifGPcolor}{%
    \newif\ifGPcolor
    \GPcolorfalse
  }{}%
  \@ifundefined{ifGPblacktext}{%
    \newif\ifGPblacktext
    \GPblacktexttrue
  }{}%
  \let\gplgaddtomacro\g@addto@macro
  \gdef\gplbacktext{}%
  \gdef\gplfronttext{}%
  \makeatother
  \ifGPblacktext
    \def\colorrgb#1{}%
    \def\colorgray#1{}%
  \else
    \ifGPcolor
      \def\colorrgb#1{\color[rgb]{#1}}%
      \def\colorgray#1{\color[gray]{#1}}%
      \expandafter\def\csname LTw\endcsname{\color{white}}%
      \expandafter\def\csname LTb\endcsname{\color{black}}%
      \expandafter\def\csname LTa\endcsname{\color{black}}%
      \expandafter\def\csname LT0\endcsname{\color[rgb]{1,0,0}}%
      \expandafter\def\csname LT1\endcsname{\color[rgb]{0,1,0}}%
      \expandafter\def\csname LT2\endcsname{\color[rgb]{0,0,1}}%
      \expandafter\def\csname LT3\endcsname{\color[rgb]{1,0,1}}%
      \expandafter\def\csname LT4\endcsname{\color[rgb]{0,1,1}}%
      \expandafter\def\csname LT5\endcsname{\color[rgb]{1,1,0}}%
      \expandafter\def\csname LT6\endcsname{\color[rgb]{0,0,0}}%
      \expandafter\def\csname LT7\endcsname{\color[rgb]{1,0.3,0}}%
      \expandafter\def\csname LT8\endcsname{\color[rgb]{0.5,0.5,0.5}}%
    \else
      \def\colorrgb#1{\color{black}}%
      \def\colorgray#1{\color[gray]{#1}}%
      \expandafter\def\csname LTw\endcsname{\color{white}}%
      \expandafter\def\csname LTb\endcsname{\color{black}}%
      \expandafter\def\csname LTa\endcsname{\color{black}}%
      \expandafter\def\csname LT0\endcsname{\color{black}}%
      \expandafter\def\csname LT1\endcsname{\color{black}}%
      \expandafter\def\csname LT2\endcsname{\color{black}}%
      \expandafter\def\csname LT3\endcsname{\color{black}}%
      \expandafter\def\csname LT4\endcsname{\color{black}}%
      \expandafter\def\csname LT5\endcsname{\color{black}}%
      \expandafter\def\csname LT6\endcsname{\color{black}}%
      \expandafter\def\csname LT7\endcsname{\color{black}}%
      \expandafter\def\csname LT8\endcsname{\color{black}}%
    \fi
  \fi
  \setlength{\unitlength}{0.0500bp}%
  \begin{picture}(7200.00,5040.00)%
    \gplgaddtomacro\gplbacktext{%
      \csname LTb\endcsname%
      \put(594,704){\makebox(0,0)[r]{\strut{}-30}}%
      \put(594,1317){\makebox(0,0)[r]{\strut{}-20}}%
      \put(594,1929){\makebox(0,0)[r]{\strut{}-10}}%
      \put(594,2542){\makebox(0,0)[r]{\strut{} 0}}%
      \put(594,3154){\makebox(0,0)[r]{\strut{} 10}}%
      \put(594,3767){\makebox(0,0)[r]{\strut{} 20}}%
      \put(594,4379){\makebox(0,0)[r]{\strut{} 30}}%
      \put(726,484){\makebox(0,0){\strut{} 8}}%
      \put(1651,484){\makebox(0,0){\strut{} 8.5}}%
      \put(2576,484){\makebox(0,0){\strut{} 9}}%
      \put(3501,484){\makebox(0,0){\strut{} 9.5}}%
      \put(4425,484){\makebox(0,0){\strut{} 10}}%
      \put(5350,484){\makebox(0,0){\strut{} 10.5}}%
      \put(6275,484){\makebox(0,0){\strut{} 11}}%
      \put(6407,704){\makebox(0,0)[l]{\strut{}-8}}%
      \put(6407,1163){\makebox(0,0)[l]{\strut{}-7}}%
      \put(6407,1623){\makebox(0,0)[l]{\strut{}-6}}%
      \put(6407,2082){\makebox(0,0)[l]{\strut{}-5}}%
      \put(6407,2542){\makebox(0,0)[l]{\strut{}-4}}%
      \put(6407,3001){\makebox(0,0)[l]{\strut{}-3}}%
      \put(6407,3460){\makebox(0,0)[l]{\strut{}-2}}%
      \put(6407,3920){\makebox(0,0)[l]{\strut{}-1}}%
      \put(6407,4379){\makebox(0,0)[l]{\strut{} 0}}%
      \put(3500,154){\makebox(0,0){\strut{}$f$ (GHz)}}%
      \put(3500,4709){\makebox(0,0){\strut{}$\epsilon^{\mathrm{eff}}_{rr}=$3.324, $\epsilon^{\mathrm{eff}}_{zz}=$3.704, $\Lambda=1.050$}}%
    }%
    \gplgaddtomacro\gplfronttext{%
      \csname LTb\endcsname%
      \put(5288,1317){\makebox(0,0)[r]{\strut{}infinite TWT}}%
      \csname LTb\endcsname%
      \put(5288,1097){\makebox(0,0)[r]{\strut{}short TWT with beam}}%
      \csname LTb\endcsname%
      \put(5288,877){\makebox(0,0)[r]{\strut{}short TWT without beam}}%
    }%
    \gplbacktext
    \put(0,0){\includegraphics{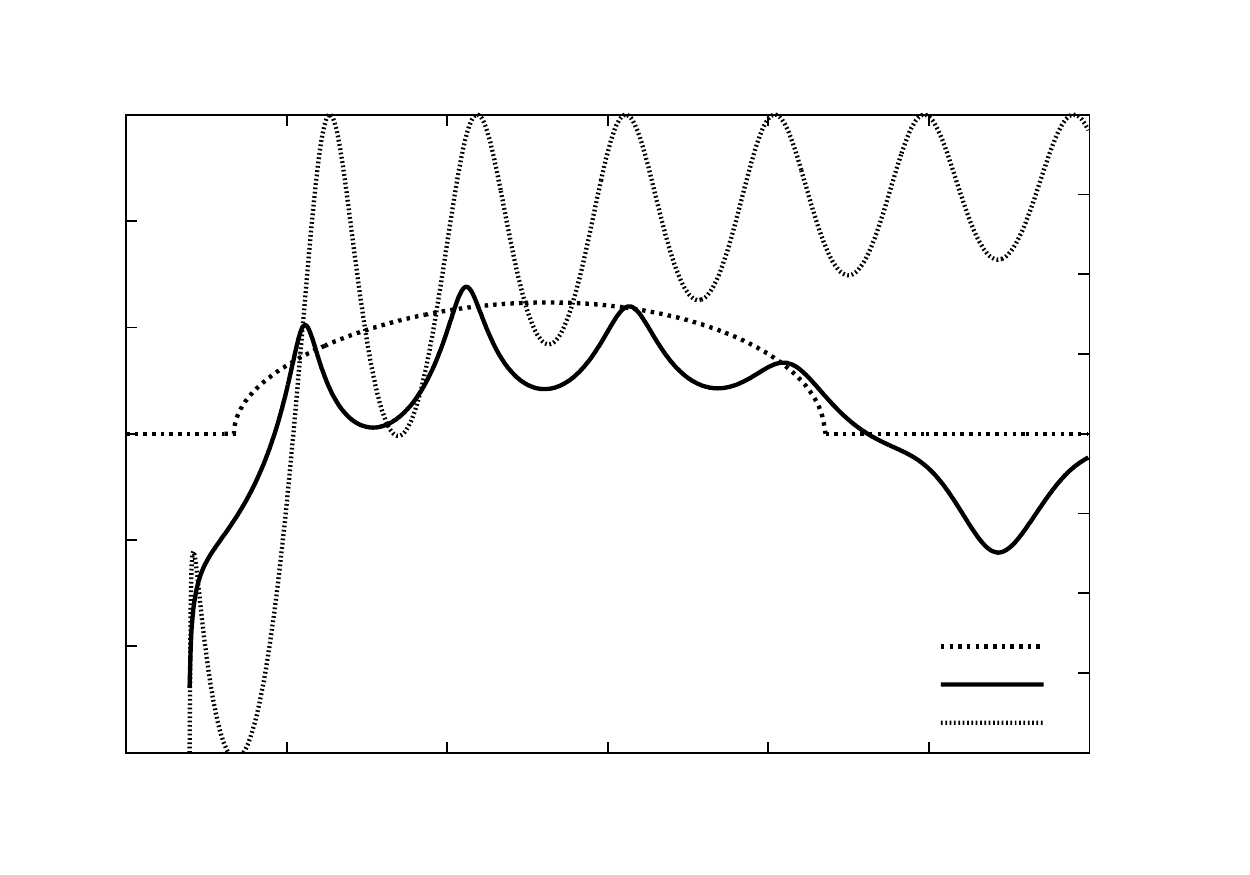}}%
    \gplfronttext
  \end{picture}%
\endgroup

%% file: LozengeLambda9.tex
\begingroup
  \makeatletter
  \providecommand\color[2][]{%
    \GenericError{(gnuplot) \space\space\space\@spaces}{%
      Package color not loaded in conjunction with
      terminal option `colourtext'%
    }{See the gnuplot documentation for explanation.%
    }{Either use 'blacktext' in gnuplot or load the package
      color.sty in LaTeX.}%
    \renewcommand\color[2][]{}%
  }%
  \providecommand\includegraphics[2][]{%
    \GenericError{(gnuplot) \space\space\space\@spaces}{%
      Package graphicx or graphics not loaded%
    }{See the gnuplot documentation for explanation.%
    }{The gnuplot epslatex terminal needs graphicx.sty or graphics.sty.}%
    \renewcommand\includegraphics[2][]{}%
  }%
  \providecommand\rotatebox[2]{#2}%
  \@ifundefined{ifGPcolor}{%
    \newif\ifGPcolor
    \GPcolorfalse
  }{}%
  \@ifundefined{ifGPblacktext}{%
    \newif\ifGPblacktext
    \GPblacktexttrue
  }{}%
  \let\gplgaddtomacro\g@addto@macro
  \gdef\gplbacktext{}%
  \gdef\gplfronttext{}%
  \makeatother
  \ifGPblacktext
    \def\colorrgb#1{}%
    \def\colorgray#1{}%
  \else
    \ifGPcolor
      \def\colorrgb#1{\color[rgb]{#1}}%
      \def\colorgray#1{\color[gray]{#1}}%
      \expandafter\def\csname LTw\endcsname{\color{white}}%
      \expandafter\def\csname LTb\endcsname{\color{black}}%
      \expandafter\def\csname LTa\endcsname{\color{black}}%
      \expandafter\def\csname LT0\endcsname{\color[rgb]{1,0,0}}%
      \expandafter\def\csname LT1\endcsname{\color[rgb]{0,1,0}}%
      \expandafter\def\csname LT2\endcsname{\color[rgb]{0,0,1}}%
      \expandafter\def\csname LT3\endcsname{\color[rgb]{1,0,1}}%
      \expandafter\def\csname LT4\endcsname{\color[rgb]{0,1,1}}%
      \expandafter\def\csname LT5\endcsname{\color[rgb]{1,1,0}}%
      \expandafter\def\csname LT6\endcsname{\color[rgb]{0,0,0}}%
      \expandafter\def\csname LT7\endcsname{\color[rgb]{1,0.3,0}}%
      \expandafter\def\csname LT8\endcsname{\color[rgb]{0.5,0.5,0.5}}%
    \else
      \def\colorrgb#1{\color{black}}%
      \def\colorgray#1{\color[gray]{#1}}%
      \expandafter\def\csname LTw\endcsname{\color{white}}%
      \expandafter\def\csname LTb\endcsname{\color{black}}%
      \expandafter\def\csname LTa\endcsname{\color{black}}%
      \expandafter\def\csname LT0\endcsname{\color{black}}%
      \expandafter\def\csname LT1\endcsname{\color{black}}%
      \expandafter\def\csname LT2\endcsname{\color{black}}%
      \expandafter\def\csname LT3\endcsname{\color{black}}%
      \expandafter\def\csname LT4\endcsname{\color{black}}%
      \expandafter\def\csname LT5\endcsname{\color{black}}%
      \expandafter\def\csname LT6\endcsname{\color{black}}%
      \expandafter\def\csname LT7\endcsname{\color{black}}%
      \expandafter\def\csname LT8\endcsname{\color{black}}%
    \fi
  \fi
  \setlength{\unitlength}{0.0500bp}%
  \begin{picture}(7200.00,5040.00)%
    \gplgaddtomacro\gplbacktext{%
      \csname LTb\endcsname%
      \put(594,704){\makebox(0,0)[r]{\strut{}-30}}%
      \put(594,1317){\makebox(0,0)[r]{\strut{}-20}}%
      \put(594,1929){\makebox(0,0)[r]{\strut{}-10}}%
      \put(594,2542){\makebox(0,0)[r]{\strut{} 0}}%
      \put(594,3154){\makebox(0,0)[r]{\strut{} 10}}%
      \put(594,3767){\makebox(0,0)[r]{\strut{} 20}}%
      \put(594,4379){\makebox(0,0)[r]{\strut{} 30}}%
      \put(726,484){\makebox(0,0){\strut{} 8}}%
      \put(1651,484){\makebox(0,0){\strut{} 8.5}}%
      \put(2576,484){\makebox(0,0){\strut{} 9}}%
      \put(3501,484){\makebox(0,0){\strut{} 9.5}}%
      \put(4425,484){\makebox(0,0){\strut{} 10}}%
      \put(5350,484){\makebox(0,0){\strut{} 10.5}}%
      \put(6275,484){\makebox(0,0){\strut{} 11}}%
      \put(6407,704){\makebox(0,0)[l]{\strut{}-8}}%
      \put(6407,1163){\makebox(0,0)[l]{\strut{}-7}}%
      \put(6407,1623){\makebox(0,0)[l]{\strut{}-6}}%
      \put(6407,2082){\makebox(0,0)[l]{\strut{}-5}}%
      \put(6407,2542){\makebox(0,0)[l]{\strut{}-4}}%
      \put(6407,3001){\makebox(0,0)[l]{\strut{}-3}}%
      \put(6407,3460){\makebox(0,0)[l]{\strut{}-2}}%
      \put(6407,3920){\makebox(0,0)[l]{\strut{}-1}}%
      \put(6407,4379){\makebox(0,0)[l]{\strut{} 0}}%
      \put(3500,154){\makebox(0,0){\strut{}$f$ (GHz)}}%
      \put(3500,4709){\makebox(0,0){\strut{}$\epsilon^{\mathrm{eff}}_{rr}=$3.050, $\epsilon^{\mathrm{eff}}_{zz}=$4.186, $\Lambda=1.150$}}%
    }%
    \gplgaddtomacro\gplfronttext{%
      \csname LTb\endcsname%
      \put(5288,1317){\makebox(0,0)[r]{\strut{}infinite TWT}}%
      \csname LTb\endcsname%
      \put(5288,1097){\makebox(0,0)[r]{\strut{}short TWT with beam}}%
      \csname LTb\endcsname%
      \put(5288,877){\makebox(0,0)[r]{\strut{}short TWT without beam}}%
    }%
    \gplbacktext
    \put(0,0){\includegraphics{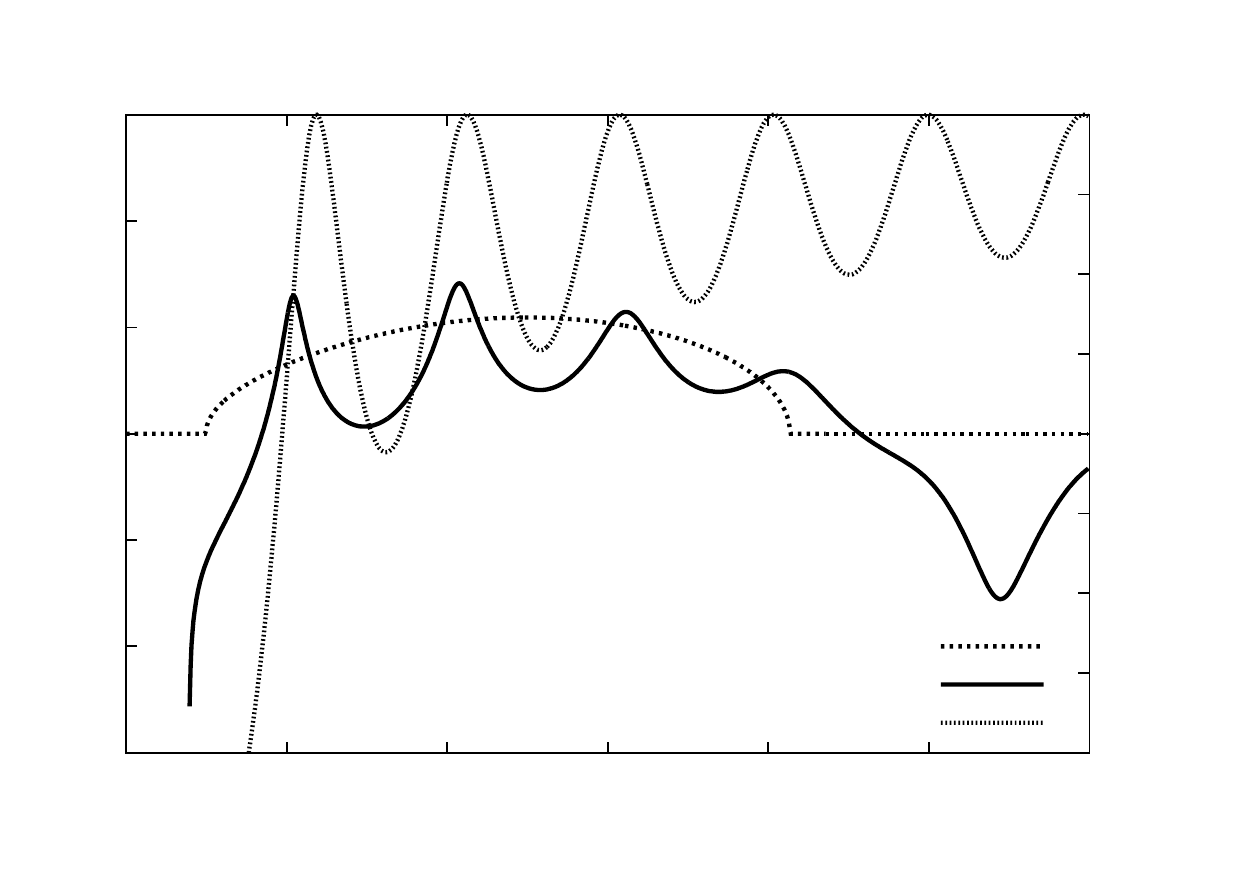}}%
    \gplfronttext
  \end{picture}%
\endgroup

%% file: LozengeLambda11.tex
\begingroup
  \makeatletter
  \providecommand\color[2][]{%
    \GenericError{(gnuplot) \space\space\space\@spaces}{%
      Package color not loaded in conjunction with
      terminal option `colourtext'%
    }{See the gnuplot documentation for explanation.%
    }{Either use 'blacktext' in gnuplot or load the package
      color.sty in LaTeX.}%
    \renewcommand\color[2][]{}%
  }%
  \providecommand\includegraphics[2][]{%
    \GenericError{(gnuplot) \space\space\space\@spaces}{%
      Package graphicx or graphics not loaded%
    }{See the gnuplot documentation for explanation.%
    }{The gnuplot epslatex terminal needs graphicx.sty or graphics.sty.}%
    \renewcommand\includegraphics[2][]{}%
  }%
  \providecommand\rotatebox[2]{#2}%
  \@ifundefined{ifGPcolor}{%
    \newif\ifGPcolor
    \GPcolorfalse
  }{}%
  \@ifundefined{ifGPblacktext}{%
    \newif\ifGPblacktext
    \GPblacktexttrue
  }{}%
  \let\gplgaddtomacro\g@addto@macro
  \gdef\gplbacktext{}%
  \gdef\gplfronttext{}%
  \makeatother
  \ifGPblacktext
    \def\colorrgb#1{}%
    \def\colorgray#1{}%
  \else
    \ifGPcolor
      \def\colorrgb#1{\color[rgb]{#1}}%
      \def\colorgray#1{\color[gray]{#1}}%
      \expandafter\def\csname LTw\endcsname{\color{white}}%
      \expandafter\def\csname LTb\endcsname{\color{black}}%
      \expandafter\def\csname LTa\endcsname{\color{black}}%
      \expandafter\def\csname LT0\endcsname{\color[rgb]{1,0,0}}%
      \expandafter\def\csname LT1\endcsname{\color[rgb]{0,1,0}}%
      \expandafter\def\csname LT2\endcsname{\color[rgb]{0,0,1}}%
      \expandafter\def\csname LT3\endcsname{\color[rgb]{1,0,1}}%
      \expandafter\def\csname LT4\endcsname{\color[rgb]{0,1,1}}%
      \expandafter\def\csname LT5\endcsname{\color[rgb]{1,1,0}}%
      \expandafter\def\csname LT6\endcsname{\color[rgb]{0,0,0}}%
      \expandafter\def\csname LT7\endcsname{\color[rgb]{1,0.3,0}}%
      \expandafter\def\csname LT8\endcsname{\color[rgb]{0.5,0.5,0.5}}%
    \else
      \def\colorrgb#1{\color{black}}%
      \def\colorgray#1{\color[gray]{#1}}%
      \expandafter\def\csname LTw\endcsname{\color{white}}%
      \expandafter\def\csname LTb\endcsname{\color{black}}%
      \expandafter\def\csname LTa\endcsname{\color{black}}%
      \expandafter\def\csname LT0\endcsname{\color{black}}%
      \expandafter\def\csname LT1\endcsname{\color{black}}%
      \expandafter\def\csname LT2\endcsname{\color{black}}%
      \expandafter\def\csname LT3\endcsname{\color{black}}%
      \expandafter\def\csname LT4\endcsname{\color{black}}%
      \expandafter\def\csname LT5\endcsname{\color{black}}%
      \expandafter\def\csname LT6\endcsname{\color{black}}%
      \expandafter\def\csname LT7\endcsname{\color{black}}%
      \expandafter\def\csname LT8\endcsname{\color{black}}%
    \fi
  \fi
  \setlength{\unitlength}{0.0500bp}%
  \begin{picture}(7200.00,5040.00)%
    \gplgaddtomacro\gplbacktext{%
      \csname LTb\endcsname%
      \put(594,704){\makebox(0,0)[r]{\strut{}-30}}%
      \put(594,1317){\makebox(0,0)[r]{\strut{}-20}}%
      \put(594,1929){\makebox(0,0)[r]{\strut{}-10}}%
      \put(594,2542){\makebox(0,0)[r]{\strut{} 0}}%
      \put(594,3154){\makebox(0,0)[r]{\strut{} 10}}%
      \put(594,3767){\makebox(0,0)[r]{\strut{} 20}}%
      \put(594,4379){\makebox(0,0)[r]{\strut{} 30}}%
      \put(726,484){\makebox(0,0){\strut{} 7.5}}%
      \put(1651,484){\makebox(0,0){\strut{} 8}}%
      \put(2576,484){\makebox(0,0){\strut{} 8.5}}%
      \put(3501,484){\makebox(0,0){\strut{} 9}}%
      \put(4425,484){\makebox(0,0){\strut{} 9.5}}%
      \put(5350,484){\makebox(0,0){\strut{} 10}}%
      \put(6275,484){\makebox(0,0){\strut{} 10.5}}%
      \put(6407,704){\makebox(0,0)[l]{\strut{}-8}}%
      \put(6407,1163){\makebox(0,0)[l]{\strut{}-7}}%
      \put(6407,1623){\makebox(0,0)[l]{\strut{}-6}}%
      \put(6407,2082){\makebox(0,0)[l]{\strut{}-5}}%
      \put(6407,2542){\makebox(0,0)[l]{\strut{}-4}}%
      \put(6407,3001){\makebox(0,0)[l]{\strut{}-3}}%
      \put(6407,3460){\makebox(0,0)[l]{\strut{}-2}}%
      \put(6407,3920){\makebox(0,0)[l]{\strut{}-1}}%
      \put(6407,4379){\makebox(0,0)[l]{\strut{} 0}}%
      \put(3500,154){\makebox(0,0){\strut{}$f$ (GHz)}}%
      \put(3500,4709){\makebox(0,0){\strut{}$\epsilon^{\mathrm{eff}}_{rr}=$2.845, $\epsilon^{\mathrm{eff}}_{zz}=$4.812, $\Lambda=1.250$}}%
    }%
    \gplgaddtomacro\gplfronttext{%
      \csname LTb\endcsname%
      \put(5288,1317){\makebox(0,0)[r]{\strut{}infinite TWT}}%
      \csname LTb\endcsname%
      \put(5288,1097){\makebox(0,0)[r]{\strut{}short TWT with beam}}%
      \csname LTb\endcsname%
      \put(5288,877){\makebox(0,0)[r]{\strut{}short TWT without beam}}%
    }%
    \gplbacktext
    \put(0,0){\includegraphics{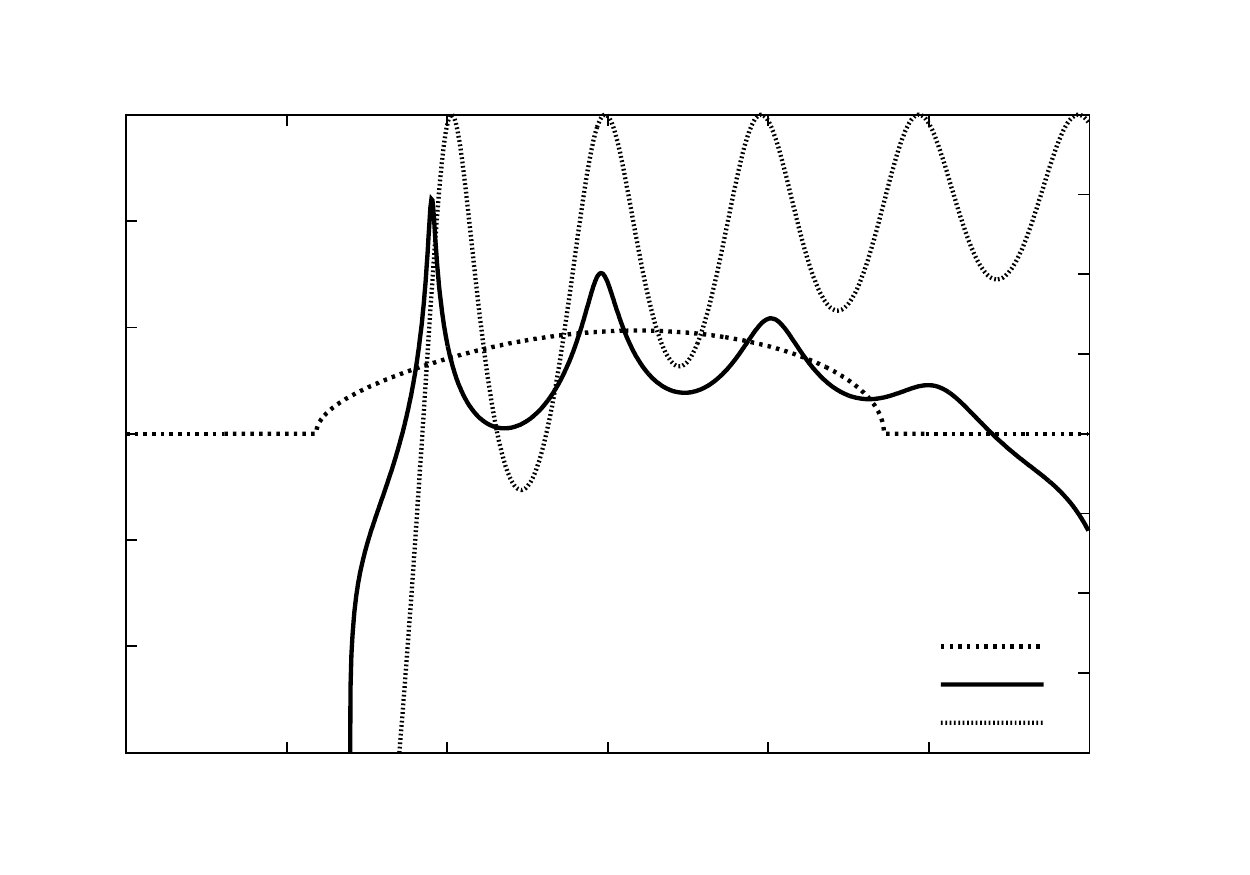}}%
    \gplfronttext
  \end{picture}%
\endgroup

%% file: EllipseLambda3.tex
\begingroup
  \makeatletter
  \providecommand\color[2][]{%
    \GenericError{(gnuplot) \space\space\space\@spaces}{%
      Package color not loaded in conjunction with
      terminal option `colourtext'%
    }{See the gnuplot documentation for explanation.%
    }{Either use 'blacktext' in gnuplot or load the package
      color.sty in LaTeX.}%
    \renewcommand\color[2][]{}%
  }%
  \providecommand\includegraphics[2][]{%
    \GenericError{(gnuplot) \space\space\space\@spaces}{%
      Package graphicx or graphics not loaded%
    }{See the gnuplot documentation for explanation.%
    }{The gnuplot epslatex terminal needs graphicx.sty or graphics.sty.}%
    \renewcommand\includegraphics[2][]{}%
  }%
  \providecommand\rotatebox[2]{#2}%
  \@ifundefined{ifGPcolor}{%
    \newif\ifGPcolor
    \GPcolorfalse
  }{}%
  \@ifundefined{ifGPblacktext}{%
    \newif\ifGPblacktext
    \GPblacktexttrue
  }{}%
  \let\gplgaddtomacro\g@addto@macro
  \gdef\gplbacktext{}%
  \gdef\gplfronttext{}%
  \makeatother
  \ifGPblacktext
    \def\colorrgb#1{}%
    \def\colorgray#1{}%
  \else
    \ifGPcolor
      \def\colorrgb#1{\color[rgb]{#1}}%
      \def\colorgray#1{\color[gray]{#1}}%
      \expandafter\def\csname LTw\endcsname{\color{white}}%
      \expandafter\def\csname LTb\endcsname{\color{black}}%
      \expandafter\def\csname LTa\endcsname{\color{black}}%
      \expandafter\def\csname LT0\endcsname{\color[rgb]{1,0,0}}%
      \expandafter\def\csname LT1\endcsname{\color[rgb]{0,1,0}}%
      \expandafter\def\csname LT2\endcsname{\color[rgb]{0,0,1}}%
      \expandafter\def\csname LT3\endcsname{\color[rgb]{1,0,1}}%
      \expandafter\def\csname LT4\endcsname{\color[rgb]{0,1,1}}%
      \expandafter\def\csname LT5\endcsname{\color[rgb]{1,1,0}}%
      \expandafter\def\csname LT6\endcsname{\color[rgb]{0,0,0}}%
      \expandafter\def\csname LT7\endcsname{\color[rgb]{1,0.3,0}}%
      \expandafter\def\csname LT8\endcsname{\color[rgb]{0.5,0.5,0.5}}%
    \else
      \def\colorrgb#1{\color{black}}%
      \def\colorgray#1{\color[gray]{#1}}%
      \expandafter\def\csname LTw\endcsname{\color{white}}%
      \expandafter\def\csname LTb\endcsname{\color{black}}%
      \expandafter\def\csname LTa\endcsname{\color{black}}%
      \expandafter\def\csname LT0\endcsname{\color{black}}%
      \expandafter\def\csname LT1\endcsname{\color{black}}%
      \expandafter\def\csname LT2\endcsname{\color{black}}%
      \expandafter\def\csname LT3\endcsname{\color{black}}%
      \expandafter\def\csname LT4\endcsname{\color{black}}%
      \expandafter\def\csname LT5\endcsname{\color{black}}%
      \expandafter\def\csname LT6\endcsname{\color{black}}%
      \expandafter\def\csname LT7\endcsname{\color{black}}%
      \expandafter\def\csname LT8\endcsname{\color{black}}%
    \fi
  \fi
  \setlength{\unitlength}{0.0500bp}%
  \begin{picture}(7200.00,5040.00)%
    \gplgaddtomacro\gplbacktext{%
      \csname LTb\endcsname%
      \put(594,704){\makebox(0,0)[r]{\strut{}-30}}%
      \put(594,1317){\makebox(0,0)[r]{\strut{}-20}}%
      \put(594,1929){\makebox(0,0)[r]{\strut{}-10}}%
      \put(594,2542){\makebox(0,0)[r]{\strut{} 0}}%
      \put(594,3154){\makebox(0,0)[r]{\strut{} 10}}%
      \put(594,3767){\makebox(0,0)[r]{\strut{} 20}}%
      \put(594,4379){\makebox(0,0)[r]{\strut{} 30}}%
      \put(1274,484){\makebox(0,0){\strut{} 8.5}}%
      \put(2465,484){\makebox(0,0){\strut{} 9}}%
      \put(3655,484){\makebox(0,0){\strut{} 9.5}}%
      \put(4846,484){\makebox(0,0){\strut{} 10}}%
      \put(6037,484){\makebox(0,0){\strut{} 10.5}}%
      \put(6407,704){\makebox(0,0)[l]{\strut{}-8}}%
      \put(6407,1163){\makebox(0,0)[l]{\strut{}-7}}%
      \put(6407,1623){\makebox(0,0)[l]{\strut{}-6}}%
      \put(6407,2082){\makebox(0,0)[l]{\strut{}-5}}%
      \put(6407,2542){\makebox(0,0)[l]{\strut{}-4}}%
      \put(6407,3001){\makebox(0,0)[l]{\strut{}-3}}%
      \put(6407,3460){\makebox(0,0)[l]{\strut{}-2}}%
      \put(6407,3920){\makebox(0,0)[l]{\strut{}-1}}%
      \put(6407,4379){\makebox(0,0)[l]{\strut{} 0}}%
      \put(3500,154){\makebox(0,0){\strut{}$f$ (GHz)}}%
      \put(3500,4709){\makebox(0,0){\strut{}$\epsilon^{\mathrm{eff}}_{rr}=$4.644, $\epsilon^{\mathrm{eff}}_{zz}=$2.936, $\Lambda=0.850$}}%
    }%
    \gplgaddtomacro\gplfronttext{%
      \csname LTb\endcsname%
      \put(5288,1317){\makebox(0,0)[r]{\strut{}infinite TWT}}%
      \csname LTb\endcsname%
      \put(5288,1097){\makebox(0,0)[r]{\strut{}short TWT with beam}}%
      \csname LTb\endcsname%
      \put(5288,877){\makebox(0,0)[r]{\strut{}short TWT without beam}}%
    }%
    \gplbacktext
    \put(0,0){\includegraphics{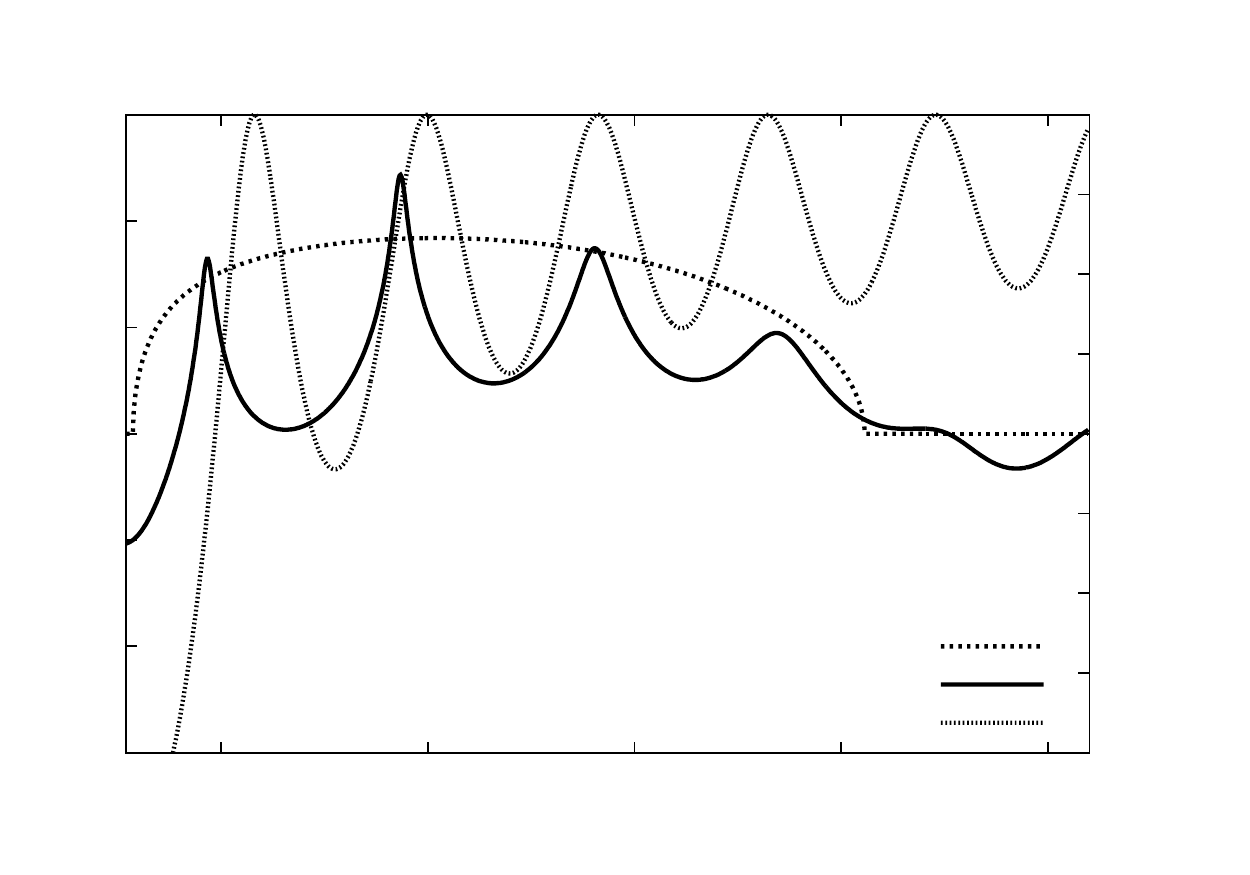}}%
    \gplfronttext
  \end{picture}%
\endgroup

%% file: EllipseLambda7.tex
\begingroup
  \makeatletter
  \providecommand\color[2][]{%
    \GenericError{(gnuplot) \space\space\space\@spaces}{%
      Package color not loaded in conjunction with
      terminal option `colourtext'%
    }{See the gnuplot documentation for explanation.%
    }{Either use 'blacktext' in gnuplot or load the package
      color.sty in LaTeX.}%
    \renewcommand\color[2][]{}%
  }%
  \providecommand\includegraphics[2][]{%
    \GenericError{(gnuplot) \space\space\space\@spaces}{%
      Package graphicx or graphics not loaded%
    }{See the gnuplot documentation for explanation.%
    }{The gnuplot epslatex terminal needs graphicx.sty or graphics.sty.}%
    \renewcommand\includegraphics[2][]{}%
  }%
  \providecommand\rotatebox[2]{#2}%
  \@ifundefined{ifGPcolor}{%
    \newif\ifGPcolor
    \GPcolorfalse
  }{}%
  \@ifundefined{ifGPblacktext}{%
    \newif\ifGPblacktext
    \GPblacktexttrue
  }{}%
  \let\gplgaddtomacro\g@addto@macro
  \gdef\gplbacktext{}%
  \gdef\gplfronttext{}%
  \makeatother
  \ifGPblacktext
    \def\colorrgb#1{}%
    \def\colorgray#1{}%
  \else
    \ifGPcolor
      \def\colorrgb#1{\color[rgb]{#1}}%
      \def\colorgray#1{\color[gray]{#1}}%
      \expandafter\def\csname LTw\endcsname{\color{white}}%
      \expandafter\def\csname LTb\endcsname{\color{black}}%
      \expandafter\def\csname LTa\endcsname{\color{black}}%
      \expandafter\def\csname LT0\endcsname{\color[rgb]{1,0,0}}%
      \expandafter\def\csname LT1\endcsname{\color[rgb]{0,1,0}}%
      \expandafter\def\csname LT2\endcsname{\color[rgb]{0,0,1}}%
      \expandafter\def\csname LT3\endcsname{\color[rgb]{1,0,1}}%
      \expandafter\def\csname LT4\endcsname{\color[rgb]{0,1,1}}%
      \expandafter\def\csname LT5\endcsname{\color[rgb]{1,1,0}}%
      \expandafter\def\csname LT6\endcsname{\color[rgb]{0,0,0}}%
      \expandafter\def\csname LT7\endcsname{\color[rgb]{1,0.3,0}}%
      \expandafter\def\csname LT8\endcsname{\color[rgb]{0.5,0.5,0.5}}%
    \else
      \def\colorrgb#1{\color{black}}%
      \def\colorgray#1{\color[gray]{#1}}%
      \expandafter\def\csname LTw\endcsname{\color{white}}%
      \expandafter\def\csname LTb\endcsname{\color{black}}%
      \expandafter\def\csname LTa\endcsname{\color{black}}%
      \expandafter\def\csname LT0\endcsname{\color{black}}%
      \expandafter\def\csname LT1\endcsname{\color{black}}%
      \expandafter\def\csname LT2\endcsname{\color{black}}%
      \expandafter\def\csname LT3\endcsname{\color{black}}%
      \expandafter\def\csname LT4\endcsname{\color{black}}%
      \expandafter\def\csname LT5\endcsname{\color{black}}%
      \expandafter\def\csname LT6\endcsname{\color{black}}%
      \expandafter\def\csname LT7\endcsname{\color{black}}%
      \expandafter\def\csname LT8\endcsname{\color{black}}%
    \fi
  \fi
  \setlength{\unitlength}{0.0500bp}%
  \begin{picture}(7200.00,5040.00)%
    \gplgaddtomacro\gplbacktext{%
      \csname LTb\endcsname%
      \put(594,704){\makebox(0,0)[r]{\strut{}-30}}%
      \put(594,1317){\makebox(0,0)[r]{\strut{}-20}}%
      \put(594,1929){\makebox(0,0)[r]{\strut{}-10}}%
      \put(594,2542){\makebox(0,0)[r]{\strut{} 0}}%
      \put(594,3154){\makebox(0,0)[r]{\strut{} 10}}%
      \put(594,3767){\makebox(0,0)[r]{\strut{} 20}}%
      \put(594,4379){\makebox(0,0)[r]{\strut{} 30}}%
      \put(726,484){\makebox(0,0){\strut{} 8}}%
      \put(1651,484){\makebox(0,0){\strut{} 8.5}}%
      \put(2576,484){\makebox(0,0){\strut{} 9}}%
      \put(3501,484){\makebox(0,0){\strut{} 9.5}}%
      \put(4425,484){\makebox(0,0){\strut{} 10}}%
      \put(5350,484){\makebox(0,0){\strut{} 10.5}}%
      \put(6275,484){\makebox(0,0){\strut{} 11}}%
      \put(6407,704){\makebox(0,0)[l]{\strut{}-8}}%
      \put(6407,1163){\makebox(0,0)[l]{\strut{}-7}}%
      \put(6407,1623){\makebox(0,0)[l]{\strut{}-6}}%
      \put(6407,2082){\makebox(0,0)[l]{\strut{}-5}}%
      \put(6407,2542){\makebox(0,0)[l]{\strut{}-4}}%
      \put(6407,3001){\makebox(0,0)[l]{\strut{}-3}}%
      \put(6407,3460){\makebox(0,0)[l]{\strut{}-2}}%
      \put(6407,3920){\makebox(0,0)[l]{\strut{}-1}}%
      \put(6407,4379){\makebox(0,0)[l]{\strut{} 0}}%
      \put(3500,154){\makebox(0,0){\strut{}$f$ (GHz)}}%
      \put(3500,4709){\makebox(0,0){\strut{}$\epsilon^{\mathrm{eff}}_{rr}=$3.298, $\epsilon^{\mathrm{eff}}_{zz}=$3.752, $\Lambda=1.050$}}%
    }%
    \gplgaddtomacro\gplfronttext{%
      \csname LTb\endcsname%
      \put(5288,1317){\makebox(0,0)[r]{\strut{}infinite TWT}}%
      \csname LTb\endcsname%
      \put(5288,1097){\makebox(0,0)[r]{\strut{}short TWT with beam}}%
      \csname LTb\endcsname%
      \put(5288,877){\makebox(0,0)[r]{\strut{}short TWT without beam}}%
    }%
    \gplbacktext
    \put(0,0){\includegraphics{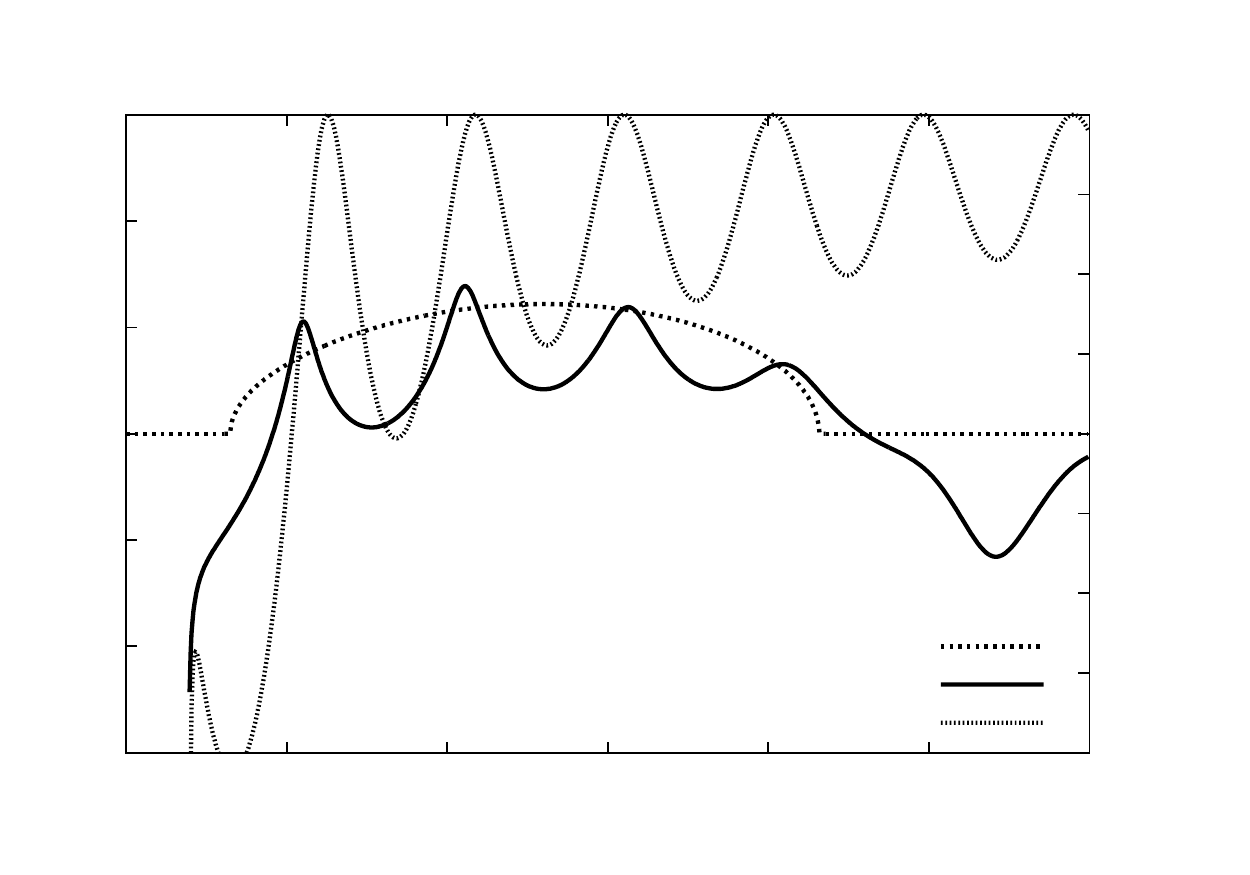}}%
    \gplfronttext
  \end{picture}%
\endgroup

%% file: EllipseLambda9.tex
\begingroup
  \makeatletter
  \providecommand\color[2][]{%
    \GenericError{(gnuplot) \space\space\space\@spaces}{%
      Package color not loaded in conjunction with
      terminal option `colourtext'%
    }{See the gnuplot documentation for explanation.%
    }{Either use 'blacktext' in gnuplot or load the package
      color.sty in LaTeX.}%
    \renewcommand\color[2][]{}%
  }%
  \providecommand\includegraphics[2][]{%
    \GenericError{(gnuplot) \space\space\space\@spaces}{%
      Package graphicx or graphics not loaded%
    }{See the gnuplot documentation for explanation.%
    }{The gnuplot epslatex terminal needs graphicx.sty or graphics.sty.}%
    \renewcommand\includegraphics[2][]{}%
  }%
  \providecommand\rotatebox[2]{#2}%
  \@ifundefined{ifGPcolor}{%
    \newif\ifGPcolor
    \GPcolorfalse
  }{}%
  \@ifundefined{ifGPblacktext}{%
    \newif\ifGPblacktext
    \GPblacktexttrue
  }{}%
  \let\gplgaddtomacro\g@addto@macro
  \gdef\gplbacktext{}%
  \gdef\gplfronttext{}%
  \makeatother
  \ifGPblacktext
    \def\colorrgb#1{}%
    \def\colorgray#1{}%
  \else
    \ifGPcolor
      \def\colorrgb#1{\color[rgb]{#1}}%
      \def\colorgray#1{\color[gray]{#1}}%
      \expandafter\def\csname LTw\endcsname{\color{white}}%
      \expandafter\def\csname LTb\endcsname{\color{black}}%
      \expandafter\def\csname LTa\endcsname{\color{black}}%
      \expandafter\def\csname LT0\endcsname{\color[rgb]{1,0,0}}%
      \expandafter\def\csname LT1\endcsname{\color[rgb]{0,1,0}}%
      \expandafter\def\csname LT2\endcsname{\color[rgb]{0,0,1}}%
      \expandafter\def\csname LT3\endcsname{\color[rgb]{1,0,1}}%
      \expandafter\def\csname LT4\endcsname{\color[rgb]{0,1,1}}%
      \expandafter\def\csname LT5\endcsname{\color[rgb]{1,1,0}}%
      \expandafter\def\csname LT6\endcsname{\color[rgb]{0,0,0}}%
      \expandafter\def\csname LT7\endcsname{\color[rgb]{1,0.3,0}}%
      \expandafter\def\csname LT8\endcsname{\color[rgb]{0.5,0.5,0.5}}%
    \else
      \def\colorrgb#1{\color{black}}%
      \def\colorgray#1{\color[gray]{#1}}%
      \expandafter\def\csname LTw\endcsname{\color{white}}%
      \expandafter\def\csname LTb\endcsname{\color{black}}%
      \expandafter\def\csname LTa\endcsname{\color{black}}%
      \expandafter\def\csname LT0\endcsname{\color{black}}%
      \expandafter\def\csname LT1\endcsname{\color{black}}%
      \expandafter\def\csname LT2\endcsname{\color{black}}%
      \expandafter\def\csname LT3\endcsname{\color{black}}%
      \expandafter\def\csname LT4\endcsname{\color{black}}%
      \expandafter\def\csname LT5\endcsname{\color{black}}%
      \expandafter\def\csname LT6\endcsname{\color{black}}%
      \expandafter\def\csname LT7\endcsname{\color{black}}%
      \expandafter\def\csname LT8\endcsname{\color{black}}%
    \fi
  \fi
  \setlength{\unitlength}{0.0500bp}%
  \begin{picture}(7200.00,5040.00)%
    \gplgaddtomacro\gplbacktext{%
      \csname LTb\endcsname%
      \put(594,704){\makebox(0,0)[r]{\strut{}-30}}%
      \put(594,1317){\makebox(0,0)[r]{\strut{}-20}}%
      \put(594,1929){\makebox(0,0)[r]{\strut{}-10}}%
      \put(594,2542){\makebox(0,0)[r]{\strut{} 0}}%
      \put(594,3154){\makebox(0,0)[r]{\strut{} 10}}%
      \put(594,3767){\makebox(0,0)[r]{\strut{} 20}}%
      \put(594,4379){\makebox(0,0)[r]{\strut{} 30}}%
      \put(726,484){\makebox(0,0){\strut{} 8}}%
      \put(1651,484){\makebox(0,0){\strut{} 8.5}}%
      \put(2576,484){\makebox(0,0){\strut{} 9}}%
      \put(3501,484){\makebox(0,0){\strut{} 9.5}}%
      \put(4425,484){\makebox(0,0){\strut{} 10}}%
      \put(5350,484){\makebox(0,0){\strut{} 10.5}}%
      \put(6275,484){\makebox(0,0){\strut{} 11}}%
      \put(6407,704){\makebox(0,0)[l]{\strut{}-8}}%
      \put(6407,1163){\makebox(0,0)[l]{\strut{}-7}}%
      \put(6407,1623){\makebox(0,0)[l]{\strut{}-6}}%
      \put(6407,2082){\makebox(0,0)[l]{\strut{}-5}}%
      \put(6407,2542){\makebox(0,0)[l]{\strut{}-4}}%
      \put(6407,3001){\makebox(0,0)[l]{\strut{}-3}}%
      \put(6407,3460){\makebox(0,0)[l]{\strut{}-2}}%
      \put(6407,3920){\makebox(0,0)[l]{\strut{}-1}}%
      \put(6407,4379){\makebox(0,0)[l]{\strut{} 0}}%
      \put(3500,154){\makebox(0,0){\strut{}$f$ (GHz)}}%
      \put(3500,4709){\makebox(0,0){\strut{}$\epsilon^{\mathrm{eff}}_{rr}=$2.998, $\epsilon^{\mathrm{eff}}_{zz}=$4.414, $\Lambda=1.150$}}%
    }%
    \gplgaddtomacro\gplfronttext{%
      \csname LTb\endcsname%
      \put(5288,1317){\makebox(0,0)[r]{\strut{}infinite TWT}}%
      \csname LTb\endcsname%
      \put(5288,1097){\makebox(0,0)[r]{\strut{}short TWT with beam}}%
      \csname LTb\endcsname%
      \put(5288,877){\makebox(0,0)[r]{\strut{}short TWT without beam}}%
    }%
    \gplbacktext
    \put(0,0){\includegraphics{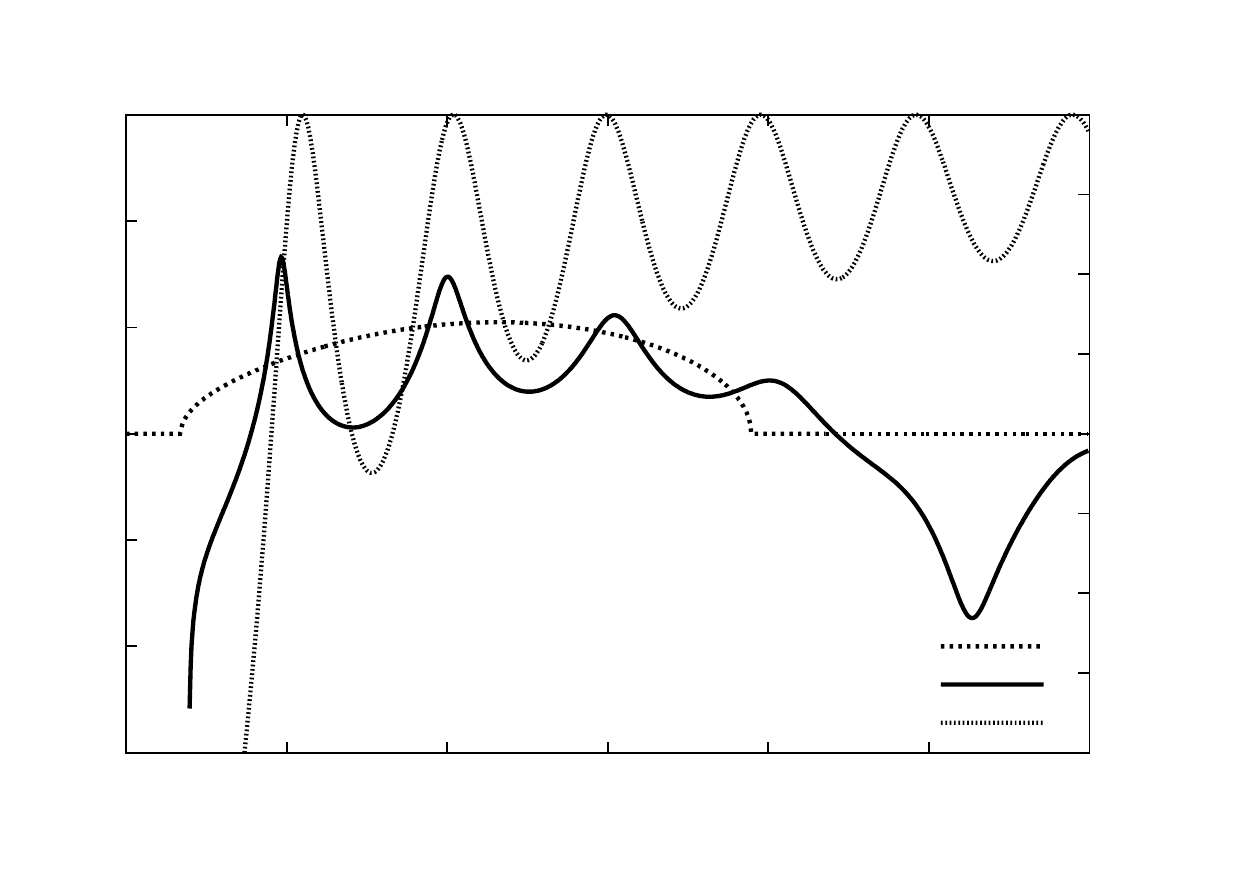}}%
    \gplfronttext
  \end{picture}%
\endgroup

%% file: EllipseLambda11.tex
\begingroup
  \makeatletter
  \providecommand\color[2][]{%
    \GenericError{(gnuplot) \space\space\space\@spaces}{%
      Package color not loaded in conjunction with
      terminal option `colourtext'%
    }{See the gnuplot documentation for explanation.%
    }{Either use 'blacktext' in gnuplot or load the package
      color.sty in LaTeX.}%
    \renewcommand\color[2][]{}%
  }%
  \providecommand\includegraphics[2][]{%
    \GenericError{(gnuplot) \space\space\space\@spaces}{%
      Package graphicx or graphics not loaded%
    }{See the gnuplot documentation for explanation.%
    }{The gnuplot epslatex terminal needs graphicx.sty or graphics.sty.}%
    \renewcommand\includegraphics[2][]{}%
  }%
  \providecommand\rotatebox[2]{#2}%
  \@ifundefined{ifGPcolor}{%
    \newif\ifGPcolor
    \GPcolorfalse
  }{}%
  \@ifundefined{ifGPblacktext}{%
    \newif\ifGPblacktext
    \GPblacktexttrue
  }{}%
  \let\gplgaddtomacro\g@addto@macro
  \gdef\gplbacktext{}%
  \gdef\gplfronttext{}%
  \makeatother
  \ifGPblacktext
    \def\colorrgb#1{}%
    \def\colorgray#1{}%
  \else
    \ifGPcolor
      \def\colorrgb#1{\color[rgb]{#1}}%
      \def\colorgray#1{\color[gray]{#1}}%
      \expandafter\def\csname LTw\endcsname{\color{white}}%
      \expandafter\def\csname LTb\endcsname{\color{black}}%
      \expandafter\def\csname LTa\endcsname{\color{black}}%
      \expandafter\def\csname LT0\endcsname{\color[rgb]{1,0,0}}%
      \expandafter\def\csname LT1\endcsname{\color[rgb]{0,1,0}}%
      \expandafter\def\csname LT2\endcsname{\color[rgb]{0,0,1}}%
      \expandafter\def\csname LT3\endcsname{\color[rgb]{1,0,1}}%
      \expandafter\def\csname LT4\endcsname{\color[rgb]{0,1,1}}%
      \expandafter\def\csname LT5\endcsname{\color[rgb]{1,1,0}}%
      \expandafter\def\csname LT6\endcsname{\color[rgb]{0,0,0}}%
      \expandafter\def\csname LT7\endcsname{\color[rgb]{1,0.3,0}}%
      \expandafter\def\csname LT8\endcsname{\color[rgb]{0.5,0.5,0.5}}%
    \else
      \def\colorrgb#1{\color{black}}%
      \def\colorgray#1{\color[gray]{#1}}%
      \expandafter\def\csname LTw\endcsname{\color{white}}%
      \expandafter\def\csname LTb\endcsname{\color{black}}%
      \expandafter\def\csname LTa\endcsname{\color{black}}%
      \expandafter\def\csname LT0\endcsname{\color{black}}%
      \expandafter\def\csname LT1\endcsname{\color{black}}%
      \expandafter\def\csname LT2\endcsname{\color{black}}%
      \expandafter\def\csname LT3\endcsname{\color{black}}%
      \expandafter\def\csname LT4\endcsname{\color{black}}%
      \expandafter\def\csname LT5\endcsname{\color{black}}%
      \expandafter\def\csname LT6\endcsname{\color{black}}%
      \expandafter\def\csname LT7\endcsname{\color{black}}%
      \expandafter\def\csname LT8\endcsname{\color{black}}%
    \fi
  \fi
  \setlength{\unitlength}{0.0500bp}%
  \begin{picture}(7200.00,5040.00)%
    \gplgaddtomacro\gplbacktext{%
      \csname LTb\endcsname%
      \put(594,704){\makebox(0,0)[r]{\strut{}-30}}%
      \put(594,1317){\makebox(0,0)[r]{\strut{}-20}}%
      \put(594,1929){\makebox(0,0)[r]{\strut{}-10}}%
      \put(594,2542){\makebox(0,0)[r]{\strut{} 0}}%
      \put(594,3154){\makebox(0,0)[r]{\strut{} 10}}%
      \put(594,3767){\makebox(0,0)[r]{\strut{} 20}}%
      \put(594,4379){\makebox(0,0)[r]{\strut{} 30}}%
      \put(726,484){\makebox(0,0){\strut{} 7.5}}%
      \put(1651,484){\makebox(0,0){\strut{} 8}}%
      \put(2576,484){\makebox(0,0){\strut{} 8.5}}%
      \put(3501,484){\makebox(0,0){\strut{} 9}}%
      \put(4425,484){\makebox(0,0){\strut{} 9.5}}%
      \put(5350,484){\makebox(0,0){\strut{} 10}}%
      \put(6275,484){\makebox(0,0){\strut{} 10.5}}%
      \put(6407,704){\makebox(0,0)[l]{\strut{}-8}}%
      \put(6407,1163){\makebox(0,0)[l]{\strut{}-7}}%
      \put(6407,1623){\makebox(0,0)[l]{\strut{}-6}}%
      \put(6407,2082){\makebox(0,0)[l]{\strut{}-5}}%
      \put(6407,2542){\makebox(0,0)[l]{\strut{}-4}}%
      \put(6407,3001){\makebox(0,0)[l]{\strut{}-3}}%
      \put(6407,3460){\makebox(0,0)[l]{\strut{}-2}}%
      \put(6407,3920){\makebox(0,0)[l]{\strut{}-1}}%
      \put(6407,4379){\makebox(0,0)[l]{\strut{} 0}}%
      \put(3500,154){\makebox(0,0){\strut{}$f$ (GHz)}}%
      \put(3500,4709){\makebox(0,0){\strut{}$\epsilon^{\mathrm{eff}}_{rr}=$2.787, $\epsilon^{\mathrm{eff}}_{zz}=$5.498, $\Lambda=1.250$}}%
    }%
    \gplgaddtomacro\gplfronttext{%
      \csname LTb\endcsname%
      \put(5288,1317){\makebox(0,0)[r]{\strut{}infinite TWT}}%
      \csname LTb\endcsname%
      \put(5288,1097){\makebox(0,0)[r]{\strut{}short TWT with beam}}%
      \csname LTb\endcsname%
      \put(5288,877){\makebox(0,0)[r]{\strut{}short TWT without beam}}%
    }%
    \gplbacktext
    \put(0,0){\includegraphics{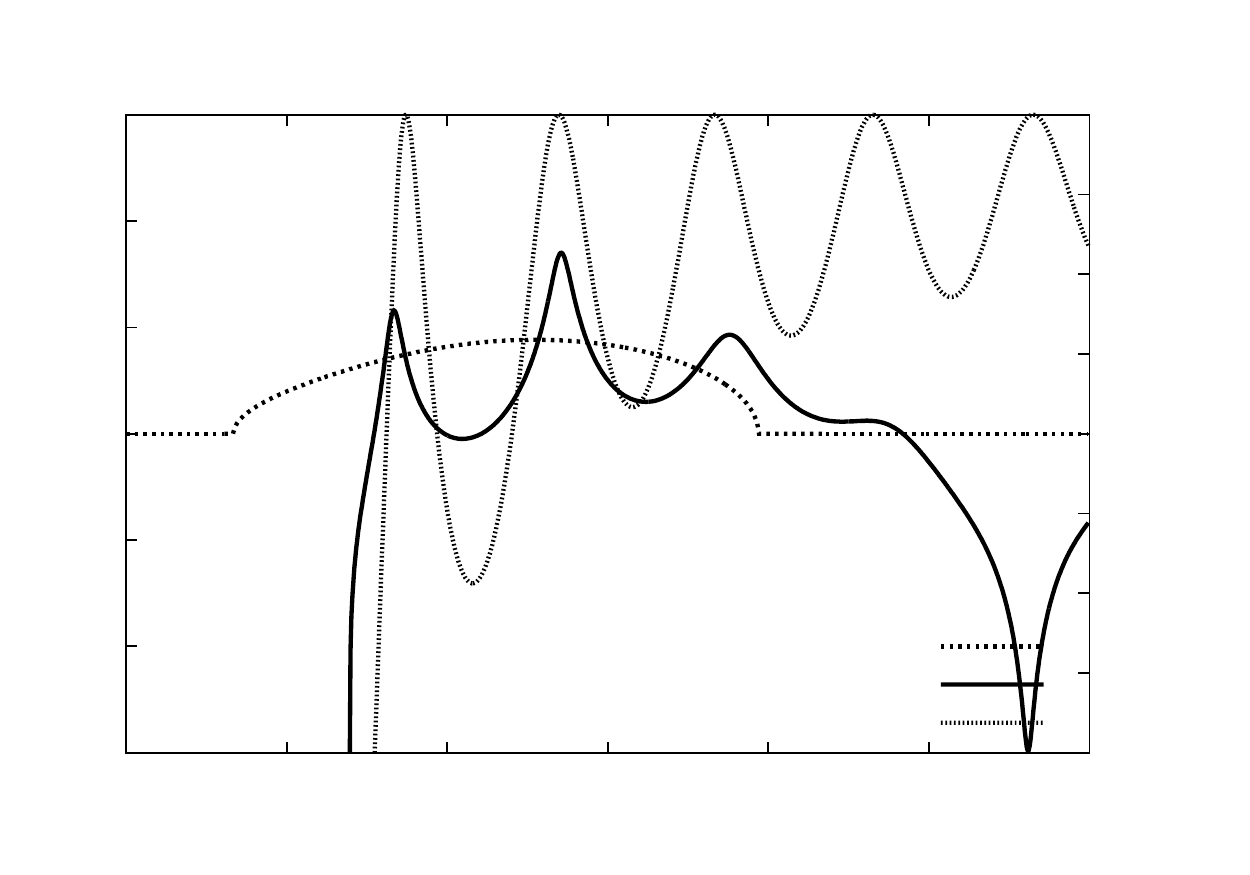}}%
    \gplfronttext
  \end{picture}%
\endgroup

%% file: LozengeTheta0.tex
\begingroup
  \makeatletter
  \providecommand\color[2][]{%
    \GenericError{(gnuplot) \space\space\space\@spaces}{%
      Package color not loaded in conjunction with
      terminal option `colourtext'%
    }{See the gnuplot documentation for explanation.%
    }{Either use 'blacktext' in gnuplot or load the package
      color.sty in LaTeX.}%
    \renewcommand\color[2][]{}%
  }%
  \providecommand\includegraphics[2][]{%
    \GenericError{(gnuplot) \space\space\space\@spaces}{%
      Package graphicx or graphics not loaded%
    }{See the gnuplot documentation for explanation.%
    }{The gnuplot epslatex terminal needs graphicx.sty or graphics.sty.}%
    \renewcommand\includegraphics[2][]{}%
  }%
  \providecommand\rotatebox[2]{#2}%
  \@ifundefined{ifGPcolor}{%
    \newif\ifGPcolor
    \GPcolorfalse
  }{}%
  \@ifundefined{ifGPblacktext}{%
    \newif\ifGPblacktext
    \GPblacktexttrue
  }{}%
  \let\gplgaddtomacro\g@addto@macro
  \gdef\gplbacktext{}%
  \gdef\gplfronttext{}%
  \makeatother
  \ifGPblacktext
    \def\colorrgb#1{}%
    \def\colorgray#1{}%
  \else
    \ifGPcolor
      \def\colorrgb#1{\color[rgb]{#1}}%
      \def\colorgray#1{\color[gray]{#1}}%
      \expandafter\def\csname LTw\endcsname{\color{white}}%
      \expandafter\def\csname LTb\endcsname{\color{black}}%
      \expandafter\def\csname LTa\endcsname{\color{black}}%
      \expandafter\def\csname LT0\endcsname{\color[rgb]{1,0,0}}%
      \expandafter\def\csname LT1\endcsname{\color[rgb]{0,1,0}}%
      \expandafter\def\csname LT2\endcsname{\color[rgb]{0,0,1}}%
      \expandafter\def\csname LT3\endcsname{\color[rgb]{1,0,1}}%
      \expandafter\def\csname LT4\endcsname{\color[rgb]{0,1,1}}%
      \expandafter\def\csname LT5\endcsname{\color[rgb]{1,1,0}}%
      \expandafter\def\csname LT6\endcsname{\color[rgb]{0,0,0}}%
      \expandafter\def\csname LT7\endcsname{\color[rgb]{1,0.3,0}}%
      \expandafter\def\csname LT8\endcsname{\color[rgb]{0.5,0.5,0.5}}%
    \else
      \def\colorrgb#1{\color{black}}%
      \def\colorgray#1{\color[gray]{#1}}%
      \expandafter\def\csname LTw\endcsname{\color{white}}%
      \expandafter\def\csname LTb\endcsname{\color{black}}%
      \expandafter\def\csname LTa\endcsname{\color{black}}%
      \expandafter\def\csname LT0\endcsname{\color{black}}%
      \expandafter\def\csname LT1\endcsname{\color{black}}%
      \expandafter\def\csname LT2\endcsname{\color{black}}%
      \expandafter\def\csname LT3\endcsname{\color{black}}%
      \expandafter\def\csname LT4\endcsname{\color{black}}%
      \expandafter\def\csname LT5\endcsname{\color{black}}%
      \expandafter\def\csname LT6\endcsname{\color{black}}%
      \expandafter\def\csname LT7\endcsname{\color{black}}%
      \expandafter\def\csname LT8\endcsname{\color{black}}%
    \fi
  \fi
  \setlength{\unitlength}{0.0500bp}%
  \begin{picture}(7200.00,5040.00)%
    \gplgaddtomacro\gplbacktext{%
      \csname LTb\endcsname%
      \put(594,704){\makebox(0,0)[r]{\strut{}-30}}%
      \put(594,1317){\makebox(0,0)[r]{\strut{}-20}}%
      \put(594,1929){\makebox(0,0)[r]{\strut{}-10}}%
      \put(594,2542){\makebox(0,0)[r]{\strut{} 0}}%
      \put(594,3154){\makebox(0,0)[r]{\strut{} 10}}%
      \put(594,3767){\makebox(0,0)[r]{\strut{} 20}}%
      \put(594,4379){\makebox(0,0)[r]{\strut{} 30}}%
      \put(1004,484){\makebox(0,0){\strut{} 11}}%
      \put(1560,484){\makebox(0,0){\strut{} 12}}%
      \put(2117,484){\makebox(0,0){\strut{} 13}}%
      \put(2673,484){\makebox(0,0){\strut{} 14}}%
      \put(3229,484){\makebox(0,0){\strut{} 15}}%
      \put(3786,484){\makebox(0,0){\strut{} 16}}%
      \put(4342,484){\makebox(0,0){\strut{} 17}}%
      \put(4898,484){\makebox(0,0){\strut{} 18}}%
      \put(5455,484){\makebox(0,0){\strut{} 19}}%
      \put(6011,484){\makebox(0,0){\strut{} 20}}%
      \put(6143,704){\makebox(0,0)[l]{\strut{}-1}}%
      \put(6143,1439){\makebox(0,0)[l]{\strut{}-0.8}}%
      \put(6143,2174){\makebox(0,0)[l]{\strut{}-0.6}}%
      \put(6143,2909){\makebox(0,0)[l]{\strut{}-0.4}}%
      \put(6143,3644){\makebox(0,0)[l]{\strut{}-0.2}}%
      \put(6143,4379){\makebox(0,0)[l]{\strut{} 0}}%
      \put(3368,154){\makebox(0,0){\strut{}$f$ (GHz)}}%
      \put(3368,4709){\makebox(0,0){\strut{}$\epsilon^{\mathrm{eff}}_{rr}=$1.890, $\epsilon^{\mathrm{eff}}_{zz}=$1.890, $\theta=$0.300}}%
    }%
    \gplgaddtomacro\gplfronttext{%
      \csname LTb\endcsname%
      \put(5024,1317){\makebox(0,0)[r]{\strut{}infinite TWT}}%
      \csname LTb\endcsname%
      \put(5024,1097){\makebox(0,0)[r]{\strut{}short TWT with beam}}%
      \csname LTb\endcsname%
      \put(5024,877){\makebox(0,0)[r]{\strut{}short TWT without beam}}%
    }%
    \gplbacktext
    \put(0,0){\includegraphics{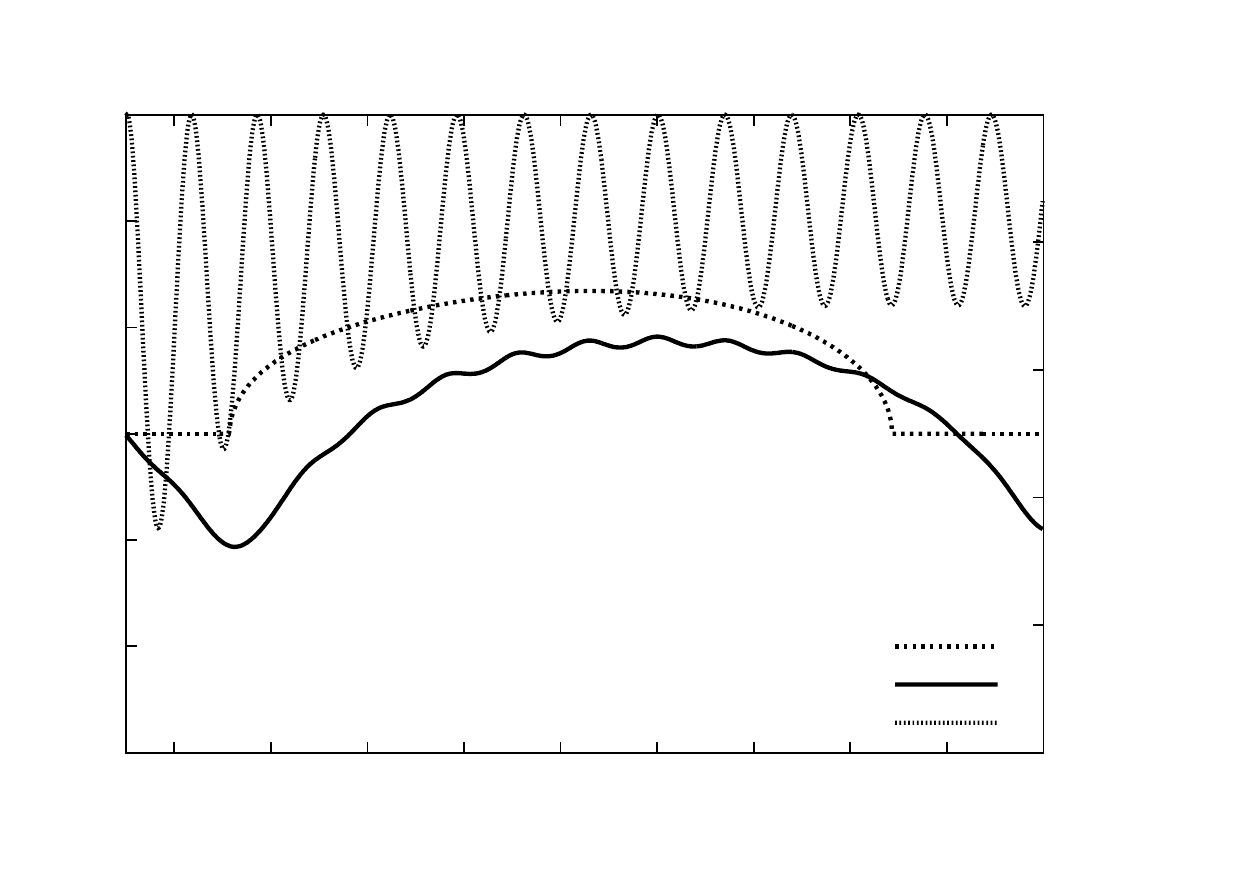}}%
    \gplfronttext
  \end{picture}%
\endgroup

%% file: LozengeTheta1.tex
\begingroup
  \makeatletter
  \providecommand\color[2][]{%
    \GenericError{(gnuplot) \space\space\space\@spaces}{%
      Package color not loaded in conjunction with
      terminal option `colourtext'%
    }{See the gnuplot documentation for explanation.%
    }{Either use 'blacktext' in gnuplot or load the package
      color.sty in LaTeX.}%
    \renewcommand\color[2][]{}%
  }%
  \providecommand\includegraphics[2][]{%
    \GenericError{(gnuplot) \space\space\space\@spaces}{%
      Package graphicx or graphics not loaded%
    }{See the gnuplot documentation for explanation.%
    }{The gnuplot epslatex terminal needs graphicx.sty or graphics.sty.}%
    \renewcommand\includegraphics[2][]{}%
  }%
  \providecommand\rotatebox[2]{#2}%
  \@ifundefined{ifGPcolor}{%
    \newif\ifGPcolor
    \GPcolorfalse
  }{}%
  \@ifundefined{ifGPblacktext}{%
    \newif\ifGPblacktext
    \GPblacktexttrue
  }{}%
  \let\gplgaddtomacro\g@addto@macro
  \gdef\gplbacktext{}%
  \gdef\gplfronttext{}%
  \makeatother
  \ifGPblacktext
    \def\colorrgb#1{}%
    \def\colorgray#1{}%
  \else
    \ifGPcolor
      \def\colorrgb#1{\color[rgb]{#1}}%
      \def\colorgray#1{\color[gray]{#1}}%
      \expandafter\def\csname LTw\endcsname{\color{white}}%
      \expandafter\def\csname LTb\endcsname{\color{black}}%
      \expandafter\def\csname LTa\endcsname{\color{black}}%
      \expandafter\def\csname LT0\endcsname{\color[rgb]{1,0,0}}%
      \expandafter\def\csname LT1\endcsname{\color[rgb]{0,1,0}}%
      \expandafter\def\csname LT2\endcsname{\color[rgb]{0,0,1}}%
      \expandafter\def\csname LT3\endcsname{\color[rgb]{1,0,1}}%
      \expandafter\def\csname LT4\endcsname{\color[rgb]{0,1,1}}%
      \expandafter\def\csname LT5\endcsname{\color[rgb]{1,1,0}}%
      \expandafter\def\csname LT6\endcsname{\color[rgb]{0,0,0}}%
      \expandafter\def\csname LT7\endcsname{\color[rgb]{1,0.3,0}}%
      \expandafter\def\csname LT8\endcsname{\color[rgb]{0.5,0.5,0.5}}%
    \else
      \def\colorrgb#1{\color{black}}%
      \def\colorgray#1{\color[gray]{#1}}%
      \expandafter\def\csname LTw\endcsname{\color{white}}%
      \expandafter\def\csname LTb\endcsname{\color{black}}%
      \expandafter\def\csname LTa\endcsname{\color{black}}%
      \expandafter\def\csname LT0\endcsname{\color{black}}%
      \expandafter\def\csname LT1\endcsname{\color{black}}%
      \expandafter\def\csname LT2\endcsname{\color{black}}%
      \expandafter\def\csname LT3\endcsname{\color{black}}%
      \expandafter\def\csname LT4\endcsname{\color{black}}%
      \expandafter\def\csname LT5\endcsname{\color{black}}%
      \expandafter\def\csname LT6\endcsname{\color{black}}%
      \expandafter\def\csname LT7\endcsname{\color{black}}%
      \expandafter\def\csname LT8\endcsname{\color{black}}%
    \fi
  \fi
  \setlength{\unitlength}{0.0500bp}%
  \begin{picture}(7200.00,5040.00)%
    \gplgaddtomacro\gplbacktext{%
      \csname LTb\endcsname%
      \put(594,704){\makebox(0,0)[r]{\strut{}-30}}%
      \put(594,1317){\makebox(0,0)[r]{\strut{}-20}}%
      \put(594,1929){\makebox(0,0)[r]{\strut{}-10}}%
      \put(594,2542){\makebox(0,0)[r]{\strut{} 0}}%
      \put(594,3154){\makebox(0,0)[r]{\strut{} 10}}%
      \put(594,3767){\makebox(0,0)[r]{\strut{} 20}}%
      \put(594,4379){\makebox(0,0)[r]{\strut{} 30}}%
      \put(1206,484){\makebox(0,0){\strut{} 10}}%
      \put(2167,484){\makebox(0,0){\strut{} 11}}%
      \put(3128,484){\makebox(0,0){\strut{} 12}}%
      \put(4089,484){\makebox(0,0){\strut{} 13}}%
      \put(5050,484){\makebox(0,0){\strut{} 14}}%
      \put(6011,484){\makebox(0,0){\strut{} 15}}%
      \put(6143,704){\makebox(0,0)[l]{\strut{}-1}}%
      \put(6143,1439){\makebox(0,0)[l]{\strut{}-0.8}}%
      \put(6143,2174){\makebox(0,0)[l]{\strut{}-0.6}}%
      \put(6143,2909){\makebox(0,0)[l]{\strut{}-0.4}}%
      \put(6143,3644){\makebox(0,0)[l]{\strut{}-0.2}}%
      \put(6143,4379){\makebox(0,0)[l]{\strut{} 0}}%
      \put(3368,154){\makebox(0,0){\strut{}$f$ (GHz)}}%
      \put(3368,4709){\makebox(0,0){\strut{}$\epsilon^{\mathrm{eff}}_{rr}=$2.378, $\epsilon^{\mathrm{eff}}_{zz}=$2.378, $\theta=$0.400}}%
    }%
    \gplgaddtomacro\gplfronttext{%
      \csname LTb\endcsname%
      \put(5024,1317){\makebox(0,0)[r]{\strut{}infinite TWT}}%
      \csname LTb\endcsname%
      \put(5024,1097){\makebox(0,0)[r]{\strut{}short TWT with beam}}%
      \csname LTb\endcsname%
      \put(5024,877){\makebox(0,0)[r]{\strut{}short TWT without beam}}%
    }%
    \gplbacktext
    \put(0,0){\includegraphics{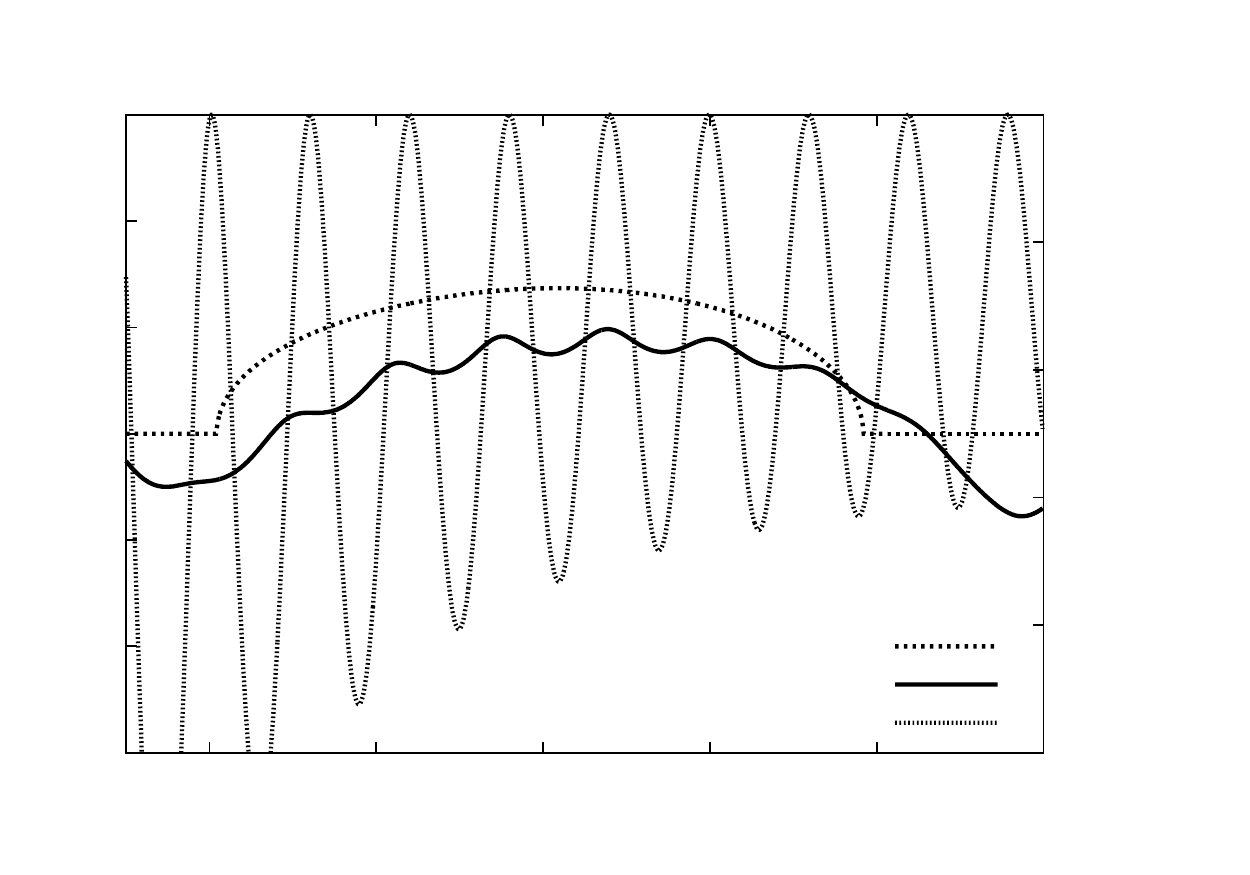}}%
    \gplfronttext
  \end{picture}%
\endgroup

%% file: LozengeTheta2.tex
\begingroup
  \makeatletter
  \providecommand\color[2][]{%
    \GenericError{(gnuplot) \space\space\space\@spaces}{%
      Package color not loaded in conjunction with
      terminal option `colourtext'%
    }{See the gnuplot documentation for explanation.%
    }{Either use 'blacktext' in gnuplot or load the package
      color.sty in LaTeX.}%
    \renewcommand\color[2][]{}%
  }%
  \providecommand\includegraphics[2][]{%
    \GenericError{(gnuplot) \space\space\space\@spaces}{%
      Package graphicx or graphics not loaded%
    }{See the gnuplot documentation for explanation.%
    }{The gnuplot epslatex terminal needs graphicx.sty or graphics.sty.}%
    \renewcommand\includegraphics[2][]{}%
  }%
  \providecommand\rotatebox[2]{#2}%
  \@ifundefined{ifGPcolor}{%
    \newif\ifGPcolor
    \GPcolorfalse
  }{}%
  \@ifundefined{ifGPblacktext}{%
    \newif\ifGPblacktext
    \GPblacktexttrue
  }{}%
  \let\gplgaddtomacro\g@addto@macro
  \gdef\gplbacktext{}%
  \gdef\gplfronttext{}%
  \makeatother
  \ifGPblacktext
    \def\colorrgb#1{}%
    \def\colorgray#1{}%
  \else
    \ifGPcolor
      \def\colorrgb#1{\color[rgb]{#1}}%
      \def\colorgray#1{\color[gray]{#1}}%
      \expandafter\def\csname LTw\endcsname{\color{white}}%
      \expandafter\def\csname LTb\endcsname{\color{black}}%
      \expandafter\def\csname LTa\endcsname{\color{black}}%
      \expandafter\def\csname LT0\endcsname{\color[rgb]{1,0,0}}%
      \expandafter\def\csname LT1\endcsname{\color[rgb]{0,1,0}}%
      \expandafter\def\csname LT2\endcsname{\color[rgb]{0,0,1}}%
      \expandafter\def\csname LT3\endcsname{\color[rgb]{1,0,1}}%
      \expandafter\def\csname LT4\endcsname{\color[rgb]{0,1,1}}%
      \expandafter\def\csname LT5\endcsname{\color[rgb]{1,1,0}}%
      \expandafter\def\csname LT6\endcsname{\color[rgb]{0,0,0}}%
      \expandafter\def\csname LT7\endcsname{\color[rgb]{1,0.3,0}}%
      \expandafter\def\csname LT8\endcsname{\color[rgb]{0.5,0.5,0.5}}%
    \else
      \def\colorrgb#1{\color{black}}%
      \def\colorgray#1{\color[gray]{#1}}%
      \expandafter\def\csname LTw\endcsname{\color{white}}%
      \expandafter\def\csname LTb\endcsname{\color{black}}%
      \expandafter\def\csname LTa\endcsname{\color{black}}%
      \expandafter\def\csname LT0\endcsname{\color{black}}%
      \expandafter\def\csname LT1\endcsname{\color{black}}%
      \expandafter\def\csname LT2\endcsname{\color{black}}%
      \expandafter\def\csname LT3\endcsname{\color{black}}%
      \expandafter\def\csname LT4\endcsname{\color{black}}%
      \expandafter\def\csname LT5\endcsname{\color{black}}%
      \expandafter\def\csname LT6\endcsname{\color{black}}%
      \expandafter\def\csname LT7\endcsname{\color{black}}%
      \expandafter\def\csname LT8\endcsname{\color{black}}%
    \fi
  \fi
  \setlength{\unitlength}{0.0500bp}%
  \begin{picture}(7200.00,5040.00)%
    \gplgaddtomacro\gplbacktext{%
      \csname LTb\endcsname%
      \put(594,704){\makebox(0,0)[r]{\strut{}-30}}%
      \put(594,1317){\makebox(0,0)[r]{\strut{}-20}}%
      \put(594,1929){\makebox(0,0)[r]{\strut{}-10}}%
      \put(594,2542){\makebox(0,0)[r]{\strut{} 0}}%
      \put(594,3154){\makebox(0,0)[r]{\strut{} 10}}%
      \put(594,3767){\makebox(0,0)[r]{\strut{} 20}}%
      \put(594,4379){\makebox(0,0)[r]{\strut{} 30}}%
      \put(726,484){\makebox(0,0){\strut{} 8}}%
      \put(1387,484){\makebox(0,0){\strut{} 8.5}}%
      \put(2047,484){\makebox(0,0){\strut{} 9}}%
      \put(2708,484){\makebox(0,0){\strut{} 9.5}}%
      \put(3369,484){\makebox(0,0){\strut{} 10}}%
      \put(4029,484){\makebox(0,0){\strut{} 10.5}}%
      \put(4690,484){\makebox(0,0){\strut{} 11}}%
      \put(5350,484){\makebox(0,0){\strut{} 11.5}}%
      \put(6011,484){\makebox(0,0){\strut{} 12}}%
      \put(6143,704){\makebox(0,0)[l]{\strut{}-3}}%
      \put(6143,1317){\makebox(0,0)[l]{\strut{}-2.5}}%
      \put(6143,1929){\makebox(0,0)[l]{\strut{}-2}}%
      \put(6143,2542){\makebox(0,0)[l]{\strut{}-1.5}}%
      \put(6143,3154){\makebox(0,0)[l]{\strut{}-1}}%
      \put(6143,3767){\makebox(0,0)[l]{\strut{}-0.5}}%
      \put(6143,4379){\makebox(0,0)[l]{\strut{} 0}}%
      \put(3368,154){\makebox(0,0){\strut{}$f$ (GHz)}}%
      \put(3368,4709){\makebox(0,0){\strut{}$\epsilon^{\mathrm{eff}}_{rr}=$3.050, $\epsilon^{\mathrm{eff}}_{zz}=$3.050, $\theta=$0.500}}%
    }%
    \gplgaddtomacro\gplfronttext{%
      \csname LTb\endcsname%
      \put(5024,1317){\makebox(0,0)[r]{\strut{}infinite TWT}}%
      \csname LTb\endcsname%
      \put(5024,1097){\makebox(0,0)[r]{\strut{}short TWT with beam}}%
      \csname LTb\endcsname%
      \put(5024,877){\makebox(0,0)[r]{\strut{}short TWT without beam}}%
    }%
    \gplbacktext
    \put(0,0){\includegraphics{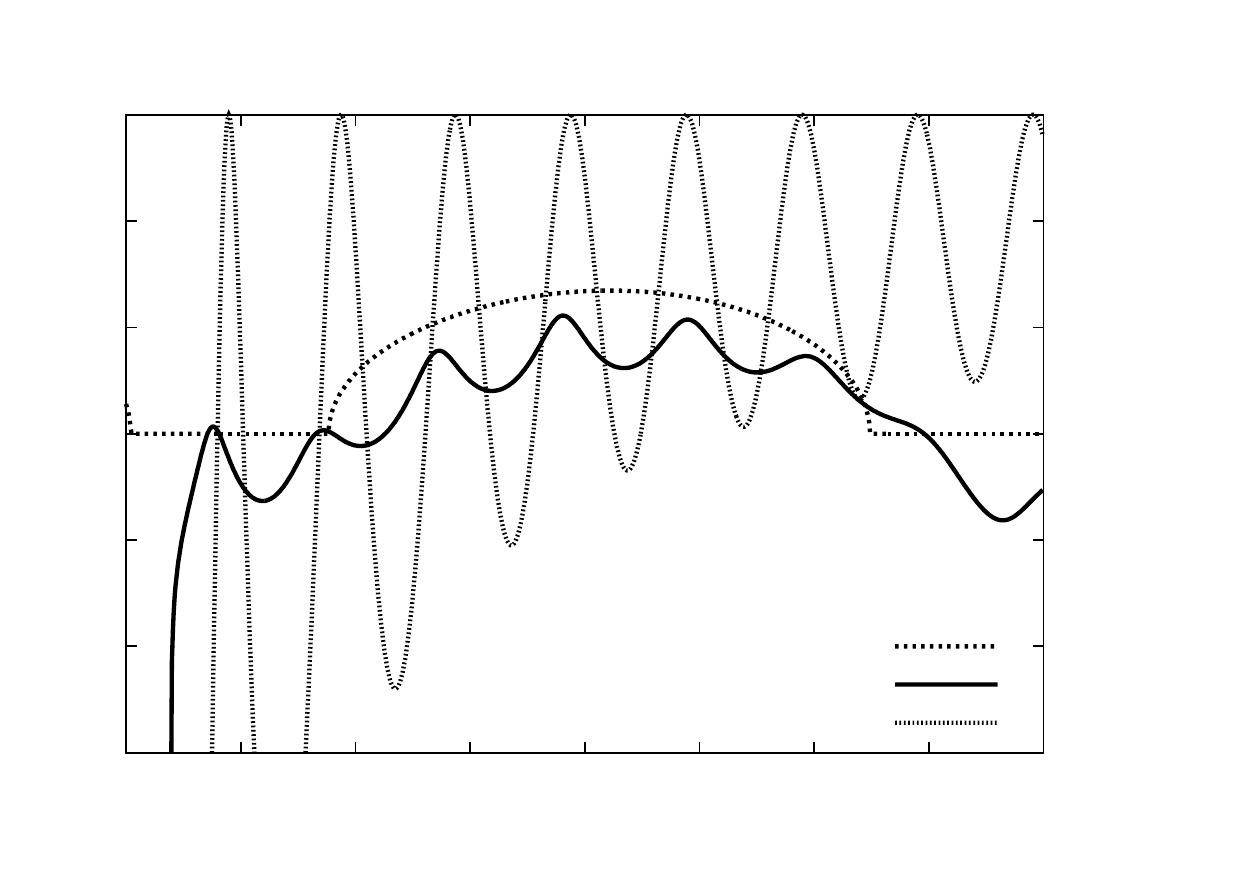}}%
    \gplfronttext
  \end{picture}%
\endgroup

%% file: Lozenge125Theta0.tex
\begingroup
  \makeatletter
  \providecommand\color[2][]{%
    \GenericError{(gnuplot) \space\space\space\@spaces}{%
      Package color not loaded in conjunction with
      terminal option `colourtext'%
    }{See the gnuplot documentation for explanation.%
    }{Either use 'blacktext' in gnuplot or load the package
      color.sty in LaTeX.}%
    \renewcommand\color[2][]{}%
  }%
  \providecommand\includegraphics[2][]{%
    \GenericError{(gnuplot) \space\space\space\@spaces}{%
      Package graphicx or graphics not loaded%
    }{See the gnuplot documentation for explanation.%
    }{The gnuplot epslatex terminal needs graphicx.sty or graphics.sty.}%
    \renewcommand\includegraphics[2][]{}%
  }%
  \providecommand\rotatebox[2]{#2}%
  \@ifundefined{ifGPcolor}{%
    \newif\ifGPcolor
    \GPcolorfalse
  }{}%
  \@ifundefined{ifGPblacktext}{%
    \newif\ifGPblacktext
    \GPblacktexttrue
  }{}%
  \let\gplgaddtomacro\g@addto@macro
  \gdef\gplbacktext{}%
  \gdef\gplfronttext{}%
  \makeatother
  \ifGPblacktext
    \def\colorrgb#1{}%
    \def\colorgray#1{}%
  \else
    \ifGPcolor
      \def\colorrgb#1{\color[rgb]{#1}}%
      \def\colorgray#1{\color[gray]{#1}}%
      \expandafter\def\csname LTw\endcsname{\color{white}}%
      \expandafter\def\csname LTb\endcsname{\color{black}}%
      \expandafter\def\csname LTa\endcsname{\color{black}}%
      \expandafter\def\csname LT0\endcsname{\color[rgb]{1,0,0}}%
      \expandafter\def\csname LT1\endcsname{\color[rgb]{0,1,0}}%
      \expandafter\def\csname LT2\endcsname{\color[rgb]{0,0,1}}%
      \expandafter\def\csname LT3\endcsname{\color[rgb]{1,0,1}}%
      \expandafter\def\csname LT4\endcsname{\color[rgb]{0,1,1}}%
      \expandafter\def\csname LT5\endcsname{\color[rgb]{1,1,0}}%
      \expandafter\def\csname LT6\endcsname{\color[rgb]{0,0,0}}%
      \expandafter\def\csname LT7\endcsname{\color[rgb]{1,0.3,0}}%
      \expandafter\def\csname LT8\endcsname{\color[rgb]{0.5,0.5,0.5}}%
    \else
      \def\colorrgb#1{\color{black}}%
      \def\colorgray#1{\color[gray]{#1}}%
      \expandafter\def\csname LTw\endcsname{\color{white}}%
      \expandafter\def\csname LTb\endcsname{\color{black}}%
      \expandafter\def\csname LTa\endcsname{\color{black}}%
      \expandafter\def\csname LT0\endcsname{\color{black}}%
      \expandafter\def\csname LT1\endcsname{\color{black}}%
      \expandafter\def\csname LT2\endcsname{\color{black}}%
      \expandafter\def\csname LT3\endcsname{\color{black}}%
      \expandafter\def\csname LT4\endcsname{\color{black}}%
      \expandafter\def\csname LT5\endcsname{\color{black}}%
      \expandafter\def\csname LT6\endcsname{\color{black}}%
      \expandafter\def\csname LT7\endcsname{\color{black}}%
      \expandafter\def\csname LT8\endcsname{\color{black}}%
    \fi
  \fi
  \setlength{\unitlength}{0.0500bp}%
  \begin{picture}(7200.00,5040.00)%
    \gplgaddtomacro\gplbacktext{%
      \csname LTb\endcsname%
      \put(594,704){\makebox(0,0)[r]{\strut{}-30}}%
      \put(594,1317){\makebox(0,0)[r]{\strut{}-20}}%
      \put(594,1929){\makebox(0,0)[r]{\strut{}-10}}%
      \put(594,2542){\makebox(0,0)[r]{\strut{} 0}}%
      \put(594,3154){\makebox(0,0)[r]{\strut{} 10}}%
      \put(594,3767){\makebox(0,0)[r]{\strut{} 20}}%
      \put(594,4379){\makebox(0,0)[r]{\strut{} 30}}%
      \put(726,484){\makebox(0,0){\strut{} 10}}%
      \put(1687,484){\makebox(0,0){\strut{} 12}}%
      \put(2648,484){\makebox(0,0){\strut{} 14}}%
      \put(3609,484){\makebox(0,0){\strut{} 16}}%
      \put(4570,484){\makebox(0,0){\strut{} 18}}%
      \put(5531,484){\makebox(0,0){\strut{} 20}}%
      \put(6143,851){\makebox(0,0)[l]{\strut{}-1.2}}%
      \put(6143,1439){\makebox(0,0)[l]{\strut{}-1}}%
      \put(6143,2027){\makebox(0,0)[l]{\strut{}-0.8}}%
      \put(6143,2615){\makebox(0,0)[l]{\strut{}-0.6}}%
      \put(6143,3203){\makebox(0,0)[l]{\strut{}-0.4}}%
      \put(6143,3791){\makebox(0,0)[l]{\strut{}-0.2}}%
      \put(6143,4379){\makebox(0,0)[l]{\strut{} 0}}%
      \put(3368,154){\makebox(0,0){\strut{}$f$ (GHz)}}%
      \put(3368,4709){\makebox(0,0){\strut{}$\epsilon^{\mathrm{eff}}_{rr}=$1.758, $\epsilon^{\mathrm{eff}}_{zz}=$2.074, $\theta=$0.300}}%
    }%
    \gplgaddtomacro\gplfronttext{%
      \csname LTb\endcsname%
      \put(5024,1317){\makebox(0,0)[r]{\strut{}infinite TWT}}%
      \csname LTb\endcsname%
      \put(5024,1097){\makebox(0,0)[r]{\strut{}short TWT with beam}}%
      \csname LTb\endcsname%
      \put(5024,877){\makebox(0,0)[r]{\strut{}short TWT without beam}}%
    }%
    \gplbacktext
    \put(0,0){\includegraphics{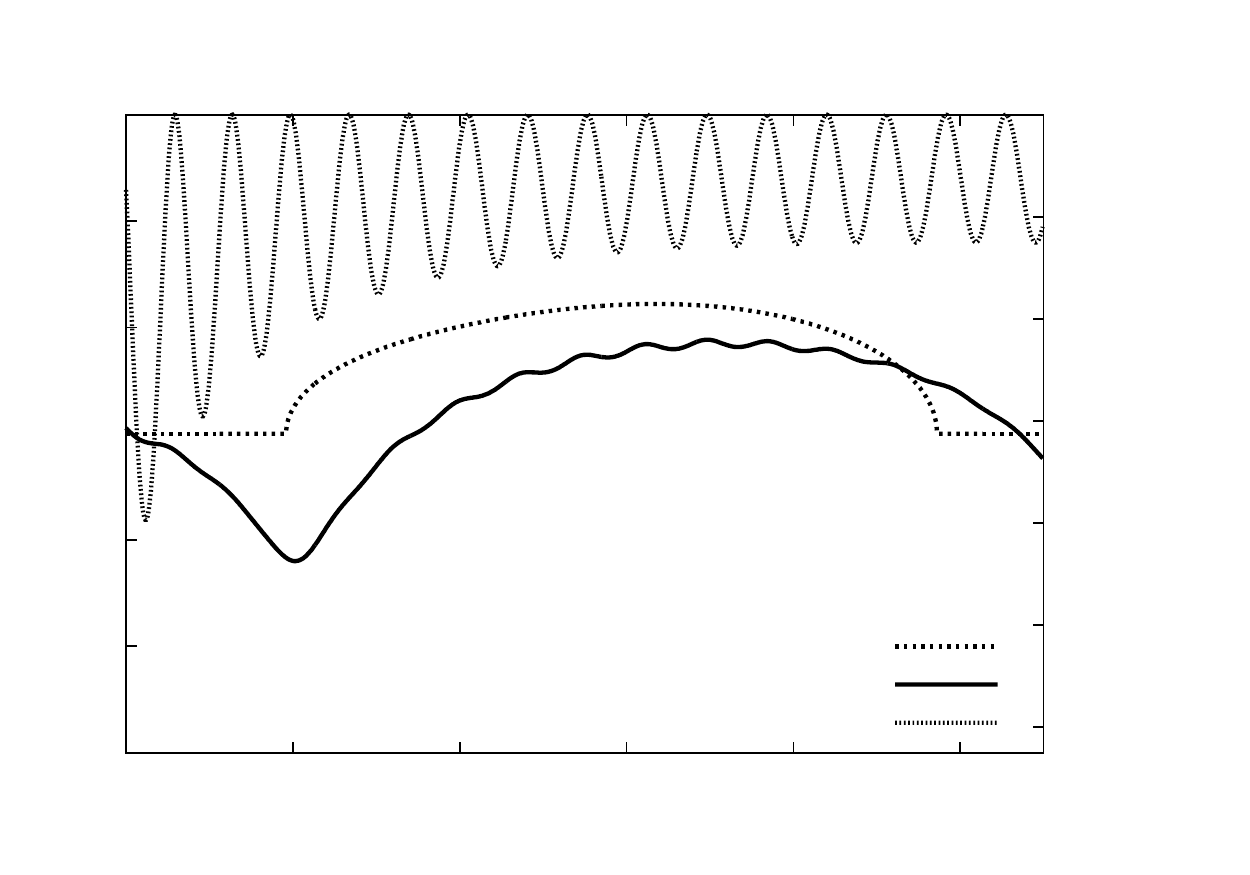}}%
    \gplfronttext
  \end{picture}%
\endgroup

%% file: Lozenge125Theta1.tex
\begingroup
  \makeatletter
  \providecommand\color[2][]{%
    \GenericError{(gnuplot) \space\space\space\@spaces}{%
      Package color not loaded in conjunction with
      terminal option `colourtext'%
    }{See the gnuplot documentation for explanation.%
    }{Either use 'blacktext' in gnuplot or load the package
      color.sty in LaTeX.}%
    \renewcommand\color[2][]{}%
  }%
  \providecommand\includegraphics[2][]{%
    \GenericError{(gnuplot) \space\space\space\@spaces}{%
      Package graphicx or graphics not loaded%
    }{See the gnuplot documentation for explanation.%
    }{The gnuplot epslatex terminal needs graphicx.sty or graphics.sty.}%
    \renewcommand\includegraphics[2][]{}%
  }%
  \providecommand\rotatebox[2]{#2}%
  \@ifundefined{ifGPcolor}{%
    \newif\ifGPcolor
    \GPcolorfalse
  }{}%
  \@ifundefined{ifGPblacktext}{%
    \newif\ifGPblacktext
    \GPblacktexttrue
  }{}%
  \let\gplgaddtomacro\g@addto@macro
  \gdef\gplbacktext{}%
  \gdef\gplfronttext{}%
  \makeatother
  \ifGPblacktext
    \def\colorrgb#1{}%
    \def\colorgray#1{}%
  \else
    \ifGPcolor
      \def\colorrgb#1{\color[rgb]{#1}}%
      \def\colorgray#1{\color[gray]{#1}}%
      \expandafter\def\csname LTw\endcsname{\color{white}}%
      \expandafter\def\csname LTb\endcsname{\color{black}}%
      \expandafter\def\csname LTa\endcsname{\color{black}}%
      \expandafter\def\csname LT0\endcsname{\color[rgb]{1,0,0}}%
      \expandafter\def\csname LT1\endcsname{\color[rgb]{0,1,0}}%
      \expandafter\def\csname LT2\endcsname{\color[rgb]{0,0,1}}%
      \expandafter\def\csname LT3\endcsname{\color[rgb]{1,0,1}}%
      \expandafter\def\csname LT4\endcsname{\color[rgb]{0,1,1}}%
      \expandafter\def\csname LT5\endcsname{\color[rgb]{1,1,0}}%
      \expandafter\def\csname LT6\endcsname{\color[rgb]{0,0,0}}%
      \expandafter\def\csname LT7\endcsname{\color[rgb]{1,0.3,0}}%
      \expandafter\def\csname LT8\endcsname{\color[rgb]{0.5,0.5,0.5}}%
    \else
      \def\colorrgb#1{\color{black}}%
      \def\colorgray#1{\color[gray]{#1}}%
      \expandafter\def\csname LTw\endcsname{\color{white}}%
      \expandafter\def\csname LTb\endcsname{\color{black}}%
      \expandafter\def\csname LTa\endcsname{\color{black}}%
      \expandafter\def\csname LT0\endcsname{\color{black}}%
      \expandafter\def\csname LT1\endcsname{\color{black}}%
      \expandafter\def\csname LT2\endcsname{\color{black}}%
      \expandafter\def\csname LT3\endcsname{\color{black}}%
      \expandafter\def\csname LT4\endcsname{\color{black}}%
      \expandafter\def\csname LT5\endcsname{\color{black}}%
      \expandafter\def\csname LT6\endcsname{\color{black}}%
      \expandafter\def\csname LT7\endcsname{\color{black}}%
      \expandafter\def\csname LT8\endcsname{\color{black}}%
    \fi
  \fi
  \setlength{\unitlength}{0.0500bp}%
  \begin{picture}(7200.00,5040.00)%
    \gplgaddtomacro\gplbacktext{%
      \csname LTb\endcsname%
      \put(594,704){\makebox(0,0)[r]{\strut{}-30}}%
      \put(594,1317){\makebox(0,0)[r]{\strut{}-20}}%
      \put(594,1929){\makebox(0,0)[r]{\strut{}-10}}%
      \put(594,2542){\makebox(0,0)[r]{\strut{} 0}}%
      \put(594,3154){\makebox(0,0)[r]{\strut{} 10}}%
      \put(594,3767){\makebox(0,0)[r]{\strut{} 20}}%
      \put(594,4379){\makebox(0,0)[r]{\strut{} 30}}%
      \put(726,484){\makebox(0,0){\strut{} 10}}%
      \put(1387,484){\makebox(0,0){\strut{} 11}}%
      \put(2047,484){\makebox(0,0){\strut{} 12}}%
      \put(2708,484){\makebox(0,0){\strut{} 13}}%
      \put(3369,484){\makebox(0,0){\strut{} 14}}%
      \put(4029,484){\makebox(0,0){\strut{} 15}}%
      \put(4690,484){\makebox(0,0){\strut{} 16}}%
      \put(5350,484){\makebox(0,0){\strut{} 17}}%
      \put(6011,484){\makebox(0,0){\strut{} 18}}%
      \put(6143,851){\makebox(0,0)[l]{\strut{}-1.2}}%
      \put(6143,1439){\makebox(0,0)[l]{\strut{}-1}}%
      \put(6143,2027){\makebox(0,0)[l]{\strut{}-0.8}}%
      \put(6143,2615){\makebox(0,0)[l]{\strut{}-0.6}}%
      \put(6143,3203){\makebox(0,0)[l]{\strut{}-0.4}}%
      \put(6143,3791){\makebox(0,0)[l]{\strut{}-0.2}}%
      \put(6143,4379){\makebox(0,0)[l]{\strut{} 0}}%
      \put(3368,154){\makebox(0,0){\strut{}$f$ (GHz)}}%
      \put(3368,4709){\makebox(0,0){\strut{}$\epsilon^{\mathrm{eff}}_{rr}=$1.932, $\epsilon^{\mathrm{eff}}_{zz}=$2.385, $\theta=$0.350}}%
    }%
    \gplgaddtomacro\gplfronttext{%
      \csname LTb\endcsname%
      \put(5024,1317){\makebox(0,0)[r]{\strut{}infinite TWT}}%
      \csname LTb\endcsname%
      \put(5024,1097){\makebox(0,0)[r]{\strut{}short TWT with beam}}%
      \csname LTb\endcsname%
      \put(5024,877){\makebox(0,0)[r]{\strut{}short TWT without beam}}%
    }%
    \gplbacktext
    \put(0,0){\includegraphics{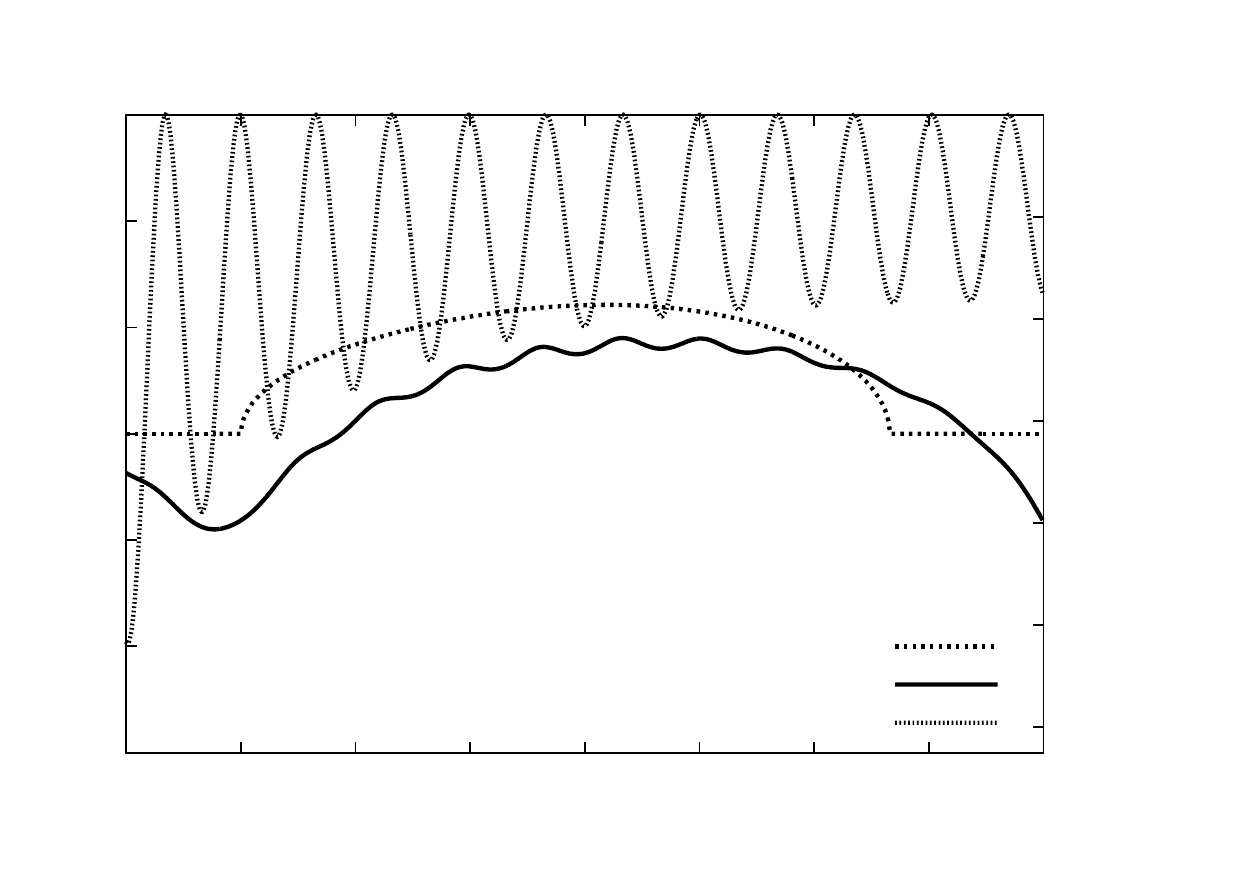}}%
    \gplfronttext
  \end{picture}%
\endgroup

%% file: Lozenge125Theta2.tex
\begingroup
  \makeatletter
  \providecommand\color[2][]{%
    \GenericError{(gnuplot) \space\space\space\@spaces}{%
      Package color not loaded in conjunction with
      terminal option `colourtext'%
    }{See the gnuplot documentation for explanation.%
    }{Either use 'blacktext' in gnuplot or load the package
      color.sty in LaTeX.}%
    \renewcommand\color[2][]{}%
  }%
  \providecommand\includegraphics[2][]{%
    \GenericError{(gnuplot) \space\space\space\@spaces}{%
      Package graphicx or graphics not loaded%
    }{See the gnuplot documentation for explanation.%
    }{The gnuplot epslatex terminal needs graphicx.sty or graphics.sty.}%
    \renewcommand\includegraphics[2][]{}%
  }%
  \providecommand\rotatebox[2]{#2}%
  \@ifundefined{ifGPcolor}{%
    \newif\ifGPcolor
    \GPcolorfalse
  }{}%
  \@ifundefined{ifGPblacktext}{%
    \newif\ifGPblacktext
    \GPblacktexttrue
  }{}%
  \let\gplgaddtomacro\g@addto@macro
  \gdef\gplbacktext{}%
  \gdef\gplfronttext{}%
  \makeatother
  \ifGPblacktext
    \def\colorrgb#1{}%
    \def\colorgray#1{}%
  \else
    \ifGPcolor
      \def\colorrgb#1{\color[rgb]{#1}}%
      \def\colorgray#1{\color[gray]{#1}}%
      \expandafter\def\csname LTw\endcsname{\color{white}}%
      \expandafter\def\csname LTb\endcsname{\color{black}}%
      \expandafter\def\csname LTa\endcsname{\color{black}}%
      \expandafter\def\csname LT0\endcsname{\color[rgb]{1,0,0}}%
      \expandafter\def\csname LT1\endcsname{\color[rgb]{0,1,0}}%
      \expandafter\def\csname LT2\endcsname{\color[rgb]{0,0,1}}%
      \expandafter\def\csname LT3\endcsname{\color[rgb]{1,0,1}}%
      \expandafter\def\csname LT4\endcsname{\color[rgb]{0,1,1}}%
      \expandafter\def\csname LT5\endcsname{\color[rgb]{1,1,0}}%
      \expandafter\def\csname LT6\endcsname{\color[rgb]{0,0,0}}%
      \expandafter\def\csname LT7\endcsname{\color[rgb]{1,0.3,0}}%
      \expandafter\def\csname LT8\endcsname{\color[rgb]{0.5,0.5,0.5}}%
    \else
      \def\colorrgb#1{\color{black}}%
      \def\colorgray#1{\color[gray]{#1}}%
      \expandafter\def\csname LTw\endcsname{\color{white}}%
      \expandafter\def\csname LTb\endcsname{\color{black}}%
      \expandafter\def\csname LTa\endcsname{\color{black}}%
      \expandafter\def\csname LT0\endcsname{\color{black}}%
      \expandafter\def\csname LT1\endcsname{\color{black}}%
      \expandafter\def\csname LT2\endcsname{\color{black}}%
      \expandafter\def\csname LT3\endcsname{\color{black}}%
      \expandafter\def\csname LT4\endcsname{\color{black}}%
      \expandafter\def\csname LT5\endcsname{\color{black}}%
      \expandafter\def\csname LT6\endcsname{\color{black}}%
      \expandafter\def\csname LT7\endcsname{\color{black}}%
      \expandafter\def\csname LT8\endcsname{\color{black}}%
    \fi
  \fi
  \setlength{\unitlength}{0.0500bp}%
  \begin{picture}(7200.00,5040.00)%
    \gplgaddtomacro\gplbacktext{%
      \csname LTb\endcsname%
      \put(594,704){\makebox(0,0)[r]{\strut{}-30}}%
      \put(594,1317){\makebox(0,0)[r]{\strut{}-20}}%
      \put(594,1929){\makebox(0,0)[r]{\strut{}-10}}%
      \put(594,2542){\makebox(0,0)[r]{\strut{} 0}}%
      \put(594,3154){\makebox(0,0)[r]{\strut{} 10}}%
      \put(594,3767){\makebox(0,0)[r]{\strut{} 20}}%
      \put(594,4379){\makebox(0,0)[r]{\strut{} 30}}%
      \put(726,484){\makebox(0,0){\strut{} 9}}%
      \put(1481,484){\makebox(0,0){\strut{} 10}}%
      \put(2236,484){\makebox(0,0){\strut{} 11}}%
      \put(2991,484){\makebox(0,0){\strut{} 12}}%
      \put(3746,484){\makebox(0,0){\strut{} 13}}%
      \put(4501,484){\makebox(0,0){\strut{} 14}}%
      \put(5256,484){\makebox(0,0){\strut{} 15}}%
      \put(6011,484){\makebox(0,0){\strut{} 16}}%
      \put(6143,851){\makebox(0,0)[l]{\strut{}-1.2}}%
      \put(6143,1439){\makebox(0,0)[l]{\strut{}-1}}%
      \put(6143,2027){\makebox(0,0)[l]{\strut{}-0.8}}%
      \put(6143,2615){\makebox(0,0)[l]{\strut{}-0.6}}%
      \put(6143,3203){\makebox(0,0)[l]{\strut{}-0.4}}%
      \put(6143,3791){\makebox(0,0)[l]{\strut{}-0.2}}%
      \put(6143,4379){\makebox(0,0)[l]{\strut{} 0}}%
      \put(3368,154){\makebox(0,0){\strut{}$f$ (GHz)}}%
      \put(3368,4709){\makebox(0,0){\strut{}$\epsilon^{\mathrm{eff}}_{rr}=$2.123, $\epsilon^{\mathrm{eff}}_{zz}=$2.769, $\theta=$0.400}}%
    }%
    \gplgaddtomacro\gplfronttext{%
      \csname LTb\endcsname%
      \put(5024,1317){\makebox(0,0)[r]{\strut{}infinite TWT}}%
      \csname LTb\endcsname%
      \put(5024,1097){\makebox(0,0)[r]{\strut{}short TWT with beam}}%
      \csname LTb\endcsname%
      \put(5024,877){\makebox(0,0)[r]{\strut{}short TWT without beam}}%
    }%
    \gplbacktext
    \put(0,0){\includegraphics{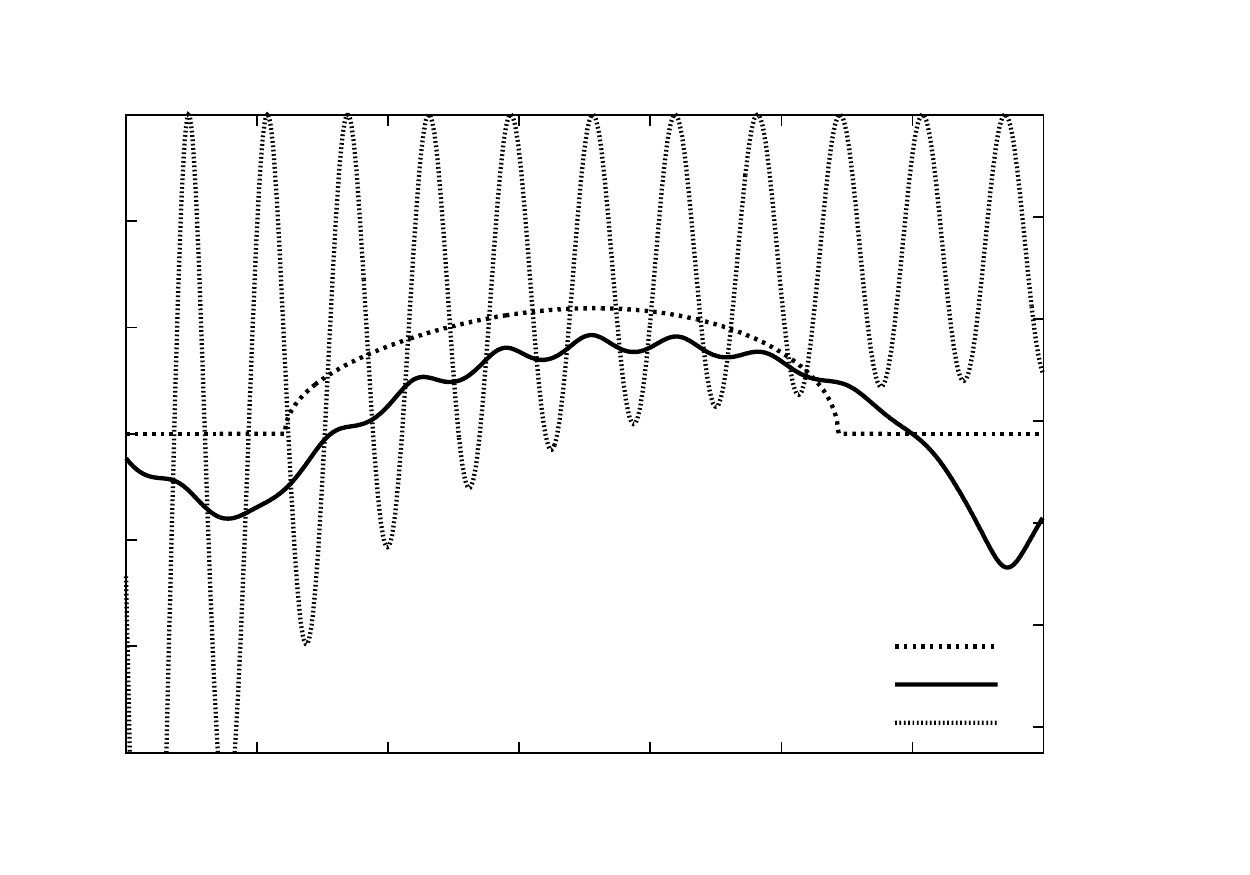}}%
    \gplfronttext
  \end{picture}%
\endgroup

%% file: Lozenge80Theta1.tex
\begingroup
  \makeatletter
  \providecommand\color[2][]{%
    \GenericError{(gnuplot) \space\space\space\@spaces}{%
      Package color not loaded in conjunction with
      terminal option `colourtext'%
    }{See the gnuplot documentation for explanation.%
    }{Either use 'blacktext' in gnuplot or load the package
      color.sty in LaTeX.}%
    \renewcommand\color[2][]{}%
  }%
  \providecommand\includegraphics[2][]{%
    \GenericError{(gnuplot) \space\space\space\@spaces}{%
      Package graphicx or graphics not loaded%
    }{See the gnuplot documentation for explanation.%
    }{The gnuplot epslatex terminal needs graphicx.sty or graphics.sty.}%
    \renewcommand\includegraphics[2][]{}%
  }%
  \providecommand\rotatebox[2]{#2}%
  \@ifundefined{ifGPcolor}{%
    \newif\ifGPcolor
    \GPcolorfalse
  }{}%
  \@ifundefined{ifGPblacktext}{%
    \newif\ifGPblacktext
    \GPblacktexttrue
  }{}%
  \let\gplgaddtomacro\g@addto@macro
  \gdef\gplbacktext{}%
  \gdef\gplfronttext{}%
  \makeatother
  \ifGPblacktext
    \def\colorrgb#1{}%
    \def\colorgray#1{}%
  \else
    \ifGPcolor
      \def\colorrgb#1{\color[rgb]{#1}}%
      \def\colorgray#1{\color[gray]{#1}}%
      \expandafter\def\csname LTw\endcsname{\color{white}}%
      \expandafter\def\csname LTb\endcsname{\color{black}}%
      \expandafter\def\csname LTa\endcsname{\color{black}}%
      \expandafter\def\csname LT0\endcsname{\color[rgb]{1,0,0}}%
      \expandafter\def\csname LT1\endcsname{\color[rgb]{0,1,0}}%
      \expandafter\def\csname LT2\endcsname{\color[rgb]{0,0,1}}%
      \expandafter\def\csname LT3\endcsname{\color[rgb]{1,0,1}}%
      \expandafter\def\csname LT4\endcsname{\color[rgb]{0,1,1}}%
      \expandafter\def\csname LT5\endcsname{\color[rgb]{1,1,0}}%
      \expandafter\def\csname LT6\endcsname{\color[rgb]{0,0,0}}%
      \expandafter\def\csname LT7\endcsname{\color[rgb]{1,0.3,0}}%
      \expandafter\def\csname LT8\endcsname{\color[rgb]{0.5,0.5,0.5}}%
    \else
      \def\colorrgb#1{\color{black}}%
      \def\colorgray#1{\color[gray]{#1}}%
      \expandafter\def\csname LTw\endcsname{\color{white}}%
      \expandafter\def\csname LTb\endcsname{\color{black}}%
      \expandafter\def\csname LTa\endcsname{\color{black}}%
      \expandafter\def\csname LT0\endcsname{\color{black}}%
      \expandafter\def\csname LT1\endcsname{\color{black}}%
      \expandafter\def\csname LT2\endcsname{\color{black}}%
      \expandafter\def\csname LT3\endcsname{\color{black}}%
      \expandafter\def\csname LT4\endcsname{\color{black}}%
      \expandafter\def\csname LT5\endcsname{\color{black}}%
      \expandafter\def\csname LT6\endcsname{\color{black}}%
      \expandafter\def\csname LT7\endcsname{\color{black}}%
      \expandafter\def\csname LT8\endcsname{\color{black}}%
    \fi
  \fi
  \setlength{\unitlength}{0.0500bp}%
  \begin{picture}(7200.00,5040.00)%
    \gplgaddtomacro\gplbacktext{%
      \csname LTb\endcsname%
      \put(594,704){\makebox(0,0)[r]{\strut{}-30}}%
      \put(594,1317){\makebox(0,0)[r]{\strut{}-20}}%
      \put(594,1929){\makebox(0,0)[r]{\strut{}-10}}%
      \put(594,2542){\makebox(0,0)[r]{\strut{} 0}}%
      \put(594,3154){\makebox(0,0)[r]{\strut{} 10}}%
      \put(594,3767){\makebox(0,0)[r]{\strut{} 20}}%
      \put(594,4379){\makebox(0,0)[r]{\strut{} 30}}%
      \put(726,484){\makebox(0,0){\strut{} 10}}%
      \put(1607,484){\makebox(0,0){\strut{} 11}}%
      \put(2488,484){\makebox(0,0){\strut{} 12}}%
      \put(3369,484){\makebox(0,0){\strut{} 13}}%
      \put(4249,484){\makebox(0,0){\strut{} 14}}%
      \put(5130,484){\makebox(0,0){\strut{} 15}}%
      \put(6011,484){\makebox(0,0){\strut{} 16}}%
      \put(6143,704){\makebox(0,0)[l]{\strut{}-2}}%
      \put(6143,1623){\makebox(0,0)[l]{\strut{}-1.5}}%
      \put(6143,2542){\makebox(0,0)[l]{\strut{}-1}}%
      \put(6143,3460){\makebox(0,0)[l]{\strut{}-0.5}}%
      \put(6143,4379){\makebox(0,0)[l]{\strut{} 0}}%
      \put(3368,154){\makebox(0,0){\strut{}$f$ (GHz)}}%
      \put(3368,4709){\makebox(0,0){\strut{}$\epsilon^{\mathrm{eff}}_{rr}=$2.385, $\epsilon^{\mathrm{eff}}_{zz}=$1.932, $\theta=$0.350}}%
    }%
    \gplgaddtomacro\gplfronttext{%
      \csname LTb\endcsname%
      \put(5024,1317){\makebox(0,0)[r]{\strut{}infinite TWT}}%
      \csname LTb\endcsname%
      \put(5024,1097){\makebox(0,0)[r]{\strut{}short TWT with beam}}%
      \csname LTb\endcsname%
      \put(5024,877){\makebox(0,0)[r]{\strut{}short TWT without beam}}%
    }%
    \gplbacktext
    \put(0,0){\includegraphics{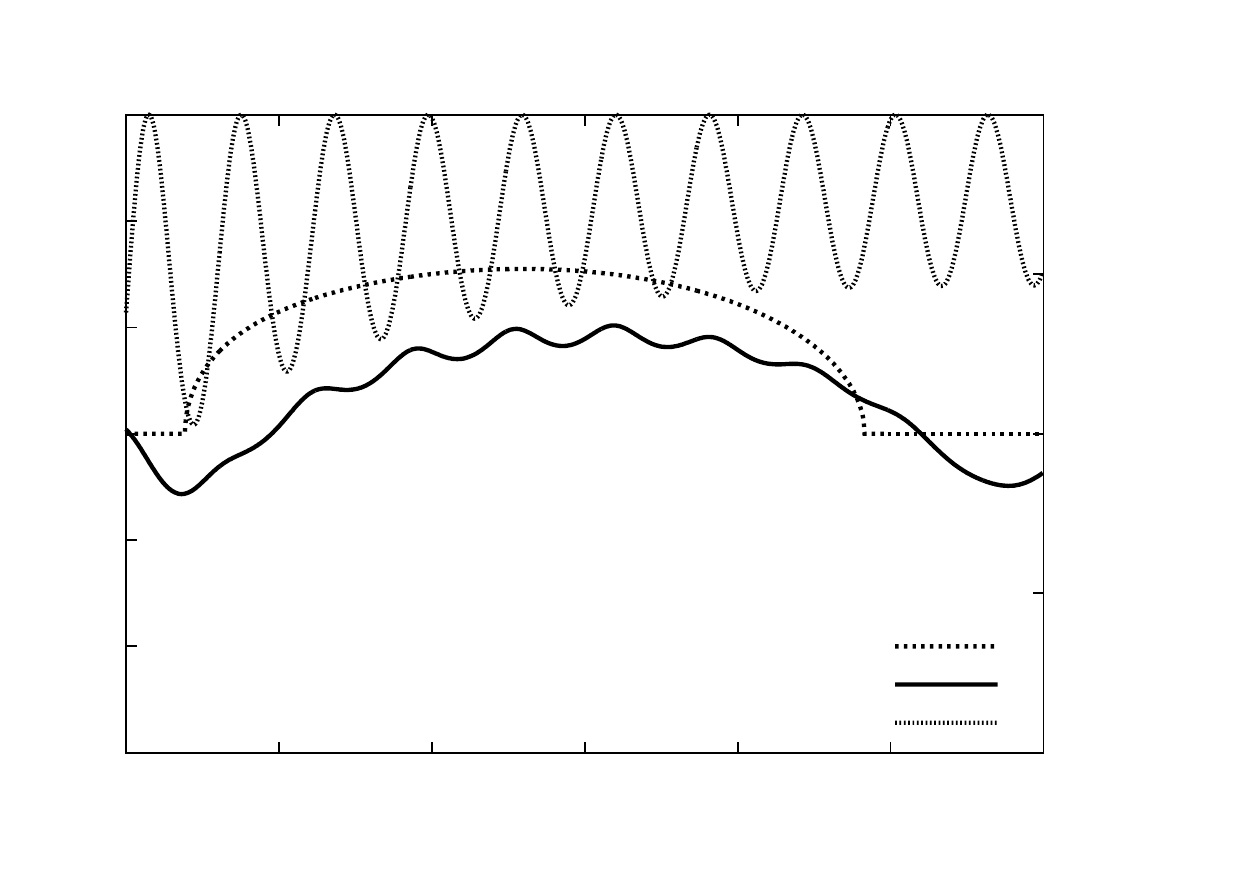}}%
    \gplfronttext
  \end{picture}%
\endgroup

%% file: Lozenge80Theta3.tex
\begingroup
  \makeatletter
  \providecommand\color[2][]{%
    \GenericError{(gnuplot) \space\space\space\@spaces}{%
      Package color not loaded in conjunction with
      terminal option `colourtext'%
    }{See the gnuplot documentation for explanation.%
    }{Either use 'blacktext' in gnuplot or load the package
      color.sty in LaTeX.}%
    \renewcommand\color[2][]{}%
  }%
  \providecommand\includegraphics[2][]{%
    \GenericError{(gnuplot) \space\space\space\@spaces}{%
      Package graphicx or graphics not loaded%
    }{See the gnuplot documentation for explanation.%
    }{The gnuplot epslatex terminal needs graphicx.sty or graphics.sty.}%
    \renewcommand\includegraphics[2][]{}%
  }%
  \providecommand\rotatebox[2]{#2}%
  \@ifundefined{ifGPcolor}{%
    \newif\ifGPcolor
    \GPcolorfalse
  }{}%
  \@ifundefined{ifGPblacktext}{%
    \newif\ifGPblacktext
    \GPblacktexttrue
  }{}%
  \let\gplgaddtomacro\g@addto@macro
  \gdef\gplbacktext{}%
  \gdef\gplfronttext{}%
  \makeatother
  \ifGPblacktext
    \def\colorrgb#1{}%
    \def\colorgray#1{}%
  \else
    \ifGPcolor
      \def\colorrgb#1{\color[rgb]{#1}}%
      \def\colorgray#1{\color[gray]{#1}}%
      \expandafter\def\csname LTw\endcsname{\color{white}}%
      \expandafter\def\csname LTb\endcsname{\color{black}}%
      \expandafter\def\csname LTa\endcsname{\color{black}}%
      \expandafter\def\csname LT0\endcsname{\color[rgb]{1,0,0}}%
      \expandafter\def\csname LT1\endcsname{\color[rgb]{0,1,0}}%
      \expandafter\def\csname LT2\endcsname{\color[rgb]{0,0,1}}%
      \expandafter\def\csname LT3\endcsname{\color[rgb]{1,0,1}}%
      \expandafter\def\csname LT4\endcsname{\color[rgb]{0,1,1}}%
      \expandafter\def\csname LT5\endcsname{\color[rgb]{1,1,0}}%
      \expandafter\def\csname LT6\endcsname{\color[rgb]{0,0,0}}%
      \expandafter\def\csname LT7\endcsname{\color[rgb]{1,0.3,0}}%
      \expandafter\def\csname LT8\endcsname{\color[rgb]{0.5,0.5,0.5}}%
    \else
      \def\colorrgb#1{\color{black}}%
      \def\colorgray#1{\color[gray]{#1}}%
      \expandafter\def\csname LTw\endcsname{\color{white}}%
      \expandafter\def\csname LTb\endcsname{\color{black}}%
      \expandafter\def\csname LTa\endcsname{\color{black}}%
      \expandafter\def\csname LT0\endcsname{\color{black}}%
      \expandafter\def\csname LT1\endcsname{\color{black}}%
      \expandafter\def\csname LT2\endcsname{\color{black}}%
      \expandafter\def\csname LT3\endcsname{\color{black}}%
      \expandafter\def\csname LT4\endcsname{\color{black}}%
      \expandafter\def\csname LT5\endcsname{\color{black}}%
      \expandafter\def\csname LT6\endcsname{\color{black}}%
      \expandafter\def\csname LT7\endcsname{\color{black}}%
      \expandafter\def\csname LT8\endcsname{\color{black}}%
    \fi
  \fi
  \setlength{\unitlength}{0.0500bp}%
  \begin{picture}(7200.00,5040.00)%
    \gplgaddtomacro\gplbacktext{%
      \csname LTb\endcsname%
      \put(594,704){\makebox(0,0)[r]{\strut{}-30}}%
      \put(594,1317){\makebox(0,0)[r]{\strut{}-20}}%
      \put(594,1929){\makebox(0,0)[r]{\strut{}-10}}%
      \put(594,2542){\makebox(0,0)[r]{\strut{} 0}}%
      \put(594,3154){\makebox(0,0)[r]{\strut{} 10}}%
      \put(594,3767){\makebox(0,0)[r]{\strut{} 20}}%
      \put(594,4379){\makebox(0,0)[r]{\strut{} 30}}%
      \put(1292,484){\makebox(0,0){\strut{} 9.5}}%
      \put(2236,484){\makebox(0,0){\strut{} 10}}%
      \put(3180,484){\makebox(0,0){\strut{} 10.5}}%
      \put(4124,484){\makebox(0,0){\strut{} 11}}%
      \put(5067,484){\makebox(0,0){\strut{} 11.5}}%
      \put(6011,484){\makebox(0,0){\strut{} 12}}%
      \put(6143,704){\makebox(0,0)[l]{\strut{}-2}}%
      \put(6143,1623){\makebox(0,0)[l]{\strut{}-1.5}}%
      \put(6143,2542){\makebox(0,0)[l]{\strut{}-1}}%
      \put(6143,3460){\makebox(0,0)[l]{\strut{}-0.5}}%
      \put(6143,4379){\makebox(0,0)[l]{\strut{} 0}}%
      \put(3368,154){\makebox(0,0){\strut{}$f$ (GHz)}}%
      \put(3368,4709){\makebox(0,0){\strut{}$\epsilon^{\mathrm{eff}}_{rr}=$3.258, $\epsilon^{\mathrm{eff}}_{zz}=$2.336, $\theta=$0.450}}%
    }%
    \gplgaddtomacro\gplfronttext{%
      \csname LTb\endcsname%
      \put(5024,1317){\makebox(0,0)[r]{\strut{}infinite TWT}}%
      \csname LTb\endcsname%
      \put(5024,1097){\makebox(0,0)[r]{\strut{}short TWT with beam}}%
      \csname LTb\endcsname%
      \put(5024,877){\makebox(0,0)[r]{\strut{}short TWT without beam}}%
    }%
    \gplbacktext
    \put(0,0){\includegraphics{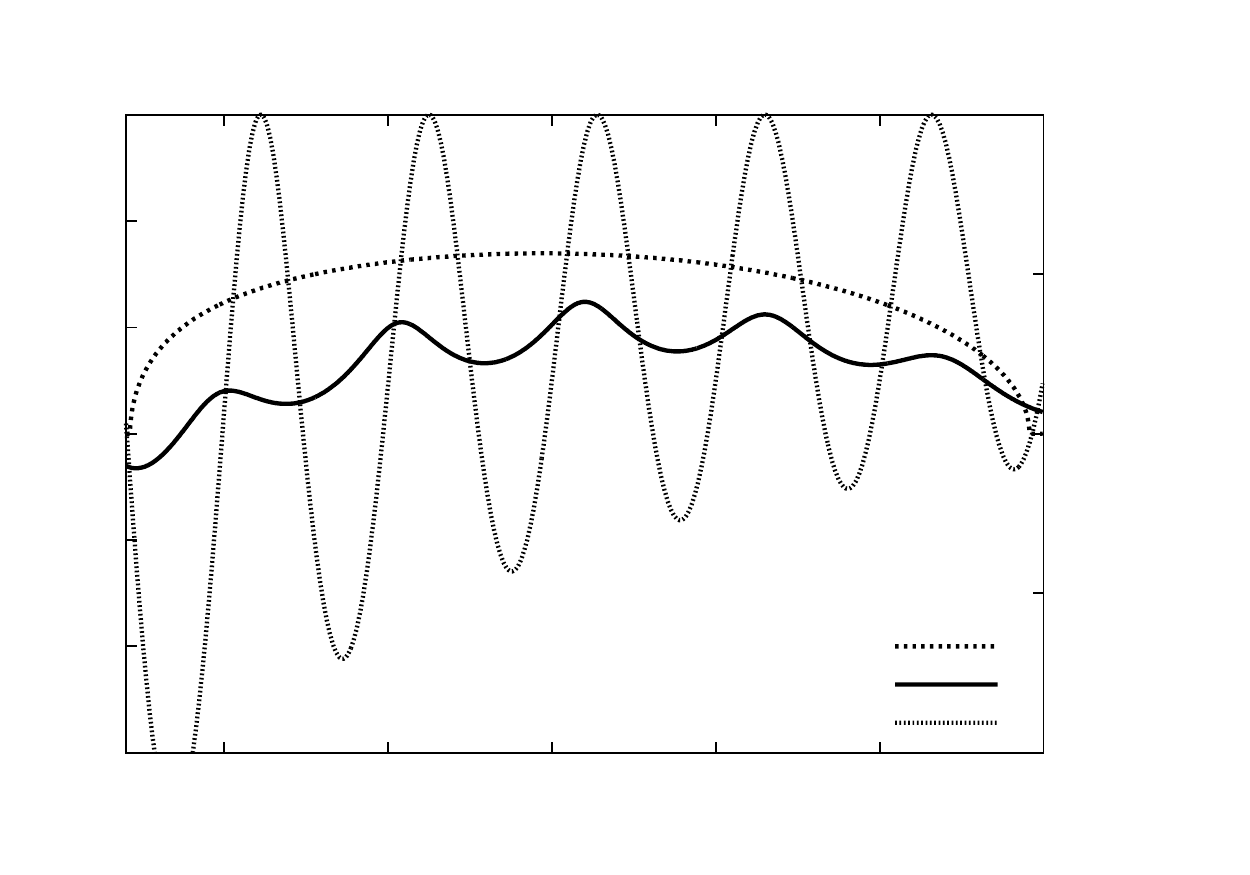}}%
    \gplfronttext
  \end{picture}%
\endgroup

%% file: Lozenge80Theta4.tex
\begingroup
  \makeatletter
  \providecommand\color[2][]{%
    \GenericError{(gnuplot) \space\space\space\@spaces}{%
      Package color not loaded in conjunction with
      terminal option `colourtext'%
    }{See the gnuplot documentation for explanation.%
    }{Either use 'blacktext' in gnuplot or load the package
      color.sty in LaTeX.}%
    \renewcommand\color[2][]{}%
  }%
  \providecommand\includegraphics[2][]{%
    \GenericError{(gnuplot) \space\space\space\@spaces}{%
      Package graphicx or graphics not loaded%
    }{See the gnuplot documentation for explanation.%
    }{The gnuplot epslatex terminal needs graphicx.sty or graphics.sty.}%
    \renewcommand\includegraphics[2][]{}%
  }%
  \providecommand\rotatebox[2]{#2}%
  \@ifundefined{ifGPcolor}{%
    \newif\ifGPcolor
    \GPcolorfalse
  }{}%
  \@ifundefined{ifGPblacktext}{%
    \newif\ifGPblacktext
    \GPblacktexttrue
  }{}%
  \let\gplgaddtomacro\g@addto@macro
  \gdef\gplbacktext{}%
  \gdef\gplfronttext{}%
  \makeatother
  \ifGPblacktext
    \def\colorrgb#1{}%
    \def\colorgray#1{}%
  \else
    \ifGPcolor
      \def\colorrgb#1{\color[rgb]{#1}}%
      \def\colorgray#1{\color[gray]{#1}}%
      \expandafter\def\csname LTw\endcsname{\color{white}}%
      \expandafter\def\csname LTb\endcsname{\color{black}}%
      \expandafter\def\csname LTa\endcsname{\color{black}}%
      \expandafter\def\csname LT0\endcsname{\color[rgb]{1,0,0}}%
      \expandafter\def\csname LT1\endcsname{\color[rgb]{0,1,0}}%
      \expandafter\def\csname LT2\endcsname{\color[rgb]{0,0,1}}%
      \expandafter\def\csname LT3\endcsname{\color[rgb]{1,0,1}}%
      \expandafter\def\csname LT4\endcsname{\color[rgb]{0,1,1}}%
      \expandafter\def\csname LT5\endcsname{\color[rgb]{1,1,0}}%
      \expandafter\def\csname LT6\endcsname{\color[rgb]{0,0,0}}%
      \expandafter\def\csname LT7\endcsname{\color[rgb]{1,0.3,0}}%
      \expandafter\def\csname LT8\endcsname{\color[rgb]{0.5,0.5,0.5}}%
    \else
      \def\colorrgb#1{\color{black}}%
      \def\colorgray#1{\color[gray]{#1}}%
      \expandafter\def\csname LTw\endcsname{\color{white}}%
      \expandafter\def\csname LTb\endcsname{\color{black}}%
      \expandafter\def\csname LTa\endcsname{\color{black}}%
      \expandafter\def\csname LT0\endcsname{\color{black}}%
      \expandafter\def\csname LT1\endcsname{\color{black}}%
      \expandafter\def\csname LT2\endcsname{\color{black}}%
      \expandafter\def\csname LT3\endcsname{\color{black}}%
      \expandafter\def\csname LT4\endcsname{\color{black}}%
      \expandafter\def\csname LT5\endcsname{\color{black}}%
      \expandafter\def\csname LT6\endcsname{\color{black}}%
      \expandafter\def\csname LT7\endcsname{\color{black}}%
      \expandafter\def\csname LT8\endcsname{\color{black}}%
    \fi
  \fi
  \setlength{\unitlength}{0.0500bp}%
  \begin{picture}(7200.00,5040.00)%
    \gplgaddtomacro\gplbacktext{%
      \csname LTb\endcsname%
      \put(594,704){\makebox(0,0)[r]{\strut{}-30}}%
      \put(594,1317){\makebox(0,0)[r]{\strut{}-20}}%
      \put(594,1929){\makebox(0,0)[r]{\strut{}-10}}%
      \put(594,2542){\makebox(0,0)[r]{\strut{} 0}}%
      \put(594,3154){\makebox(0,0)[r]{\strut{} 10}}%
      \put(594,3767){\makebox(0,0)[r]{\strut{} 20}}%
      \put(594,4379){\makebox(0,0)[r]{\strut{} 30}}%
      \put(1360,484){\makebox(0,0){\strut{} 9}}%
      \put(2417,484){\makebox(0,0){\strut{} 9.5}}%
      \put(3474,484){\makebox(0,0){\strut{} 10}}%
      \put(4531,484){\makebox(0,0){\strut{} 10.5}}%
      \put(5588,484){\makebox(0,0){\strut{} 11}}%
      \put(6143,704){\makebox(0,0)[l]{\strut{}-2}}%
      \put(6143,1623){\makebox(0,0)[l]{\strut{}-1.5}}%
      \put(6143,2542){\makebox(0,0)[l]{\strut{}-1}}%
      \put(6143,3460){\makebox(0,0)[l]{\strut{}-0.5}}%
      \put(6143,4379){\makebox(0,0)[l]{\strut{} 0}}%
      \put(3368,154){\makebox(0,0){\strut{}$f$ (GHz)}}%
      \put(3368,4709){\makebox(0,0){\strut{}$\epsilon^{\mathrm{eff}}_{rr}=$3.902, $\epsilon^{\mathrm{eff}}_{zz}=$2.573, $\theta=$0.500}}%
    }%
    \gplgaddtomacro\gplfronttext{%
      \csname LTb\endcsname%
      \put(5024,1317){\makebox(0,0)[r]{\strut{}infinite TWT}}%
      \csname LTb\endcsname%
      \put(5024,1097){\makebox(0,0)[r]{\strut{}short TWT with beam}}%
      \csname LTb\endcsname%
      \put(5024,877){\makebox(0,0)[r]{\strut{}short TWT without beam}}%
    }%
    \gplbacktext
    \put(0,0){\includegraphics{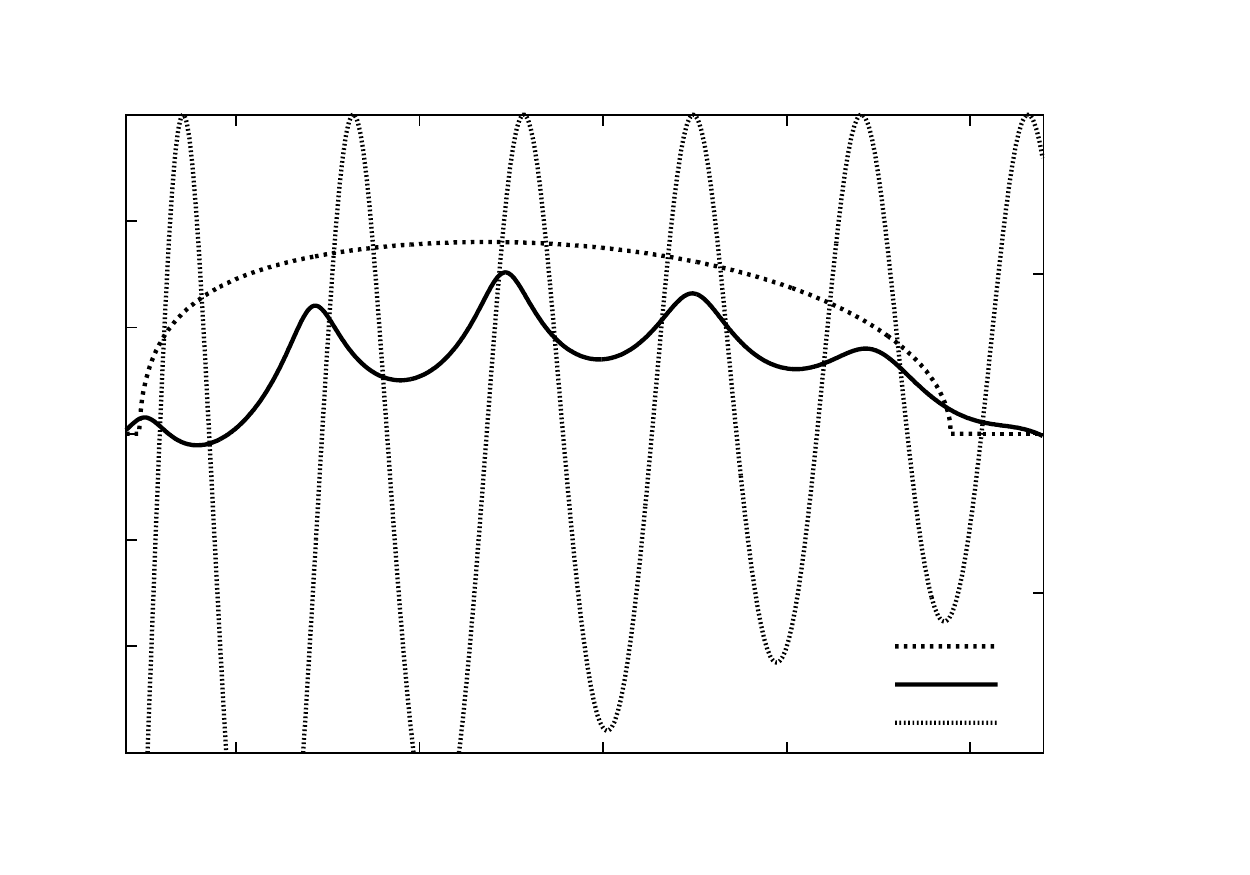}}%
    \gplfronttext
  \end{picture}%
\endgroup

%% file: EllipseTheta0.tex
\begingroup
  \makeatletter
  \providecommand\color[2][]{%
    \GenericError{(gnuplot) \space\space\space\@spaces}{%
      Package color not loaded in conjunction with
      terminal option `colourtext'%
    }{See the gnuplot documentation for explanation.%
    }{Either use 'blacktext' in gnuplot or load the package
      color.sty in LaTeX.}%
    \renewcommand\color[2][]{}%
  }%
  \providecommand\includegraphics[2][]{%
    \GenericError{(gnuplot) \space\space\space\@spaces}{%
      Package graphicx or graphics not loaded%
    }{See the gnuplot documentation for explanation.%
    }{The gnuplot epslatex terminal needs graphicx.sty or graphics.sty.}%
    \renewcommand\includegraphics[2][]{}%
  }%
  \providecommand\rotatebox[2]{#2}%
  \@ifundefined{ifGPcolor}{%
    \newif\ifGPcolor
    \GPcolorfalse
  }{}%
  \@ifundefined{ifGPblacktext}{%
    \newif\ifGPblacktext
    \GPblacktexttrue
  }{}%
  \let\gplgaddtomacro\g@addto@macro
  \gdef\gplbacktext{}%
  \gdef\gplfronttext{}%
  \makeatother
  \ifGPblacktext
    \def\colorrgb#1{}%
    \def\colorgray#1{}%
  \else
    \ifGPcolor
      \def\colorrgb#1{\color[rgb]{#1}}%
      \def\colorgray#1{\color[gray]{#1}}%
      \expandafter\def\csname LTw\endcsname{\color{white}}%
      \expandafter\def\csname LTb\endcsname{\color{black}}%
      \expandafter\def\csname LTa\endcsname{\color{black}}%
      \expandafter\def\csname LT0\endcsname{\color[rgb]{1,0,0}}%
      \expandafter\def\csname LT1\endcsname{\color[rgb]{0,1,0}}%
      \expandafter\def\csname LT2\endcsname{\color[rgb]{0,0,1}}%
      \expandafter\def\csname LT3\endcsname{\color[rgb]{1,0,1}}%
      \expandafter\def\csname LT4\endcsname{\color[rgb]{0,1,1}}%
      \expandafter\def\csname LT5\endcsname{\color[rgb]{1,1,0}}%
      \expandafter\def\csname LT6\endcsname{\color[rgb]{0,0,0}}%
      \expandafter\def\csname LT7\endcsname{\color[rgb]{1,0.3,0}}%
      \expandafter\def\csname LT8\endcsname{\color[rgb]{0.5,0.5,0.5}}%
    \else
      \def\colorrgb#1{\color{black}}%
      \def\colorgray#1{\color[gray]{#1}}%
      \expandafter\def\csname LTw\endcsname{\color{white}}%
      \expandafter\def\csname LTb\endcsname{\color{black}}%
      \expandafter\def\csname LTa\endcsname{\color{black}}%
      \expandafter\def\csname LT0\endcsname{\color{black}}%
      \expandafter\def\csname LT1\endcsname{\color{black}}%
      \expandafter\def\csname LT2\endcsname{\color{black}}%
      \expandafter\def\csname LT3\endcsname{\color{black}}%
      \expandafter\def\csname LT4\endcsname{\color{black}}%
      \expandafter\def\csname LT5\endcsname{\color{black}}%
      \expandafter\def\csname LT6\endcsname{\color{black}}%
      \expandafter\def\csname LT7\endcsname{\color{black}}%
      \expandafter\def\csname LT8\endcsname{\color{black}}%
    \fi
  \fi
  \setlength{\unitlength}{0.0500bp}%
  \begin{picture}(7200.00,5040.00)%
    \gplgaddtomacro\gplbacktext{%
      \csname LTb\endcsname%
      \put(594,704){\makebox(0,0)[r]{\strut{}-30}}%
      \put(594,1229){\makebox(0,0)[r]{\strut{}-20}}%
      \put(594,1754){\makebox(0,0)[r]{\strut{}-10}}%
      \put(594,2279){\makebox(0,0)[r]{\strut{} 0}}%
      \put(594,2804){\makebox(0,0)[r]{\strut{} 10}}%
      \put(594,3329){\makebox(0,0)[r]{\strut{} 20}}%
      \put(594,3854){\makebox(0,0)[r]{\strut{} 30}}%
      \put(594,4379){\makebox(0,0)[r]{\strut{} 40}}%
      \put(929,484){\makebox(0,0){\strut{} 10}}%
      \put(1946,484){\makebox(0,0){\strut{} 11}}%
      \put(2962,484){\makebox(0,0){\strut{} 12}}%
      \put(3978,484){\makebox(0,0){\strut{} 13}}%
      \put(4995,484){\makebox(0,0){\strut{} 14}}%
      \put(6011,484){\makebox(0,0){\strut{} 15}}%
      \put(6143,949){\makebox(0,0)[l]{\strut{}-1.4}}%
      \put(6143,1439){\makebox(0,0)[l]{\strut{}-1.2}}%
      \put(6143,1929){\makebox(0,0)[l]{\strut{}-1}}%
      \put(6143,2419){\makebox(0,0)[l]{\strut{}-0.8}}%
      \put(6143,2909){\makebox(0,0)[l]{\strut{}-0.6}}%
      \put(6143,3399){\makebox(0,0)[l]{\strut{}-0.4}}%
      \put(6143,3889){\makebox(0,0)[l]{\strut{}-0.2}}%
      \put(6143,4379){\makebox(0,0)[l]{\strut{} 0}}%
      \put(3368,154){\makebox(0,0){\strut{}$f$ (GHz)}}%
      \put(3368,4709){\makebox(0,0){\strut{}$\epsilon^{\mathrm{eff}}_{rr}=$2.351, $\epsilon^{\mathrm{eff}}_{zz}=$2.351, $\theta=0.400$}}%
    }%
    \gplgaddtomacro\gplfronttext{%
      \csname LTb\endcsname%
      \put(5024,1317){\makebox(0,0)[r]{\strut{}infinite TWT}}%
      \csname LTb\endcsname%
      \put(5024,1097){\makebox(0,0)[r]{\strut{}short TWT with beam}}%
      \csname LTb\endcsname%
      \put(5024,877){\makebox(0,0)[r]{\strut{}short TWT without beam}}%
    }%
    \gplbacktext
    \put(0,0){\includegraphics{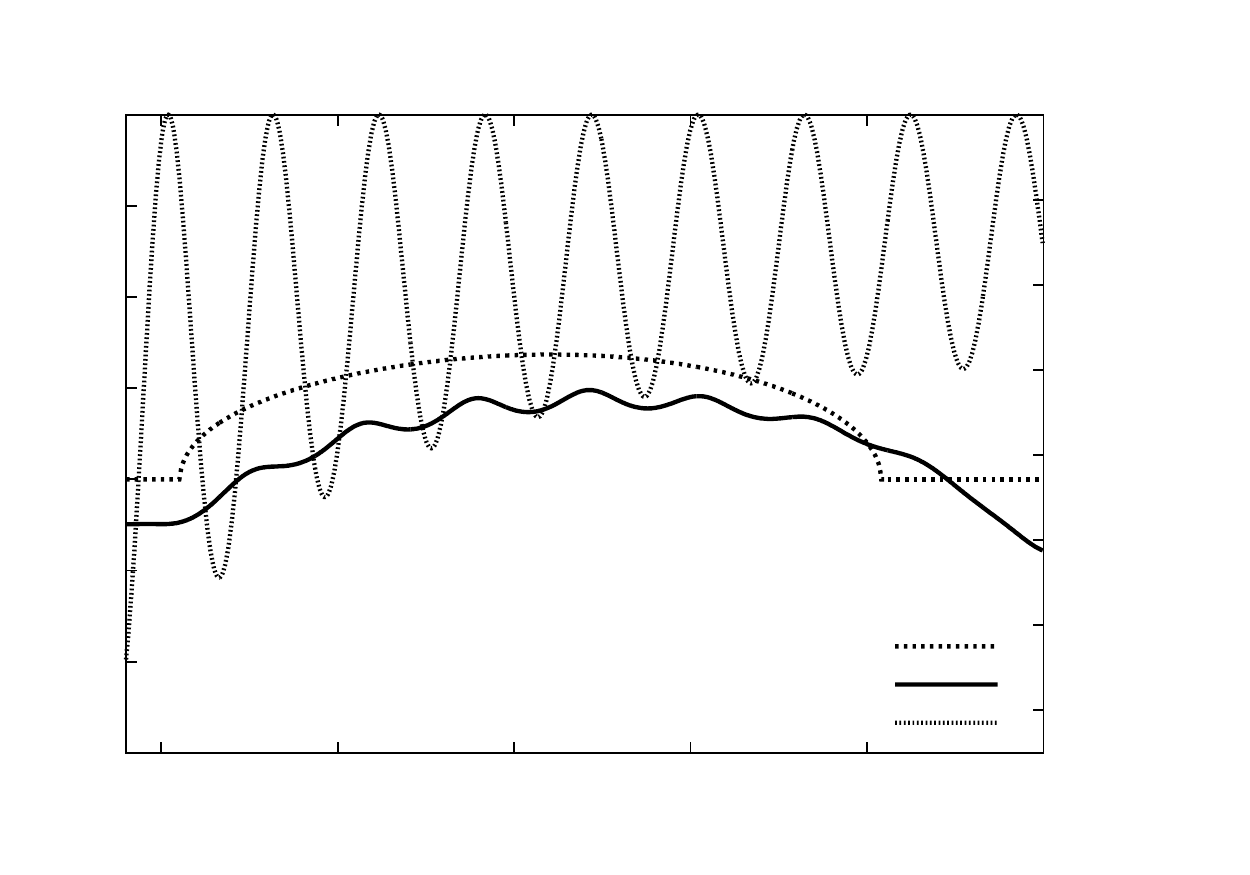}}%
    \gplfronttext
  \end{picture}%
\endgroup

%% file: EllipseTheta1.tex
\begingroup
  \makeatletter
  \providecommand\color[2][]{%
    \GenericError{(gnuplot) \space\space\space\@spaces}{%
      Package color not loaded in conjunction with
      terminal option `colourtext'%
    }{See the gnuplot documentation for explanation.%
    }{Either use 'blacktext' in gnuplot or load the package
      color.sty in LaTeX.}%
    \renewcommand\color[2][]{}%
  }%
  \providecommand\includegraphics[2][]{%
    \GenericError{(gnuplot) \space\space\space\@spaces}{%
      Package graphicx or graphics not loaded%
    }{See the gnuplot documentation for explanation.%
    }{The gnuplot epslatex terminal needs graphicx.sty or graphics.sty.}%
    \renewcommand\includegraphics[2][]{}%
  }%
  \providecommand\rotatebox[2]{#2}%
  \@ifundefined{ifGPcolor}{%
    \newif\ifGPcolor
    \GPcolorfalse
  }{}%
  \@ifundefined{ifGPblacktext}{%
    \newif\ifGPblacktext
    \GPblacktexttrue
  }{}%
  \let\gplgaddtomacro\g@addto@macro
  \gdef\gplbacktext{}%
  \gdef\gplfronttext{}%
  \makeatother
  \ifGPblacktext
    \def\colorrgb#1{}%
    \def\colorgray#1{}%
  \else
    \ifGPcolor
      \def\colorrgb#1{\color[rgb]{#1}}%
      \def\colorgray#1{\color[gray]{#1}}%
      \expandafter\def\csname LTw\endcsname{\color{white}}%
      \expandafter\def\csname LTb\endcsname{\color{black}}%
      \expandafter\def\csname LTa\endcsname{\color{black}}%
      \expandafter\def\csname LT0\endcsname{\color[rgb]{1,0,0}}%
      \expandafter\def\csname LT1\endcsname{\color[rgb]{0,1,0}}%
      \expandafter\def\csname LT2\endcsname{\color[rgb]{0,0,1}}%
      \expandafter\def\csname LT3\endcsname{\color[rgb]{1,0,1}}%
      \expandafter\def\csname LT4\endcsname{\color[rgb]{0,1,1}}%
      \expandafter\def\csname LT5\endcsname{\color[rgb]{1,1,0}}%
      \expandafter\def\csname LT6\endcsname{\color[rgb]{0,0,0}}%
      \expandafter\def\csname LT7\endcsname{\color[rgb]{1,0.3,0}}%
      \expandafter\def\csname LT8\endcsname{\color[rgb]{0.5,0.5,0.5}}%
    \else
      \def\colorrgb#1{\color{black}}%
      \def\colorgray#1{\color[gray]{#1}}%
      \expandafter\def\csname LTw\endcsname{\color{white}}%
      \expandafter\def\csname LTb\endcsname{\color{black}}%
      \expandafter\def\csname LTa\endcsname{\color{black}}%
      \expandafter\def\csname LT0\endcsname{\color{black}}%
      \expandafter\def\csname LT1\endcsname{\color{black}}%
      \expandafter\def\csname LT2\endcsname{\color{black}}%
      \expandafter\def\csname LT3\endcsname{\color{black}}%
      \expandafter\def\csname LT4\endcsname{\color{black}}%
      \expandafter\def\csname LT5\endcsname{\color{black}}%
      \expandafter\def\csname LT6\endcsname{\color{black}}%
      \expandafter\def\csname LT7\endcsname{\color{black}}%
      \expandafter\def\csname LT8\endcsname{\color{black}}%
    \fi
  \fi
  \setlength{\unitlength}{0.0500bp}%
  \begin{picture}(7200.00,5040.00)%
    \gplgaddtomacro\gplbacktext{%
      \csname LTb\endcsname%
      \put(594,704){\makebox(0,0)[r]{\strut{}-30}}%
      \put(594,1229){\makebox(0,0)[r]{\strut{}-20}}%
      \put(594,1754){\makebox(0,0)[r]{\strut{}-10}}%
      \put(594,2279){\makebox(0,0)[r]{\strut{} 0}}%
      \put(594,2804){\makebox(0,0)[r]{\strut{} 10}}%
      \put(594,3329){\makebox(0,0)[r]{\strut{} 20}}%
      \put(594,3854){\makebox(0,0)[r]{\strut{} 30}}%
      \put(594,4379){\makebox(0,0)[r]{\strut{} 40}}%
      \put(726,484){\makebox(0,0){\strut{} 8.5}}%
      \put(1519,484){\makebox(0,0){\strut{} 9}}%
      \put(2311,484){\makebox(0,0){\strut{} 9.5}}%
      \put(3104,484){\makebox(0,0){\strut{} 10}}%
      \put(3897,484){\makebox(0,0){\strut{} 10.5}}%
      \put(4690,484){\makebox(0,0){\strut{} 11}}%
      \put(5482,484){\makebox(0,0){\strut{} 11.5}}%
      \put(6275,484){\makebox(0,0){\strut{} 12}}%
      \put(6407,704){\makebox(0,0)[l]{\strut{}-5}}%
      \put(6407,1439){\makebox(0,0)[l]{\strut{}-4}}%
      \put(6407,2174){\makebox(0,0)[l]{\strut{}-3}}%
      \put(6407,2909){\makebox(0,0)[l]{\strut{}-2}}%
      \put(6407,3644){\makebox(0,0)[l]{\strut{}-1}}%
      \put(6407,4379){\makebox(0,0)[l]{\strut{} 0}}%
      \put(3500,154){\makebox(0,0){\strut{}$f$ (GHz)}}%
      \put(3500,4709){\makebox(0,0){\strut{}$\epsilon^{\mathrm{eff}}_{rr}=$3.080, $\epsilon^{\mathrm{eff}}_{zz}=$3.080, $\theta=0.500$}}%
    }%
    \gplgaddtomacro\gplfronttext{%
      \csname LTb\endcsname%
      \put(5288,1317){\makebox(0,0)[r]{\strut{}infinite TWT}}%
      \csname LTb\endcsname%
      \put(5288,1097){\makebox(0,0)[r]{\strut{}short TWT with beam}}%
      \csname LTb\endcsname%
      \put(5288,877){\makebox(0,0)[r]{\strut{}short TWT without beam}}%
    }%
    \gplbacktext
    \put(0,0){\includegraphics{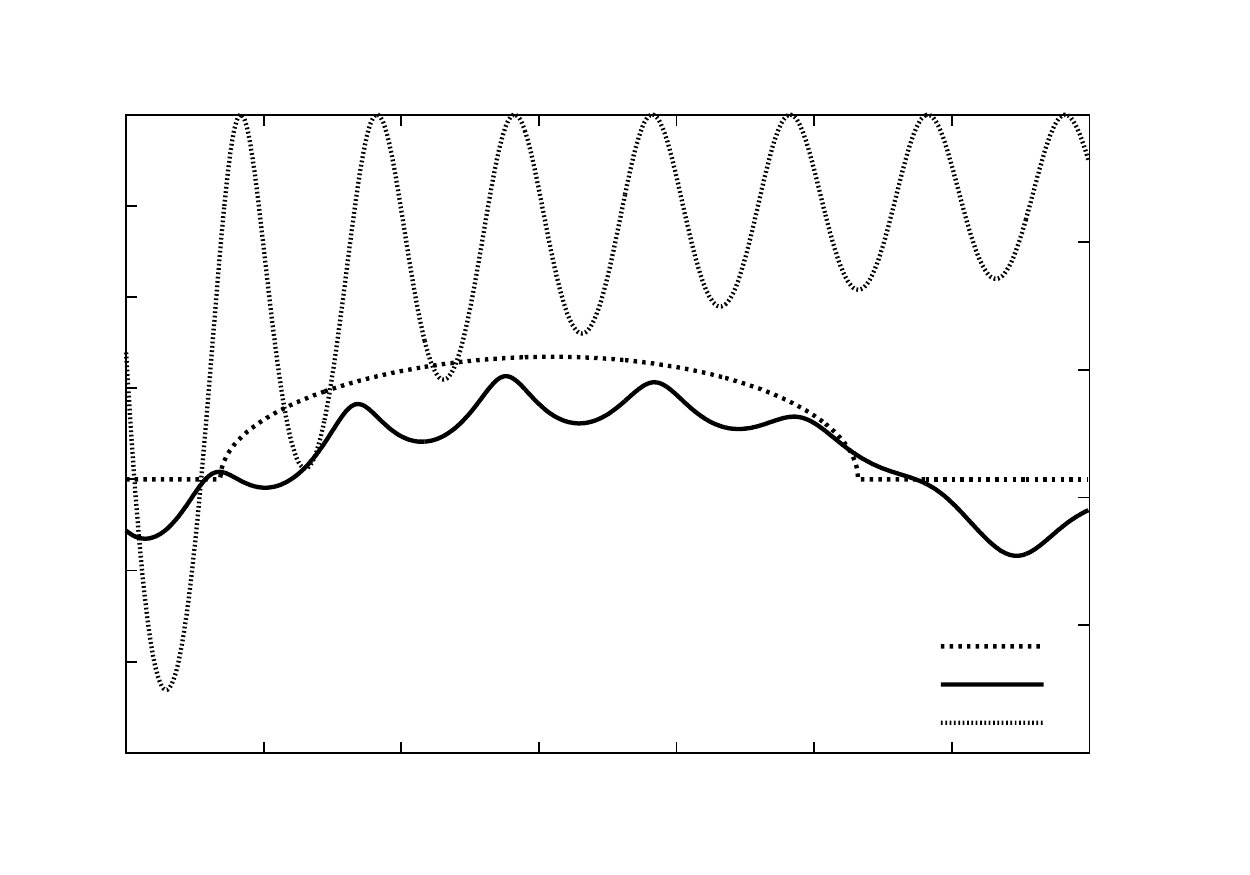}}%
    \gplfronttext
  \end{picture}%
\endgroup

%% file: EllipseTheta2.tex
\begingroup
  \makeatletter
  \providecommand\color[2][]{%
    \GenericError{(gnuplot) \space\space\space\@spaces}{%
      Package color not loaded in conjunction with
      terminal option `colourtext'%
    }{See the gnuplot documentation for explanation.%
    }{Either use 'blacktext' in gnuplot or load the package
      color.sty in LaTeX.}%
    \renewcommand\color[2][]{}%
  }%
  \providecommand\includegraphics[2][]{%
    \GenericError{(gnuplot) \space\space\space\@spaces}{%
      Package graphicx or graphics not loaded%
    }{See the gnuplot documentation for explanation.%
    }{The gnuplot epslatex terminal needs graphicx.sty or graphics.sty.}%
    \renewcommand\includegraphics[2][]{}%
  }%
  \providecommand\rotatebox[2]{#2}%
  \@ifundefined{ifGPcolor}{%
    \newif\ifGPcolor
    \GPcolorfalse
  }{}%
  \@ifundefined{ifGPblacktext}{%
    \newif\ifGPblacktext
    \GPblacktexttrue
  }{}%
  \let\gplgaddtomacro\g@addto@macro
  \gdef\gplbacktext{}%
  \gdef\gplfronttext{}%
  \makeatother
  \ifGPblacktext
    \def\colorrgb#1{}%
    \def\colorgray#1{}%
  \else
    \ifGPcolor
      \def\colorrgb#1{\color[rgb]{#1}}%
      \def\colorgray#1{\color[gray]{#1}}%
      \expandafter\def\csname LTw\endcsname{\color{white}}%
      \expandafter\def\csname LTb\endcsname{\color{black}}%
      \expandafter\def\csname LTa\endcsname{\color{black}}%
      \expandafter\def\csname LT0\endcsname{\color[rgb]{1,0,0}}%
      \expandafter\def\csname LT1\endcsname{\color[rgb]{0,1,0}}%
      \expandafter\def\csname LT2\endcsname{\color[rgb]{0,0,1}}%
      \expandafter\def\csname LT3\endcsname{\color[rgb]{1,0,1}}%
      \expandafter\def\csname LT4\endcsname{\color[rgb]{0,1,1}}%
      \expandafter\def\csname LT5\endcsname{\color[rgb]{1,1,0}}%
      \expandafter\def\csname LT6\endcsname{\color[rgb]{0,0,0}}%
      \expandafter\def\csname LT7\endcsname{\color[rgb]{1,0.3,0}}%
      \expandafter\def\csname LT8\endcsname{\color[rgb]{0.5,0.5,0.5}}%
    \else
      \def\colorrgb#1{\color{black}}%
      \def\colorgray#1{\color[gray]{#1}}%
      \expandafter\def\csname LTw\endcsname{\color{white}}%
      \expandafter\def\csname LTb\endcsname{\color{black}}%
      \expandafter\def\csname LTa\endcsname{\color{black}}%
      \expandafter\def\csname LT0\endcsname{\color{black}}%
      \expandafter\def\csname LT1\endcsname{\color{black}}%
      \expandafter\def\csname LT2\endcsname{\color{black}}%
      \expandafter\def\csname LT3\endcsname{\color{black}}%
      \expandafter\def\csname LT4\endcsname{\color{black}}%
      \expandafter\def\csname LT5\endcsname{\color{black}}%
      \expandafter\def\csname LT6\endcsname{\color{black}}%
      \expandafter\def\csname LT7\endcsname{\color{black}}%
      \expandafter\def\csname LT8\endcsname{\color{black}}%
    \fi
  \fi
  \setlength{\unitlength}{0.0500bp}%
  \begin{picture}(7200.00,5040.00)%
    \gplgaddtomacro\gplbacktext{%
      \csname LTb\endcsname%
      \put(594,704){\makebox(0,0)[r]{\strut{}-30}}%
      \put(594,1229){\makebox(0,0)[r]{\strut{}-20}}%
      \put(594,1754){\makebox(0,0)[r]{\strut{}-10}}%
      \put(594,2279){\makebox(0,0)[r]{\strut{} 0}}%
      \put(594,2804){\makebox(0,0)[r]{\strut{} 10}}%
      \put(594,3329){\makebox(0,0)[r]{\strut{} 20}}%
      \put(594,3854){\makebox(0,0)[r]{\strut{} 30}}%
      \put(594,4379){\makebox(0,0)[r]{\strut{} 40}}%
      \put(726,484){\makebox(0,0){\strut{} 7.5}}%
      \put(2113,484){\makebox(0,0){\strut{} 8}}%
      \put(3501,484){\makebox(0,0){\strut{} 8.5}}%
      \put(4888,484){\makebox(0,0){\strut{} 9}}%
      \put(6275,484){\makebox(0,0){\strut{} 9.5}}%
      \put(6407,704){\makebox(0,0)[l]{\strut{}-8}}%
      \put(6407,1163){\makebox(0,0)[l]{\strut{}-7}}%
      \put(6407,1623){\makebox(0,0)[l]{\strut{}-6}}%
      \put(6407,2082){\makebox(0,0)[l]{\strut{}-5}}%
      \put(6407,2542){\makebox(0,0)[l]{\strut{}-4}}%
      \put(6407,3001){\makebox(0,0)[l]{\strut{}-3}}%
      \put(6407,3460){\makebox(0,0)[l]{\strut{}-2}}%
      \put(6407,3920){\makebox(0,0)[l]{\strut{}-1}}%
      \put(6407,4379){\makebox(0,0)[l]{\strut{} 0}}%
      \put(3500,154){\makebox(0,0){\strut{}$f$ (GHz)}}%
      \put(3500,4709){\makebox(0,0){\strut{}$\epsilon^{\mathrm{eff}}_{rr}=$4.342, $\epsilon^{\mathrm{eff}}_{zz}=$4.342, $\theta=0.600$}}%
    }%
    \gplgaddtomacro\gplfronttext{%
      \csname LTb\endcsname%
      \put(5288,1317){\makebox(0,0)[r]{\strut{}infinite TWT}}%
      \csname LTb\endcsname%
      \put(5288,1097){\makebox(0,0)[r]{\strut{}short TWT with beam}}%
      \csname LTb\endcsname%
      \put(5288,877){\makebox(0,0)[r]{\strut{}short TWT without beam}}%
    }%
    \gplbacktext
    \put(0,0){\includegraphics{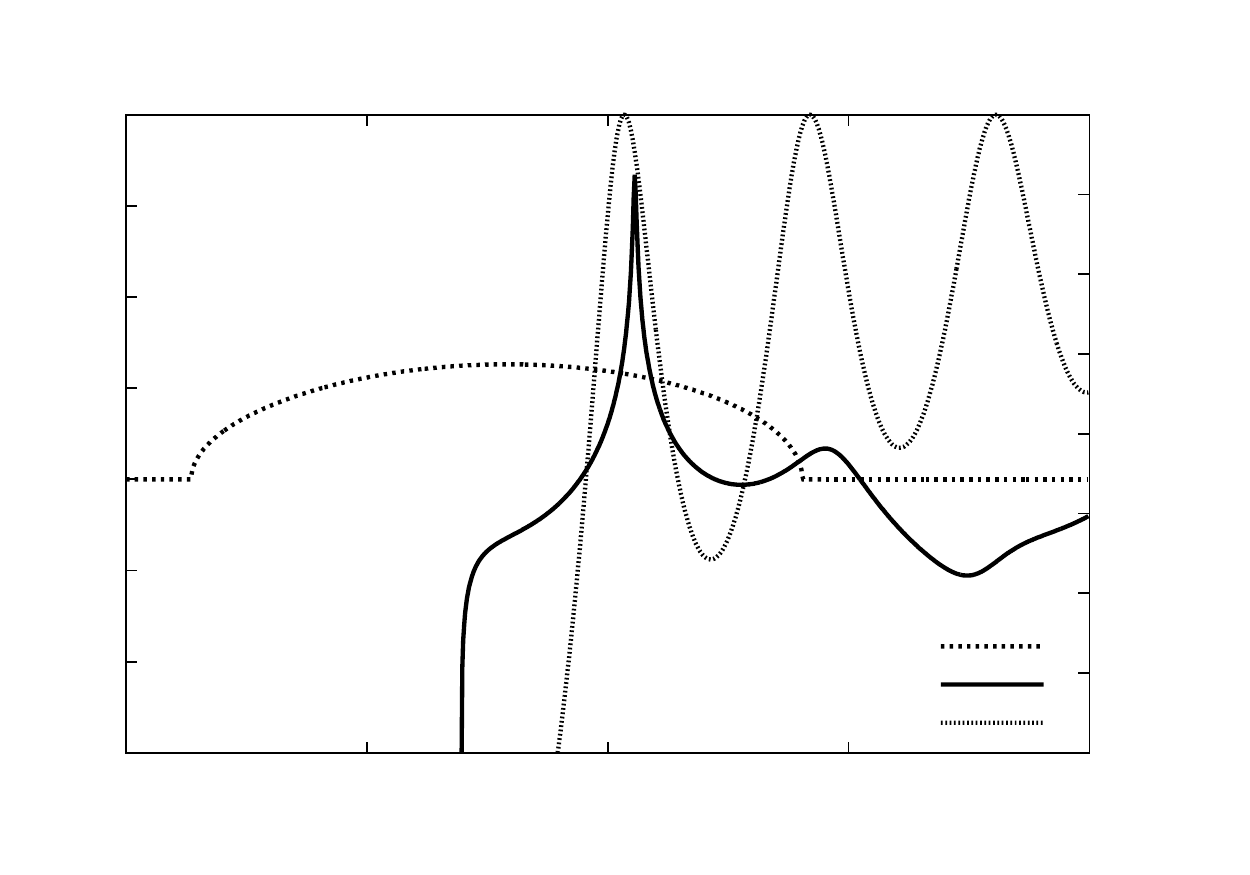}}%
    \gplfronttext
  \end{picture}%
\endgroup

%% file: Ellipse125Theta1.tex
\begingroup
  \makeatletter
  \providecommand\color[2][]{%
    \GenericError{(gnuplot) \space\space\space\@spaces}{%
      Package color not loaded in conjunction with
      terminal option `colourtext'%
    }{See the gnuplot documentation for explanation.%
    }{Either use 'blacktext' in gnuplot or load the package
      color.sty in LaTeX.}%
    \renewcommand\color[2][]{}%
  }%
  \providecommand\includegraphics[2][]{%
    \GenericError{(gnuplot) \space\space\space\@spaces}{%
      Package graphicx or graphics not loaded%
    }{See the gnuplot documentation for explanation.%
    }{The gnuplot epslatex terminal needs graphicx.sty or graphics.sty.}%
    \renewcommand\includegraphics[2][]{}%
  }%
  \providecommand\rotatebox[2]{#2}%
  \@ifundefined{ifGPcolor}{%
    \newif\ifGPcolor
    \GPcolorfalse
  }{}%
  \@ifundefined{ifGPblacktext}{%
    \newif\ifGPblacktext
    \GPblacktexttrue
  }{}%
  \let\gplgaddtomacro\g@addto@macro
  \gdef\gplbacktext{}%
  \gdef\gplfronttext{}%
  \makeatother
  \ifGPblacktext
    \def\colorrgb#1{}%
    \def\colorgray#1{}%
  \else
    \ifGPcolor
      \def\colorrgb#1{\color[rgb]{#1}}%
      \def\colorgray#1{\color[gray]{#1}}%
      \expandafter\def\csname LTw\endcsname{\color{white}}%
      \expandafter\def\csname LTb\endcsname{\color{black}}%
      \expandafter\def\csname LTa\endcsname{\color{black}}%
      \expandafter\def\csname LT0\endcsname{\color[rgb]{1,0,0}}%
      \expandafter\def\csname LT1\endcsname{\color[rgb]{0,1,0}}%
      \expandafter\def\csname LT2\endcsname{\color[rgb]{0,0,1}}%
      \expandafter\def\csname LT3\endcsname{\color[rgb]{1,0,1}}%
      \expandafter\def\csname LT4\endcsname{\color[rgb]{0,1,1}}%
      \expandafter\def\csname LT5\endcsname{\color[rgb]{1,1,0}}%
      \expandafter\def\csname LT6\endcsname{\color[rgb]{0,0,0}}%
      \expandafter\def\csname LT7\endcsname{\color[rgb]{1,0.3,0}}%
      \expandafter\def\csname LT8\endcsname{\color[rgb]{0.5,0.5,0.5}}%
    \else
      \def\colorrgb#1{\color{black}}%
      \def\colorgray#1{\color[gray]{#1}}%
      \expandafter\def\csname LTw\endcsname{\color{white}}%
      \expandafter\def\csname LTb\endcsname{\color{black}}%
      \expandafter\def\csname LTa\endcsname{\color{black}}%
      \expandafter\def\csname LT0\endcsname{\color{black}}%
      \expandafter\def\csname LT1\endcsname{\color{black}}%
      \expandafter\def\csname LT2\endcsname{\color{black}}%
      \expandafter\def\csname LT3\endcsname{\color{black}}%
      \expandafter\def\csname LT4\endcsname{\color{black}}%
      \expandafter\def\csname LT5\endcsname{\color{black}}%
      \expandafter\def\csname LT6\endcsname{\color{black}}%
      \expandafter\def\csname LT7\endcsname{\color{black}}%
      \expandafter\def\csname LT8\endcsname{\color{black}}%
    \fi
  \fi
  \setlength{\unitlength}{0.0500bp}%
  \begin{picture}(7200.00,5040.00)%
    \gplgaddtomacro\gplbacktext{%
      \csname LTb\endcsname%
      \put(594,704){\makebox(0,0)[r]{\strut{}-30}}%
      \put(594,1317){\makebox(0,0)[r]{\strut{}-20}}%
      \put(594,1929){\makebox(0,0)[r]{\strut{}-10}}%
      \put(594,2542){\makebox(0,0)[r]{\strut{} 0}}%
      \put(594,3154){\makebox(0,0)[r]{\strut{} 10}}%
      \put(594,3767){\makebox(0,0)[r]{\strut{} 20}}%
      \put(594,4379){\makebox(0,0)[r]{\strut{} 30}}%
      \put(726,484){\makebox(0,0){\strut{} 10}}%
      \put(1313,484){\makebox(0,0){\strut{} 11}}%
      \put(1900,484){\makebox(0,0){\strut{} 12}}%
      \put(2488,484){\makebox(0,0){\strut{} 13}}%
      \put(3075,484){\makebox(0,0){\strut{} 14}}%
      \put(3662,484){\makebox(0,0){\strut{} 15}}%
      \put(4249,484){\makebox(0,0){\strut{} 16}}%
      \put(4837,484){\makebox(0,0){\strut{} 17}}%
      \put(5424,484){\makebox(0,0){\strut{} 18}}%
      \put(6011,484){\makebox(0,0){\strut{} 19}}%
      \put(6143,704){\makebox(0,0)[l]{\strut{}-1}}%
      \put(6143,1439){\makebox(0,0)[l]{\strut{}-0.8}}%
      \put(6143,2174){\makebox(0,0)[l]{\strut{}-0.6}}%
      \put(6143,2909){\makebox(0,0)[l]{\strut{}-0.4}}%
      \put(6143,3644){\makebox(0,0)[l]{\strut{}-0.2}}%
      \put(6143,4379){\makebox(0,0)[l]{\strut{} 0}}%
      \put(3368,154){\makebox(0,0){\strut{}$f$ (GHz)}}%
      \put(3368,4709){\makebox(0,0){\strut{}$\epsilon^{\mathrm{eff}}_{rr}=$1.894, $\epsilon^{\mathrm{eff}}_{zz}=$2.381, $\theta=$0.350}}%
    }%
    \gplgaddtomacro\gplfronttext{%
      \csname LTb\endcsname%
      \put(5024,1317){\makebox(0,0)[r]{\strut{}infinite TWT}}%
      \csname LTb\endcsname%
      \put(5024,1097){\makebox(0,0)[r]{\strut{}short TWT with beam}}%
      \csname LTb\endcsname%
      \put(5024,877){\makebox(0,0)[r]{\strut{}short TWT without beam}}%
    }%
    \gplbacktext
    \put(0,0){\includegraphics{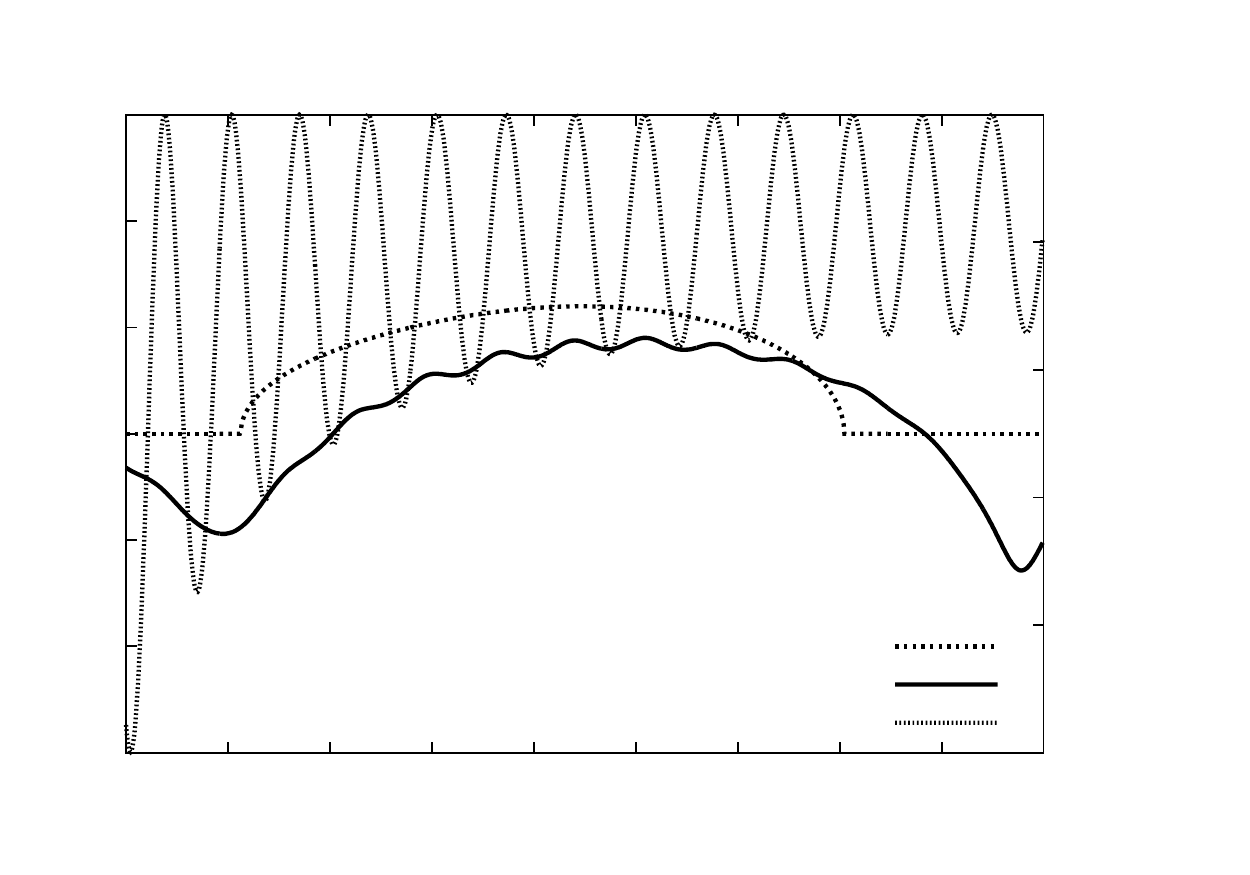}}%
    \gplfronttext
  \end{picture}%
\endgroup

%% file: Ellipse125Theta3.tex
\begingroup
  \makeatletter
  \providecommand\color[2][]{%
    \GenericError{(gnuplot) \space\space\space\@spaces}{%
      Package color not loaded in conjunction with
      terminal option `colourtext'%
    }{See the gnuplot documentation for explanation.%
    }{Either use 'blacktext' in gnuplot or load the package
      color.sty in LaTeX.}%
    \renewcommand\color[2][]{}%
  }%
  \providecommand\includegraphics[2][]{%
    \GenericError{(gnuplot) \space\space\space\@spaces}{%
      Package graphicx or graphics not loaded%
    }{See the gnuplot documentation for explanation.%
    }{The gnuplot epslatex terminal needs graphicx.sty or graphics.sty.}%
    \renewcommand\includegraphics[2][]{}%
  }%
  \providecommand\rotatebox[2]{#2}%
  \@ifundefined{ifGPcolor}{%
    \newif\ifGPcolor
    \GPcolorfalse
  }{}%
  \@ifundefined{ifGPblacktext}{%
    \newif\ifGPblacktext
    \GPblacktexttrue
  }{}%
  \let\gplgaddtomacro\g@addto@macro
  \gdef\gplbacktext{}%
  \gdef\gplfronttext{}%
  \makeatother
  \ifGPblacktext
    \def\colorrgb#1{}%
    \def\colorgray#1{}%
  \else
    \ifGPcolor
      \def\colorrgb#1{\color[rgb]{#1}}%
      \def\colorgray#1{\color[gray]{#1}}%
      \expandafter\def\csname LTw\endcsname{\color{white}}%
      \expandafter\def\csname LTb\endcsname{\color{black}}%
      \expandafter\def\csname LTa\endcsname{\color{black}}%
      \expandafter\def\csname LT0\endcsname{\color[rgb]{1,0,0}}%
      \expandafter\def\csname LT1\endcsname{\color[rgb]{0,1,0}}%
      \expandafter\def\csname LT2\endcsname{\color[rgb]{0,0,1}}%
      \expandafter\def\csname LT3\endcsname{\color[rgb]{1,0,1}}%
      \expandafter\def\csname LT4\endcsname{\color[rgb]{0,1,1}}%
      \expandafter\def\csname LT5\endcsname{\color[rgb]{1,1,0}}%
      \expandafter\def\csname LT6\endcsname{\color[rgb]{0,0,0}}%
      \expandafter\def\csname LT7\endcsname{\color[rgb]{1,0.3,0}}%
      \expandafter\def\csname LT8\endcsname{\color[rgb]{0.5,0.5,0.5}}%
    \else
      \def\colorrgb#1{\color{black}}%
      \def\colorgray#1{\color[gray]{#1}}%
      \expandafter\def\csname LTw\endcsname{\color{white}}%
      \expandafter\def\csname LTb\endcsname{\color{black}}%
      \expandafter\def\csname LTa\endcsname{\color{black}}%
      \expandafter\def\csname LT0\endcsname{\color{black}}%
      \expandafter\def\csname LT1\endcsname{\color{black}}%
      \expandafter\def\csname LT2\endcsname{\color{black}}%
      \expandafter\def\csname LT3\endcsname{\color{black}}%
      \expandafter\def\csname LT4\endcsname{\color{black}}%
      \expandafter\def\csname LT5\endcsname{\color{black}}%
      \expandafter\def\csname LT6\endcsname{\color{black}}%
      \expandafter\def\csname LT7\endcsname{\color{black}}%
      \expandafter\def\csname LT8\endcsname{\color{black}}%
    \fi
  \fi
  \setlength{\unitlength}{0.0500bp}%
  \begin{picture}(7200.00,5040.00)%
    \gplgaddtomacro\gplbacktext{%
      \csname LTb\endcsname%
      \put(594,704){\makebox(0,0)[r]{\strut{}-30}}%
      \put(594,1317){\makebox(0,0)[r]{\strut{}-20}}%
      \put(594,1929){\makebox(0,0)[r]{\strut{}-10}}%
      \put(594,2542){\makebox(0,0)[r]{\strut{} 0}}%
      \put(594,3154){\makebox(0,0)[r]{\strut{} 10}}%
      \put(594,3767){\makebox(0,0)[r]{\strut{} 20}}%
      \put(594,4379){\makebox(0,0)[r]{\strut{} 30}}%
      \put(726,484){\makebox(0,0){\strut{} 8}}%
      \put(1607,484){\makebox(0,0){\strut{} 9}}%
      \put(2488,484){\makebox(0,0){\strut{} 10}}%
      \put(3369,484){\makebox(0,0){\strut{} 11}}%
      \put(4249,484){\makebox(0,0){\strut{} 12}}%
      \put(5130,484){\makebox(0,0){\strut{} 13}}%
      \put(6011,484){\makebox(0,0){\strut{} 14}}%
      \put(6143,704){\makebox(0,0)[l]{\strut{}-2}}%
      \put(6143,1623){\makebox(0,0)[l]{\strut{}-1.5}}%
      \put(6143,2542){\makebox(0,0)[l]{\strut{}-1}}%
      \put(6143,3460){\makebox(0,0)[l]{\strut{}-0.5}}%
      \put(6143,4379){\makebox(0,0)[l]{\strut{} 0}}%
      \put(3368,154){\makebox(0,0){\strut{}$f$ (GHz)}}%
      \put(3368,4709){\makebox(0,0){\strut{}$\epsilon^{\mathrm{eff}}_{rr}=$2.301, $\epsilon^{\mathrm{eff}}_{zz}=$3.383, $\theta=0.450$}}%
    }%
    \gplgaddtomacro\gplfronttext{%
      \csname LTb\endcsname%
      \put(5024,1317){\makebox(0,0)[r]{\strut{}infinite TWT}}%
      \csname LTb\endcsname%
      \put(5024,1097){\makebox(0,0)[r]{\strut{}short TWT with beam}}%
      \csname LTb\endcsname%
      \put(5024,877){\makebox(0,0)[r]{\strut{}short TWT without beam}}%
    }%
    \gplbacktext
    \put(0,0){\includegraphics{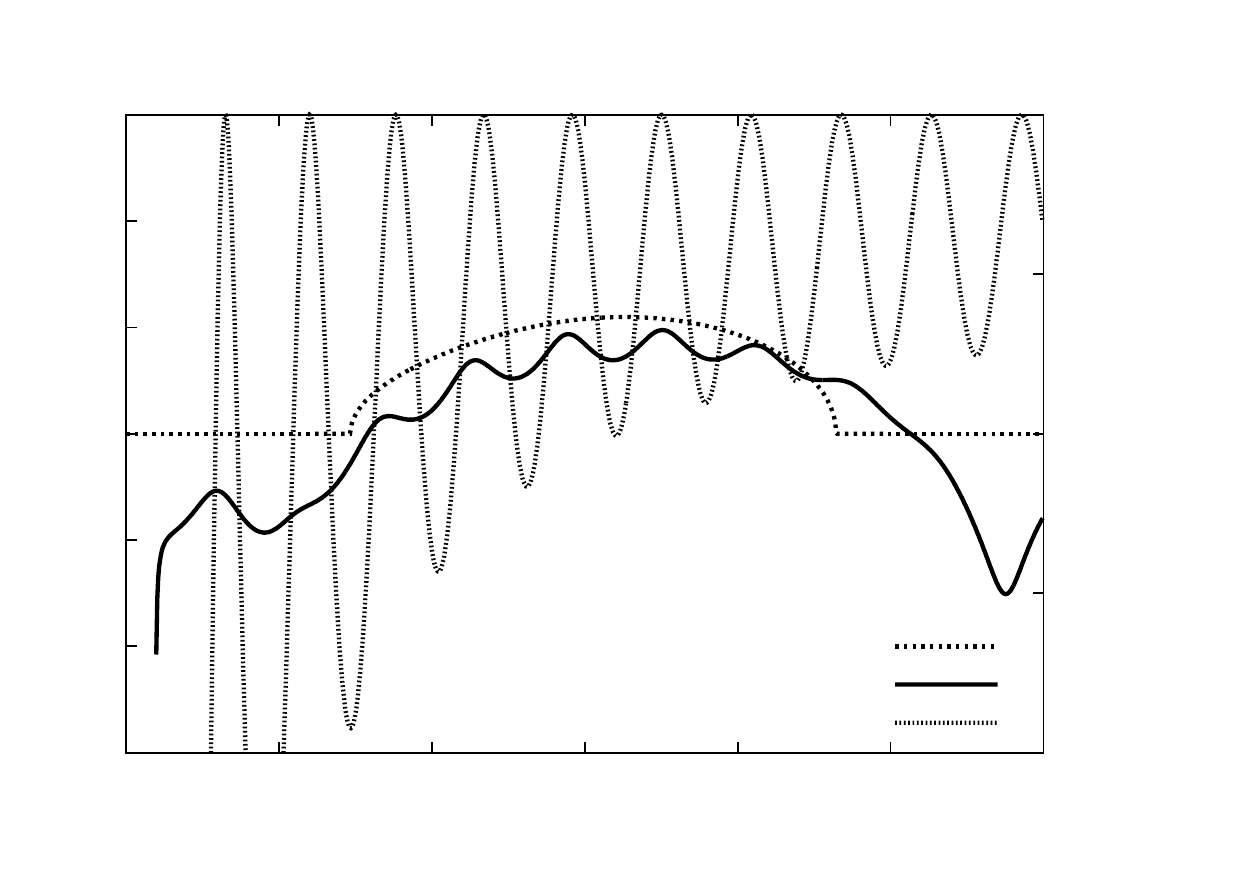}}%
    \gplfronttext
  \end{picture}%
\endgroup

%% file: Ellipse125Theta4.tex
\begingroup
  \makeatletter
  \providecommand\color[2][]{%
    \GenericError{(gnuplot) \space\space\space\@spaces}{%
      Package color not loaded in conjunction with
      terminal option `colourtext'%
    }{See the gnuplot documentation for explanation.%
    }{Either use 'blacktext' in gnuplot or load the package
      color.sty in LaTeX.}%
    \renewcommand\color[2][]{}%
  }%
  \providecommand\includegraphics[2][]{%
    \GenericError{(gnuplot) \space\space\space\@spaces}{%
      Package graphicx or graphics not loaded%
    }{See the gnuplot documentation for explanation.%
    }{The gnuplot epslatex terminal needs graphicx.sty or graphics.sty.}%
    \renewcommand\includegraphics[2][]{}%
  }%
  \providecommand\rotatebox[2]{#2}%
  \@ifundefined{ifGPcolor}{%
    \newif\ifGPcolor
    \GPcolorfalse
  }{}%
  \@ifundefined{ifGPblacktext}{%
    \newif\ifGPblacktext
    \GPblacktexttrue
  }{}%
  \let\gplgaddtomacro\g@addto@macro
  \gdef\gplbacktext{}%
  \gdef\gplfronttext{}%
  \makeatother
  \ifGPblacktext
    \def\colorrgb#1{}%
    \def\colorgray#1{}%
  \else
    \ifGPcolor
      \def\colorrgb#1{\color[rgb]{#1}}%
      \def\colorgray#1{\color[gray]{#1}}%
      \expandafter\def\csname LTw\endcsname{\color{white}}%
      \expandafter\def\csname LTb\endcsname{\color{black}}%
      \expandafter\def\csname LTa\endcsname{\color{black}}%
      \expandafter\def\csname LT0\endcsname{\color[rgb]{1,0,0}}%
      \expandafter\def\csname LT1\endcsname{\color[rgb]{0,1,0}}%
      \expandafter\def\csname LT2\endcsname{\color[rgb]{0,0,1}}%
      \expandafter\def\csname LT3\endcsname{\color[rgb]{1,0,1}}%
      \expandafter\def\csname LT4\endcsname{\color[rgb]{0,1,1}}%
      \expandafter\def\csname LT5\endcsname{\color[rgb]{1,1,0}}%
      \expandafter\def\csname LT6\endcsname{\color[rgb]{0,0,0}}%
      \expandafter\def\csname LT7\endcsname{\color[rgb]{1,0.3,0}}%
      \expandafter\def\csname LT8\endcsname{\color[rgb]{0.5,0.5,0.5}}%
    \else
      \def\colorrgb#1{\color{black}}%
      \def\colorgray#1{\color[gray]{#1}}%
      \expandafter\def\csname LTw\endcsname{\color{white}}%
      \expandafter\def\csname LTb\endcsname{\color{black}}%
      \expandafter\def\csname LTa\endcsname{\color{black}}%
      \expandafter\def\csname LT0\endcsname{\color{black}}%
      \expandafter\def\csname LT1\endcsname{\color{black}}%
      \expandafter\def\csname LT2\endcsname{\color{black}}%
      \expandafter\def\csname LT3\endcsname{\color{black}}%
      \expandafter\def\csname LT4\endcsname{\color{black}}%
      \expandafter\def\csname LT5\endcsname{\color{black}}%
      \expandafter\def\csname LT6\endcsname{\color{black}}%
      \expandafter\def\csname LT7\endcsname{\color{black}}%
      \expandafter\def\csname LT8\endcsname{\color{black}}%
    \fi
  \fi
  \setlength{\unitlength}{0.0500bp}%
  \begin{picture}(7200.00,5040.00)%
    \gplgaddtomacro\gplbacktext{%
      \csname LTb\endcsname%
      \put(594,704){\makebox(0,0)[r]{\strut{}-30}}%
      \put(594,1317){\makebox(0,0)[r]{\strut{}-20}}%
      \put(594,1929){\makebox(0,0)[r]{\strut{}-10}}%
      \put(594,2542){\makebox(0,0)[r]{\strut{} 0}}%
      \put(594,3154){\makebox(0,0)[r]{\strut{} 10}}%
      \put(594,3767){\makebox(0,0)[r]{\strut{} 20}}%
      \put(594,4379){\makebox(0,0)[r]{\strut{} 30}}%
      \put(726,484){\makebox(0,0){\strut{} 8}}%
      \put(1420,484){\makebox(0,0){\strut{} 8.5}}%
      \put(2113,484){\makebox(0,0){\strut{} 9}}%
      \put(2807,484){\makebox(0,0){\strut{} 9.5}}%
      \put(3501,484){\makebox(0,0){\strut{} 10}}%
      \put(4194,484){\makebox(0,0){\strut{} 10.5}}%
      \put(4888,484){\makebox(0,0){\strut{} 11}}%
      \put(5581,484){\makebox(0,0){\strut{} 11.5}}%
      \put(6275,484){\makebox(0,0){\strut{} 12}}%
      \put(6407,704){\makebox(0,0)[l]{\strut{}-5}}%
      \put(6407,1439){\makebox(0,0)[l]{\strut{}-4}}%
      \put(6407,2174){\makebox(0,0)[l]{\strut{}-3}}%
      \put(6407,2909){\makebox(0,0)[l]{\strut{}-2}}%
      \put(6407,3644){\makebox(0,0)[l]{\strut{}-1}}%
      \put(6407,4379){\makebox(0,0)[l]{\strut{} 0}}%
      \put(3500,154){\makebox(0,0){\strut{}$f$ (GHz)}}%
      \put(3500,4709){\makebox(0,0){\strut{}$\epsilon^{\mathrm{eff}}_{rr}=$2.551, $\epsilon^{\mathrm{eff}}_{zz}=$4.269, $\theta=$0.500}}%
    }%
    \gplgaddtomacro\gplfronttext{%
      \csname LTb\endcsname%
      \put(5288,1317){\makebox(0,0)[r]{\strut{}infinite TWT}}%
      \csname LTb\endcsname%
      \put(5288,1097){\makebox(0,0)[r]{\strut{}short TWT with beam}}%
      \csname LTb\endcsname%
      \put(5288,877){\makebox(0,0)[r]{\strut{}short TWT without beam}}%
    }%
    \gplbacktext
    \put(0,0){\includegraphics{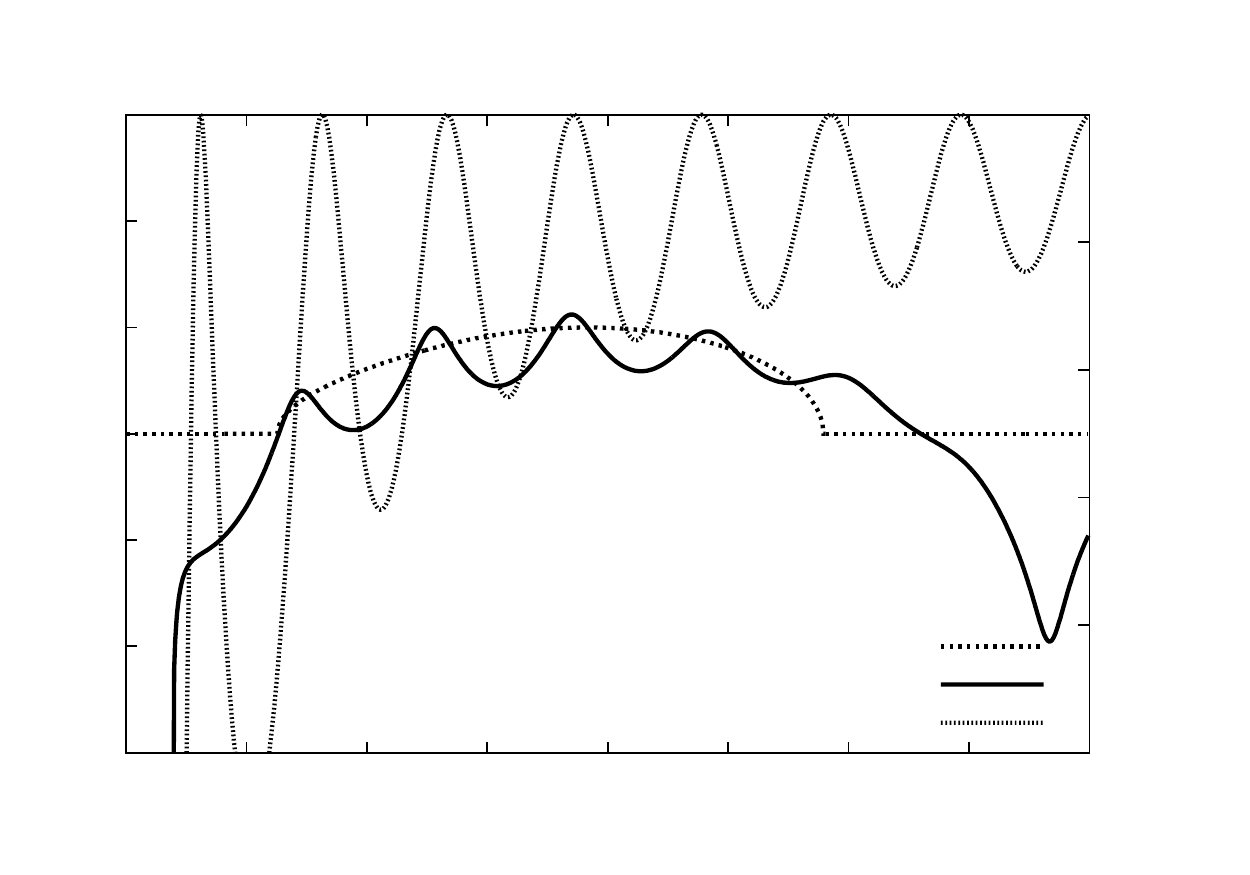}}%
    \gplfronttext
  \end{picture}%
\endgroup

%% file: Ellipse80Theta1.tex
\begingroup
  \makeatletter
  \providecommand\color[2][]{%
    \GenericError{(gnuplot) \space\space\space\@spaces}{%
      Package color not loaded in conjunction with
      terminal option `colourtext'%
    }{See the gnuplot documentation for explanation.%
    }{Either use 'blacktext' in gnuplot or load the package
      color.sty in LaTeX.}%
    \renewcommand\color[2][]{}%
  }%
  \providecommand\includegraphics[2][]{%
    \GenericError{(gnuplot) \space\space\space\@spaces}{%
      Package graphicx or graphics not loaded%
    }{See the gnuplot documentation for explanation.%
    }{The gnuplot epslatex terminal needs graphicx.sty or graphics.sty.}%
    \renewcommand\includegraphics[2][]{}%
  }%
  \providecommand\rotatebox[2]{#2}%
  \@ifundefined{ifGPcolor}{%
    \newif\ifGPcolor
    \GPcolorfalse
  }{}%
  \@ifundefined{ifGPblacktext}{%
    \newif\ifGPblacktext
    \GPblacktexttrue
  }{}%
  \let\gplgaddtomacro\g@addto@macro
  \gdef\gplbacktext{}%
  \gdef\gplfronttext{}%
  \makeatother
  \ifGPblacktext
    \def\colorrgb#1{}%
    \def\colorgray#1{}%
  \else
    \ifGPcolor
      \def\colorrgb#1{\color[rgb]{#1}}%
      \def\colorgray#1{\color[gray]{#1}}%
      \expandafter\def\csname LTw\endcsname{\color{white}}%
      \expandafter\def\csname LTb\endcsname{\color{black}}%
      \expandafter\def\csname LTa\endcsname{\color{black}}%
      \expandafter\def\csname LT0\endcsname{\color[rgb]{1,0,0}}%
      \expandafter\def\csname LT1\endcsname{\color[rgb]{0,1,0}}%
      \expandafter\def\csname LT2\endcsname{\color[rgb]{0,0,1}}%
      \expandafter\def\csname LT3\endcsname{\color[rgb]{1,0,1}}%
      \expandafter\def\csname LT4\endcsname{\color[rgb]{0,1,1}}%
      \expandafter\def\csname LT5\endcsname{\color[rgb]{1,1,0}}%
      \expandafter\def\csname LT6\endcsname{\color[rgb]{0,0,0}}%
      \expandafter\def\csname LT7\endcsname{\color[rgb]{1,0.3,0}}%
      \expandafter\def\csname LT8\endcsname{\color[rgb]{0.5,0.5,0.5}}%
    \else
      \def\colorrgb#1{\color{black}}%
      \def\colorgray#1{\color[gray]{#1}}%
      \expandafter\def\csname LTw\endcsname{\color{white}}%
      \expandafter\def\csname LTb\endcsname{\color{black}}%
      \expandafter\def\csname LTa\endcsname{\color{black}}%
      \expandafter\def\csname LT0\endcsname{\color{black}}%
      \expandafter\def\csname LT1\endcsname{\color{black}}%
      \expandafter\def\csname LT2\endcsname{\color{black}}%
      \expandafter\def\csname LT3\endcsname{\color{black}}%
      \expandafter\def\csname LT4\endcsname{\color{black}}%
      \expandafter\def\csname LT5\endcsname{\color{black}}%
      \expandafter\def\csname LT6\endcsname{\color{black}}%
      \expandafter\def\csname LT7\endcsname{\color{black}}%
      \expandafter\def\csname LT8\endcsname{\color{black}}%
    \fi
  \fi
  \setlength{\unitlength}{0.0500bp}%
  \begin{picture}(7200.00,5040.00)%
    \gplgaddtomacro\gplbacktext{%
      \csname LTb\endcsname%
      \put(594,704){\makebox(0,0)[r]{\strut{}-30}}%
      \put(594,1439){\makebox(0,0)[r]{\strut{}-20}}%
      \put(594,2174){\makebox(0,0)[r]{\strut{}-10}}%
      \put(594,2909){\makebox(0,0)[r]{\strut{} 0}}%
      \put(594,3644){\makebox(0,0)[r]{\strut{} 10}}%
      \put(594,4379){\makebox(0,0)[r]{\strut{} 20}}%
      \put(1455,484){\makebox(0,0){\strut{} 11}}%
      \put(2366,484){\makebox(0,0){\strut{} 12}}%
      \put(3277,484){\makebox(0,0){\strut{} 13}}%
      \put(4189,484){\makebox(0,0){\strut{} 14}}%
      \put(5100,484){\makebox(0,0){\strut{} 15}}%
      \put(6011,484){\makebox(0,0){\strut{} 16}}%
      \put(6143,949){\makebox(0,0)[l]{\strut{}-1.4}}%
      \put(6143,1439){\makebox(0,0)[l]{\strut{}-1.2}}%
      \put(6143,1929){\makebox(0,0)[l]{\strut{}-1}}%
      \put(6143,2419){\makebox(0,0)[l]{\strut{}-0.8}}%
      \put(6143,2909){\makebox(0,0)[l]{\strut{}-0.6}}%
      \put(6143,3399){\makebox(0,0)[l]{\strut{}-0.4}}%
      \put(6143,3889){\makebox(0,0)[l]{\strut{}-0.2}}%
      \put(6143,4379){\makebox(0,0)[l]{\strut{} 0}}%
      \put(3368,154){\makebox(0,0){\strut{}$f$ (GHz)}}%
      \put(3368,4709){\makebox(0,0){\strut{}$\epsilon^{\mathrm{eff}}_{rr}=$2.381, $\epsilon^{\mathrm{eff}}_{zz}=$1.894, $\theta=0.350$}}%
    }%
    \gplgaddtomacro\gplfronttext{%
      \csname LTb\endcsname%
      \put(5024,1317){\makebox(0,0)[r]{\strut{}infinite TWT}}%
      \csname LTb\endcsname%
      \put(5024,1097){\makebox(0,0)[r]{\strut{}short TWT with beam}}%
      \csname LTb\endcsname%
      \put(5024,877){\makebox(0,0)[r]{\strut{}short TWT without beam}}%
    }%
    \gplbacktext
    \put(0,0){\includegraphics{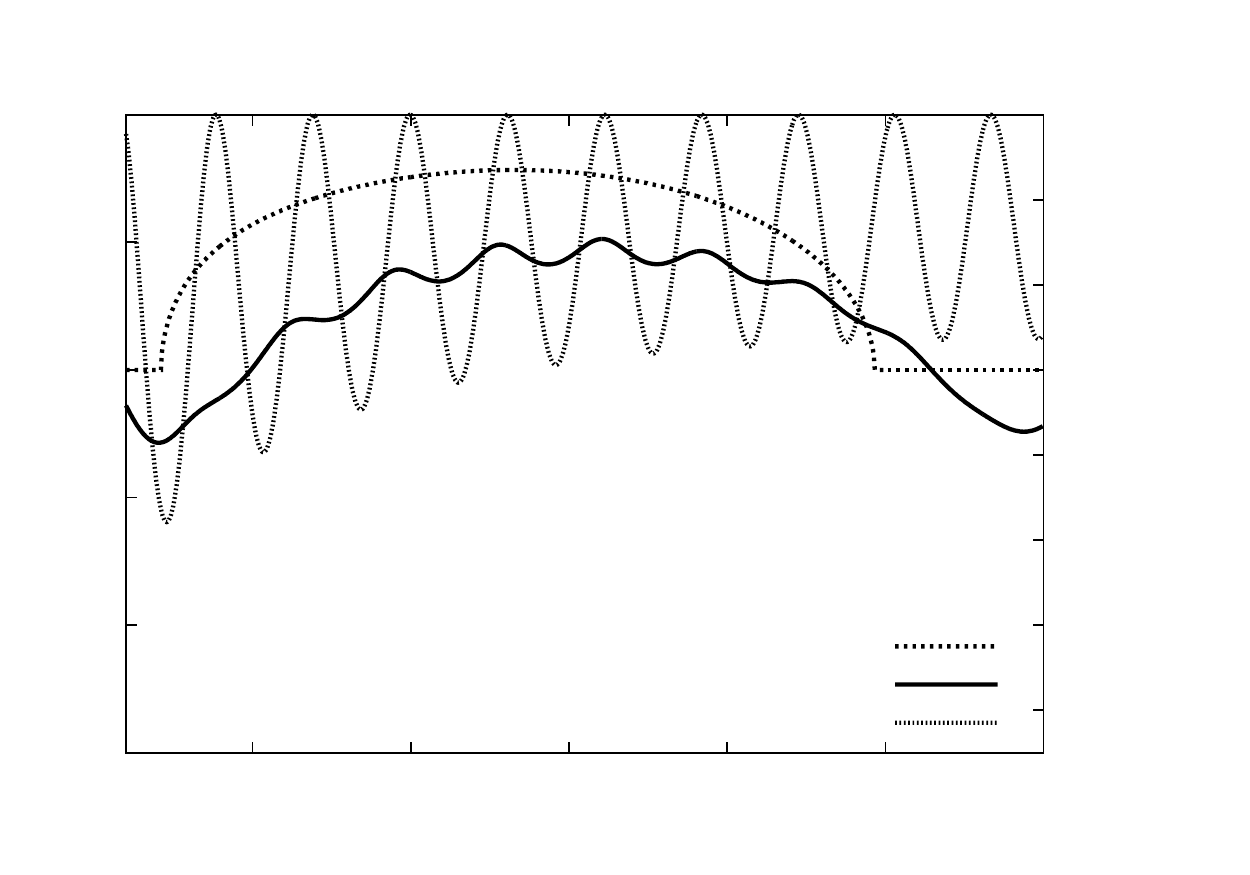}}%
    \gplfronttext
  \end{picture}%
\endgroup

%% file: Ellipse80Theta3.tex
\begingroup
  \makeatletter
  \providecommand\color[2][]{%
    \GenericError{(gnuplot) \space\space\space\@spaces}{%
      Package color not loaded in conjunction with
      terminal option `colourtext'%
    }{See the gnuplot documentation for explanation.%
    }{Either use 'blacktext' in gnuplot or load the package
      color.sty in LaTeX.}%
    \renewcommand\color[2][]{}%
  }%
  \providecommand\includegraphics[2][]{%
    \GenericError{(gnuplot) \space\space\space\@spaces}{%
      Package graphicx or graphics not loaded%
    }{See the gnuplot documentation for explanation.%
    }{The gnuplot epslatex terminal needs graphicx.sty or graphics.sty.}%
    \renewcommand\includegraphics[2][]{}%
  }%
  \providecommand\rotatebox[2]{#2}%
  \@ifundefined{ifGPcolor}{%
    \newif\ifGPcolor
    \GPcolorfalse
  }{}%
  \@ifundefined{ifGPblacktext}{%
    \newif\ifGPblacktext
    \GPblacktexttrue
  }{}%
  \let\gplgaddtomacro\g@addto@macro
  \gdef\gplbacktext{}%
  \gdef\gplfronttext{}%
  \makeatother
  \ifGPblacktext
    \def\colorrgb#1{}%
    \def\colorgray#1{}%
  \else
    \ifGPcolor
      \def\colorrgb#1{\color[rgb]{#1}}%
      \def\colorgray#1{\color[gray]{#1}}%
      \expandafter\def\csname LTw\endcsname{\color{white}}%
      \expandafter\def\csname LTb\endcsname{\color{black}}%
      \expandafter\def\csname LTa\endcsname{\color{black}}%
      \expandafter\def\csname LT0\endcsname{\color[rgb]{1,0,0}}%
      \expandafter\def\csname LT1\endcsname{\color[rgb]{0,1,0}}%
      \expandafter\def\csname LT2\endcsname{\color[rgb]{0,0,1}}%
      \expandafter\def\csname LT3\endcsname{\color[rgb]{1,0,1}}%
      \expandafter\def\csname LT4\endcsname{\color[rgb]{0,1,1}}%
      \expandafter\def\csname LT5\endcsname{\color[rgb]{1,1,0}}%
      \expandafter\def\csname LT6\endcsname{\color[rgb]{0,0,0}}%
      \expandafter\def\csname LT7\endcsname{\color[rgb]{1,0.3,0}}%
      \expandafter\def\csname LT8\endcsname{\color[rgb]{0.5,0.5,0.5}}%
    \else
      \def\colorrgb#1{\color{black}}%
      \def\colorgray#1{\color[gray]{#1}}%
      \expandafter\def\csname LTw\endcsname{\color{white}}%
      \expandafter\def\csname LTb\endcsname{\color{black}}%
      \expandafter\def\csname LTa\endcsname{\color{black}}%
      \expandafter\def\csname LT0\endcsname{\color{black}}%
      \expandafter\def\csname LT1\endcsname{\color{black}}%
      \expandafter\def\csname LT2\endcsname{\color{black}}%
      \expandafter\def\csname LT3\endcsname{\color{black}}%
      \expandafter\def\csname LT4\endcsname{\color{black}}%
      \expandafter\def\csname LT5\endcsname{\color{black}}%
      \expandafter\def\csname LT6\endcsname{\color{black}}%
      \expandafter\def\csname LT7\endcsname{\color{black}}%
      \expandafter\def\csname LT8\endcsname{\color{black}}%
    \fi
  \fi
  \setlength{\unitlength}{0.0500bp}%
  \begin{picture}(7200.00,5040.00)%
    \gplgaddtomacro\gplbacktext{%
      \csname LTb\endcsname%
      \put(594,704){\makebox(0,0)[r]{\strut{}-30}}%
      \put(594,1439){\makebox(0,0)[r]{\strut{}-20}}%
      \put(594,2174){\makebox(0,0)[r]{\strut{}-10}}%
      \put(594,2909){\makebox(0,0)[r]{\strut{} 0}}%
      \put(594,3644){\makebox(0,0)[r]{\strut{} 10}}%
      \put(594,4379){\makebox(0,0)[r]{\strut{} 20}}%
      \put(1348,484){\makebox(0,0){\strut{} 9.5}}%
      \put(2125,484){\makebox(0,0){\strut{} 10}}%
      \put(2902,484){\makebox(0,0){\strut{} 10.5}}%
      \put(3679,484){\makebox(0,0){\strut{} 11}}%
      \put(4457,484){\makebox(0,0){\strut{} 11.5}}%
      \put(5234,484){\makebox(0,0){\strut{} 12}}%
      \put(6011,484){\makebox(0,0){\strut{} 12.5}}%
      \put(6143,704){\makebox(0,0)[l]{\strut{}-3}}%
      \put(6143,1317){\makebox(0,0)[l]{\strut{}-2.5}}%
      \put(6143,1929){\makebox(0,0)[l]{\strut{}-2}}%
      \put(6143,2542){\makebox(0,0)[l]{\strut{}-1.5}}%
      \put(6143,3154){\makebox(0,0)[l]{\strut{}-1}}%
      \put(6143,3767){\makebox(0,0)[l]{\strut{}-0.5}}%
      \put(6143,4379){\makebox(0,0)[l]{\strut{} 0}}%
      \put(3368,154){\makebox(0,0){\strut{}$f$ (GHz)}}%
      \put(3368,4709){\makebox(0,0){\strut{}$\epsilon^{\mathrm{eff}}_{rr}=$3.383, $\epsilon^{\mathrm{eff}}_{zz}=$2.301, $\theta=$0.450}}%
    }%
    \gplgaddtomacro\gplfronttext{%
      \csname LTb\endcsname%
      \put(5024,1317){\makebox(0,0)[r]{\strut{}infinite TWT}}%
      \csname LTb\endcsname%
      \put(5024,1097){\makebox(0,0)[r]{\strut{}short TWT with beam}}%
      \csname LTb\endcsname%
      \put(5024,877){\makebox(0,0)[r]{\strut{}short TWT without beam}}%
    }%
    \gplbacktext
    \put(0,0){\includegraphics{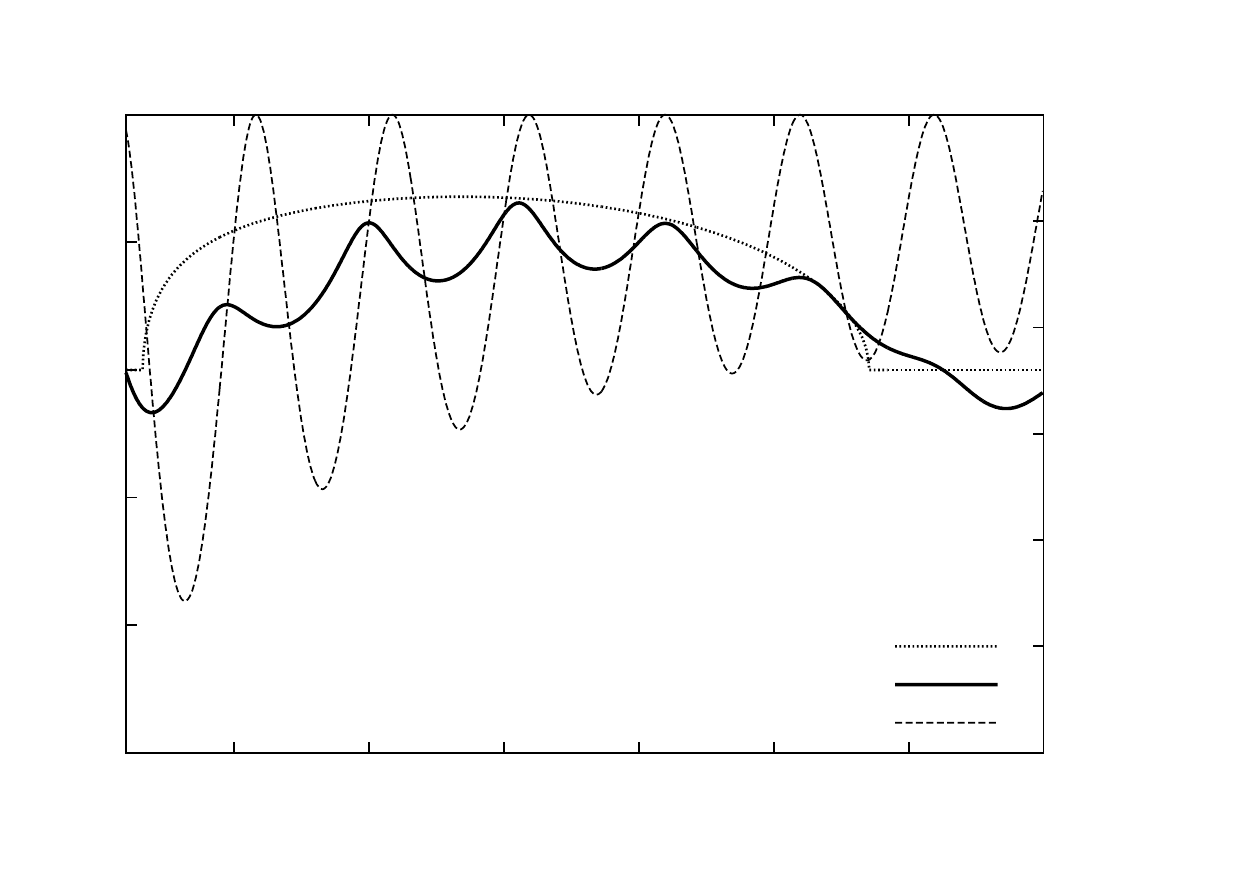}}%
    \gplfronttext
  \end{picture}%
\endgroup

%% file: Ellipse80Theta4.tex
\begingroup
  \makeatletter
  \providecommand\color[2][]{%
    \GenericError{(gnuplot) \space\space\space\@spaces}{%
      Package color not loaded in conjunction with
      terminal option `colourtext'%
    }{See the gnuplot documentation for explanation.%
    }{Either use 'blacktext' in gnuplot or load the package
      color.sty in LaTeX.}%
    \renewcommand\color[2][]{}%
  }%
  \providecommand\includegraphics[2][]{%
    \GenericError{(gnuplot) \space\space\space\@spaces}{%
      Package graphicx or graphics not loaded%
    }{See the gnuplot documentation for explanation.%
    }{The gnuplot epslatex terminal needs graphicx.sty or graphics.sty.}%
    \renewcommand\includegraphics[2][]{}%
  }%
  \providecommand\rotatebox[2]{#2}%
  \@ifundefined{ifGPcolor}{%
    \newif\ifGPcolor
    \GPcolorfalse
  }{}%
  \@ifundefined{ifGPblacktext}{%
    \newif\ifGPblacktext
    \GPblacktexttrue
  }{}%
  \let\gplgaddtomacro\g@addto@macro
  \gdef\gplbacktext{}%
  \gdef\gplfronttext{}%
  \makeatother
  \ifGPblacktext
    \def\colorrgb#1{}%
    \def\colorgray#1{}%
  \else
    \ifGPcolor
      \def\colorrgb#1{\color[rgb]{#1}}%
      \def\colorgray#1{\color[gray]{#1}}%
      \expandafter\def\csname LTw\endcsname{\color{white}}%
      \expandafter\def\csname LTb\endcsname{\color{black}}%
      \expandafter\def\csname LTa\endcsname{\color{black}}%
      \expandafter\def\csname LT0\endcsname{\color[rgb]{1,0,0}}%
      \expandafter\def\csname LT1\endcsname{\color[rgb]{0,1,0}}%
      \expandafter\def\csname LT2\endcsname{\color[rgb]{0,0,1}}%
      \expandafter\def\csname LT3\endcsname{\color[rgb]{1,0,1}}%
      \expandafter\def\csname LT4\endcsname{\color[rgb]{0,1,1}}%
      \expandafter\def\csname LT5\endcsname{\color[rgb]{1,1,0}}%
      \expandafter\def\csname LT6\endcsname{\color[rgb]{0,0,0}}%
      \expandafter\def\csname LT7\endcsname{\color[rgb]{1,0.3,0}}%
      \expandafter\def\csname LT8\endcsname{\color[rgb]{0.5,0.5,0.5}}%
    \else
      \def\colorrgb#1{\color{black}}%
      \def\colorgray#1{\color[gray]{#1}}%
      \expandafter\def\csname LTw\endcsname{\color{white}}%
      \expandafter\def\csname LTb\endcsname{\color{black}}%
      \expandafter\def\csname LTa\endcsname{\color{black}}%
      \expandafter\def\csname LT0\endcsname{\color{black}}%
      \expandafter\def\csname LT1\endcsname{\color{black}}%
      \expandafter\def\csname LT2\endcsname{\color{black}}%
      \expandafter\def\csname LT3\endcsname{\color{black}}%
      \expandafter\def\csname LT4\endcsname{\color{black}}%
      \expandafter\def\csname LT5\endcsname{\color{black}}%
      \expandafter\def\csname LT6\endcsname{\color{black}}%
      \expandafter\def\csname LT7\endcsname{\color{black}}%
      \expandafter\def\csname LT8\endcsname{\color{black}}%
    \fi
  \fi
  \setlength{\unitlength}{0.0500bp}%
  \begin{picture}(7200.00,5040.00)%
    \gplgaddtomacro\gplbacktext{%
      \csname LTb\endcsname%
      \put(594,704){\makebox(0,0)[r]{\strut{}-30}}%
      \put(594,1439){\makebox(0,0)[r]{\strut{}-20}}%
      \put(594,2174){\makebox(0,0)[r]{\strut{}-10}}%
      \put(594,2909){\makebox(0,0)[r]{\strut{} 0}}%
      \put(594,3644){\makebox(0,0)[r]{\strut{} 10}}%
      \put(594,4379){\makebox(0,0)[r]{\strut{} 20}}%
      \put(1588,484){\makebox(0,0){\strut{} 9}}%
      \put(2694,484){\makebox(0,0){\strut{} 9.5}}%
      \put(3800,484){\makebox(0,0){\strut{} 10}}%
      \put(4905,484){\makebox(0,0){\strut{} 10.5}}%
      \put(6011,484){\makebox(0,0){\strut{} 11}}%
      \put(6143,704){\makebox(0,0)[l]{\strut{}-3}}%
      \put(6143,1317){\makebox(0,0)[l]{\strut{}-2.5}}%
      \put(6143,1929){\makebox(0,0)[l]{\strut{}-2}}%
      \put(6143,2542){\makebox(0,0)[l]{\strut{}-1.5}}%
      \put(6143,3154){\makebox(0,0)[l]{\strut{}-1}}%
      \put(6143,3767){\makebox(0,0)[l]{\strut{}-0.5}}%
      \put(6143,4379){\makebox(0,0)[l]{\strut{} 0}}%
      \put(3368,154){\makebox(0,0){\strut{}$f$ (GHz)}}%
      \put(3368,4709){\makebox(0,0){\strut{}$\epsilon^{\mathrm{eff}}_{rr}=$4.269, $\epsilon^{\mathrm{eff}}_{zz}=$2.551, $\theta=$0.500}}%
    }%
    \gplgaddtomacro\gplfronttext{%
      \csname LTb\endcsname%
      \put(5024,1317){\makebox(0,0)[r]{\strut{}infinite TWT}}%
      \csname LTb\endcsname%
      \put(5024,1097){\makebox(0,0)[r]{\strut{}short TWT with beam}}%
      \csname LTb\endcsname%
      \put(5024,877){\makebox(0,0)[r]{\strut{}short TWT without beam}}%
    }%
    \gplbacktext
    \put(0,0){\includegraphics{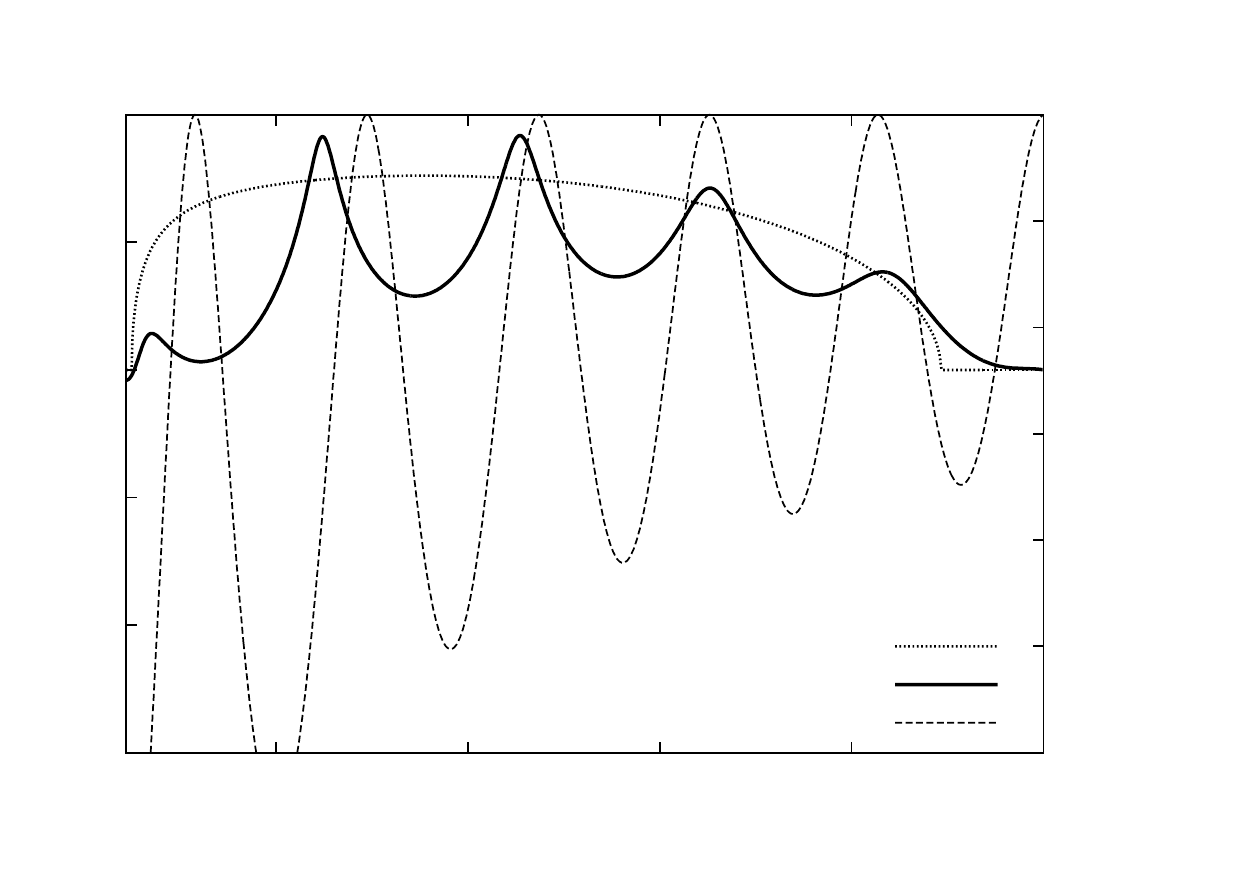}}%
    \gplfronttext
  \end{picture}%
\endgroup

%% file: Paper-Preprint.bbl
\begin{thebibliography}{8}%
\makeatletter
\providecommand \@ifxundefined [1]{%
 \@ifx{#1\undefined}
}%
\providecommand \@ifnum [1]{%
 \ifnum #1\expandafter \@firstoftwo
 \else \expandafter \@secondoftwo
 \fi
}%
\providecommand \@ifx [1]{%
 \ifx #1\expandafter \@firstoftwo
 \else \expandafter \@secondoftwo
 \fi
}%
\providecommand \natexlab [1]{#1}%
\providecommand \enquote  [1]{``#1''}%
\providecommand \bibnamefont  [1]{#1}%
\providecommand \bibfnamefont [1]{#1}%
\providecommand \citenamefont [1]{#1}%
\providecommand \href@noop [0]{\@secondoftwo}%
\providecommand \href [0]{\begingroup \@sanitize@url \@href}%
\providecommand \@href[1]{\@@startlink{#1}\@@href}%
\providecommand \@@href[1]{\endgroup#1\@@endlink}%
\providecommand \@sanitize@url [0]{\catcode `\\12\catcode `\$12\catcode
  `\&12\catcode `\#12\catcode `\^12\catcode `\_12\catcode `\%12\relax}%
\providecommand \@@startlink[1]{}%
\providecommand \@@endlink[0]{}%
\providecommand \url  [0]{\begingroup\@sanitize@url \@url }%
\providecommand \@url [1]{\endgroup\@href {#1}{\urlprefix }}%
\providecommand \urlprefix  [0]{URL }%
\providecommand \Eprint [0]{\href }%
\providecommand \doibase [0]{http://dx.doi.org/}%
\providecommand \selectlanguage [0]{\@gobble}%
\providecommand \bibinfo  [0]{\@secondoftwo}%
\providecommand \bibfield  [0]{\@secondoftwo}%
\providecommand \translation [1]{[#1]}%
\providecommand \BibitemOpen [0]{}%
\providecommand \bibitemStop [0]{}%
\providecommand \bibitemNoStop [0]{.\EOS\space}%
\providecommand \EOS [0]{\spacefactor3000\relax}%
\providecommand \BibitemShut  [1]{\csname bibitem#1\endcsname}%
\let\auto@bib@innerbib\@empty
\bibitem [{\citenamefont {Sch\"achter}(2011)}]{Schachter}%
  \BibitemOpen
  \bibfield  {author} {\bibinfo {author} {\bibfnamefont {L.}~\bibnamefont
  {Sch\"achter}},\ }\href {\doibase 10.1007/978-3-642-19848-9} {\emph {\bibinfo
  {title} {Beam-wave interaction in periodic and quasi-periodic structures}}},\
  \bibinfo {edition} {2nd}\ ed.,\ Particle acceleration and detection\
  (\bibinfo  {publisher} {Springer},\ \bibinfo {address} {Berlin},\ \bibinfo
  {year} {2011})\BibitemShut {NoStop}%
\bibitem [{\citenamefont {Sch\"achter}, \citenamefont {Nation},\ and\
  \citenamefont {Kerslick}(1990)}]{Nation}%
  \BibitemOpen
  \bibfield  {author} {\bibinfo {author} {\bibfnamefont {L.}~\bibnamefont
  {Sch\"achter}}, \bibinfo {author} {\bibfnamefont {J.~A.}\ \bibnamefont
  {Nation}}, \ and\ \bibinfo {author} {\bibfnamefont {G.}~\bibnamefont
  {Kerslick}},\ }\href {\doibase http://dx.doi.org/10.1063/1.346963} {\bibfield
   {journal} {\bibinfo  {journal} {Journal of Applied Physics}\ }\textbf
  {\bibinfo {volume} {68}},\ \bibinfo {pages} {5874} (\bibinfo {year}
  {1990})}\BibitemShut {NoStop}%
\bibitem [{\citenamefont {Shiffler}\ \emph {et~al.}(2010)\citenamefont
  {Shiffler}, \citenamefont {Luginsland}, \citenamefont {French},\ and\
  \citenamefont {Watrous}}]{Shiffler}%
  \BibitemOpen
  \bibfield  {author} {\bibinfo {author} {\bibfnamefont {D.}~\bibnamefont
  {Shiffler}}, \bibinfo {author} {\bibfnamefont {J.}~\bibnamefont
  {Luginsland}}, \bibinfo {author} {\bibfnamefont {D.}~\bibnamefont {French}},
  \ and\ \bibinfo {author} {\bibfnamefont {J.}~\bibnamefont {Watrous}},\ }\href
  {\doibase 10.1109/TPS.2010.2046914} {\bibfield  {journal} {\bibinfo
  {journal} {IEEE Transactions on Plasma Science}\ }\textbf {\bibinfo {volume}
  {38}},\ \bibinfo {pages} {1462} (\bibinfo {year} {2010})}\BibitemShut
  {NoStop}%
\bibitem [{\citenamefont {Bensoussan}, \citenamefont {Lions},\ and\
  \citenamefont {Papanicolaou}(1978)}]{papanicolaou}%
  \BibitemOpen
  \bibfield  {author} {\bibinfo {author} {\bibfnamefont {A.}~\bibnamefont
  {Bensoussan}}, \bibinfo {author} {\bibfnamefont {J.-L.}\ \bibnamefont
  {Lions}}, \ and\ \bibinfo {author} {\bibfnamefont {G.}~\bibnamefont
  {Papanicolaou}},\ }\href@noop {} {\emph {\bibinfo {title} {Asymptotic
  Analysis for Periodic Structures}}}\ (\bibinfo  {publisher} {North-Holland
  Pub. Co.},\ \bibinfo {address} {Amsterdam, New York},\ \bibinfo {year}
  {1978})\BibitemShut {NoStop}%
\bibitem [{\citenamefont {Sanchez-Palencia}(1980)}]{sanchezpalencia}%
  \BibitemOpen
  \bibfield  {author} {\bibinfo {author} {\bibfnamefont {E.}~\bibnamefont
  {Sanchez-Palencia}},\ }\href@noop {} {\emph {\bibinfo {title} {Non
  Homogeneous Media and Vibration Theory}}},\ Lecture Notes in Physics, Vol.
  127\ (\bibinfo  {publisher} {Springer},\ \bibinfo {address} {Heidelberg},\
  \bibinfo {year} {1980})\BibitemShut {NoStop}%
\bibitem [{\citenamefont {Chu}\ and\ \citenamefont {Jackson}(1947)}]{Chu}%
  \BibitemOpen
  \bibfield  {author} {\bibinfo {author} {\bibfnamefont {L.~J.}\ \bibnamefont
  {Chu}}\ and\ \bibinfo {author} {\bibfnamefont {D.}~\bibnamefont {Jackson}},\
  }\href@noop {} {\enquote {\bibinfo {title} {Field theory of traveling wave
  tubes},}\ }\bibinfo {type} {Tech. Rep.}\ \bibinfo {number} {38}\ (\bibinfo
  {institution} {Research Laboratory of Electronics Massachusetts Institute of
  Technology},\ \bibinfo {year} {1947})\BibitemShut {NoStop}%
\bibitem [{\citenamefont {Pierce}(1951)}]{Pierce}%
  \BibitemOpen
  \bibfield  {author} {\bibinfo {author} {\bibfnamefont {J.~R.}\ \bibnamefont
  {Pierce}},\ }\href {\doibase 10.1002/j.1538-7305.1951.tb03672.x} {\bibfield
  {journal} {\bibinfo  {journal} {Bell System Technical Journal}\ }\textbf
  {\bibinfo {volume} {30}},\ \bibinfo {pages} {626} (\bibinfo {year}
  {1951})}\BibitemShut {NoStop}%
\bibitem [{\citenamefont {Pierce}(1947)}]{Pierce2}%
  \BibitemOpen
  \bibfield  {author} {\bibinfo {author} {\bibfnamefont {J.~R.}\ \bibnamefont
  {Pierce}},\ }\href@noop {} {\bibfield  {journal} {\bibinfo  {journal} {Proc.
  IRE}\ }\textbf {\bibinfo {volume} {35}},\ \bibinfo {pages} {111} (\bibinfo
  {year} {1947})}\BibitemShut {NoStop}%
\end{thebibliography}%


%
